\documentclass[a4paper,11pt]{article}
\pdfoutput=1 

\usepackage{jcappub} 

\usepackage[T1]{fontenc} 
\usepackage{todonotes}

\usepackage{subcaption}
\usepackage{tikz}
\usetikzlibrary{positioning}
\usepackage{pgfplots}
\usetikzlibrary{decorations}
\usetikzlibrary {shapes.geometric} 
\usetikzlibrary{calc}

\usepackage{aas_macros}
\usepackage{multirow}

\title{\boldmath Learning Optimal and Interpretable Summary Statistics of Galaxy Catalogs with SBI}

\author[a,b]{Kai Lehman,}
\author[c]{Sven Krippendorf,}
\author[a,b,d]{Jochen Weller,}
\author[a,e]{and Klaus Dolag}

\affiliation[a]{Universit\"ats-Sternwarte M\"unchen, Fakult\"at f\"ur Physik, Ludwig-Maximilians-Universit\"at, Scheinerstr.~1, 81679 M\"unchen, Germany}
\affiliation[b]{Excellence Cluster ORIGINS, Boltzmannstr.~2, 85748 Garching, Germany}
\affiliation[c]{University of Cambridge, Cavendish Laboratory and Department of Applied Mathematics and Theoretical Physics, Cambridge CB3 0WA, United Kingdom}
\affiliation[d]{Max-Planck-Institut f\"ur extraterrestrische Physik, Giessenbachstr.~1, 85748 Garching, Germany}
\affiliation[e]{Max-Planck-Institut f\"ur Astrophysik, Karl-Schwarzschild-Str.~1, 85741 Garching, Germany}

\emailAdd{kai.lehman@physik.lmu.de}
\emailAdd{slk38@cam.ac.uk}
\emailAdd{jochen.weller@lmu.de}
\emailAdd{dolag@usm.lmu.de}

\abstract{How much cosmological information can we reliably extract from existing and upcoming large-scale structure observations? Many summary statistics fall short in describing the non-Gaussian nature of the late-time Universe in comparison to existing and upcoming measurements. In this article we demonstrate that we can identify optimal summary statistics and that we can link them with existing summary statistics. Using simulation based inference (SBI) with automatic data-compression, we learn summary statistics for galaxy catalogs in the context of cosmological parameter estimation. By construction these summary statistics do not require the ability to write down an explicit likelihood. We demonstrate that they can be used for efficient parameter inference. These summary statistics offer a new avenue for analyzing different simulation models for baryonic physics with respect to their relevance for the resulting cosmological features. The learned summary statistics are low-dimensional, feature the underlying simulation parameters, and are similar across different network architectures. To link our models, we identify the relevant scales associated to our summary statistics (e.g. in the range of modes between $k= 5 - 30 h/\mathrm{Mpc}$) and we are able to match the summary statistics to underlying simulation parameters across various simulation models.}

\begin{document}
\maketitle
\flushbottom

\section{Introduction}
\label{sec:intro}

The field of Cosmology has firmly entered the data rich era. Current~\cite{sptsz_15, hetdex_08, kids_13, hsc_15, des_16, boss_13, eboss_16,erosita_21,desi_13} and future surveys~\citep{lsst_09,euclid_11,4most_12,pfs_14,ska_15,roman_15} will reach immense statistical precision. Due to this development, systematics become the dominant source of error for cosmological inference in these surveys. This brings up a key question: Can we reliably use the entirety of the cosmological information captured by future surveys? One issue in answering this question is given by baryonic feedback~\citep{rudd_08}. Baryonic physics in active galactic nuclei (AGN) and stars can produce strong outflows of energetic gas, which changes the matter distribution on scales up to a few Mpc~\citep{chisari_19}. As these processes originate on scales much smaller than what is usually considered in cosmological simulations, these effects prove to be hard to model and the resulting suppression of the matter power spectrum is uncertain~\citep{vandaalen_20,villaescusa_21}. Hydrodynamic simulations which attempt to implement these effects via stochastic so-called subgrid physics models, therefore vary widely in the effect that baryonic physics has on the matter distribution~\citep{chisari_19}. 

In light of this, many efforts have been led to identify observables in hydrodynamic simulations, which are indicative of feedback strength. Promising probes include the (unobservable) baryon fraction in halos~\citep{vandaalen_20}, the thermal and kinetic Sunyaev-Zel'dovich effect~\citep{moser_22,wadekar_23,pandey_23,delgado_23,to_24,bigwood_24,mccarthy_24}, X-ray observations of clusters~\citep{bahar_24}, and recently the dispersion measure of fast radio bursts~\citep{theis_24,medlock_24}. While these directions of research seem very promising, it is not clear that these observables (or some linear combination) will be the {\it optimal} choice for constraining feedback in the setting of cosmological inference. It is more likely that an optimal observable would be created from a non-linear combination of different parts of the data vector.
Likewise, for the extraction of cosmological information, human-selected statistics capturing some non-Gaussian features are also not guaranteed to be optimal (see Refs.~\cite{chiang_14,chiang_15,halder_21,halder_23,friedrich_18,gruen_18} for examples). In the case of galaxy clustering, Ref.~\cite{nguyen_24} have recently shown that lower order n-point functions fail to capture a significant amount of cosmological information, which calls for the search for highly informative summaries.

The question of the ideal representation of some data can be asymptotically solved by means of an optimization problem: Given a set of observables, for instance a galaxy catalog, we can pose an inference problem to automatically identify important features in the data relevant for cosmological inference. This optimization task would be naturally solvable with the help of neural networks. Furthermore, this can be set up in such a way to guarantee maximum information extraction which results in the tightest constraints on the parameters of interest. We explore this approach in this paper.

On the machine learning side, advances over the last few years, especially in the field of simulation-based inference (SBI)~\citep{alsing_19,cranmer_20,papamakarios_17} have laid the groundwork for such a task. In this framework inference is possible solely with the help of a simulator and a probability density estimator (e.g.~to estimate the posterior of cosmological parameters). The likelihood is implicitly shifted into the simulation and need not be specified in the inference process. This is of utmost importance for machine-learned summary statistics, for which we cannot model the likelihood. Even in the case of hand-crafted summaries, the inability to write down a likelihood severely limits cosmological inference tasks. The detail of the employed density estimator may vary. Early implementations of SBI made use of traditional density estimators such as approximate bayesian computation (ABC)~\citep{rubin_84, beaumont_02, akaret_15}. However, ABC is lacking in computational efficiency and is especially struggling with higher dimensional targets. As a result, these older methods have largely been replaced by neural network accelerated density estimators such as normalizing flows~\citep{papamakarios_19}. The SBI framework has been successfully applied to astrophysical ~\citep{legin_21,anau_22,hahn_22b,khullar_22,alvey_23,alvey_23b,bhardwaj_23,crisostomi_23,darc_23,dax_23,dupourque_23,gebhard_23,graber_23,hahn_23c,prelogovic_23,vasist_23,wang_23,zhao_23,alvey_24,christy_24,coogan_24,moser_24,sun_24,xiong_24} and cosmological~\citep{dimitriou_22,hahn_22,reza_22, zhang_22, zhao_22, anau_23, hahn_23a, hahn_23b, blancard_23, gagnon_23, jo_23, karchev_23, lemos_23, list_23, modi_23, modi_23b, saxena_23, zheng_23, baxter_24, hou_24, massara_24, tucci_23, yongseok_23, davies_24, gatti_24, gatti_24b, jeffrey_24, schosser_24, wietersheim_24} inference problems to great detail showing great overall performance and efficiency. In this work we utilize SBI to address cosmological inference on galaxy catalogs where standard perturbative EFT approaches can no longer be applied.

Beyond increasing the amount of information which can be used for cosmological inference, our approach also has advantages with respect to hydrodynamical modeling itself. Our approach provides a novel evaluation tool for simulations as it enables us to compare the differences among different simulation models. This is particularly relevant with respect to selecting relevant simulation parameters, reducing the number of simulations necessary for cosmological inference. In practice, this method provides a quantitative handle on priors of baryonic feedback and galaxy formation models. It can also be used for calibration of simulations using observations as discussed in Ref.~\cite{yongseok_23}.

In order to perform inference within this framework, the data is compressed to a summary vector~\citep{blum_12}. As proposed by Ref.~\cite{blum_08}, feed forward neural network have proven to be a powerful way of achieving compression for SBI. A popular way to perform this compression consists of training a neural network regressor with a mean square error loss function to predict the input parameters of a simulation as in Refs.~\cite{lemos_23,jeffrey_24}. The approach is based on previous work by Ref.~\cite{fearnhead_10} in the context of linear regression which was extended to a neural network based approach by Ref.~\cite{jiang_15}. This prediction of the input simulation parameters would then be considered a ``summary statistic'' with which SBI can be performed. This way of compression has some immediate disadvantages. The resulting data vector is not necessarily information optimal which restricts the statistical power of the subsequent inference process. Secondly, the dimensionality of the summary vector is fixed to the dimensionality of the input parameter vector, hindering the network from finding a potentially simpler higher dimensional embedding space. Furthermore, the mean square error is only the maximum likelihood estimate of normally distributed variables~\citep{murphy_22}. This attenuates the power of SBI, which lies in its ability to perform inference for explicitly unknown likelihoods. Ref.~\cite{charnock_18} have proposed information maximizing neural networks, a way of compression in which the Gaussian approximation of the Fisher information is maximized at one or multiple fiducial points in parameter space. This is a non-linear extension of the \texttt{MOPED} algorithm~\citep{heavens_00}. While this approach does maximize the information contained in the summary vector by construction, it is computationally expensive if trained at multiple fiducial points. This is necessary as Ref.~\cite{prelogovic_24} have explicitly shown that if trained at only a single point, the information content of the learned summary varies widely over the entire prior volume. Another drawback to this compression scheme is that it requires covariance estimates and finite differences to compute the Fisher information. In order to enable these computations additional simulations have to be run. Recently, Ref.~\cite{makinen_24} have applied such a scheme to find information maximizing summary statistics complementary to the power spectrum from weak lensing convergence maps, and found a significant improvement in constraining power. In comparison to their work we attempt to discover summaries which are not only optimal at a single point in parameter space, but for the entire simulation range.

Another alternative is proposed by Ref.~\cite{jeffrey_21}, who maximize the variational mutual information between the summary and the cosmological parameters. Effectively, this follows the previously published framework by Ref.~\cite{radev_20}. The approach has been applied recently by Refs.~\cite{sharma_24,lanzieri_24} who found this method to be very performative, even yielding sufficient summaries in idealized settings. Ref.~\cite{makinen_24b} have also used this approach to find summaries complementary to human-made statistics. In all these works, the two-dimensional input data is given to a convolutional neural network which performs the compression. The compressed data vector is then directly fed into a simulation-based inference pipeline which computes the variational mutual information by means of the expected Kullback-Leibler divergence. In this work we extend on these efforts and combine a graph neural network for compression with the traditional SBI framework. Such a graph perspective is natural in the context of survey catalogs which are straight-forwardly translated into a graph where each catalog entry denotes a node and the relative position is encoded in the edges. A key advantage of this approach is that we gain access to the compressed data vector resulting from the graph, i.e.~a machine-learned summary statistic. We investigate the inference process itself by examining this summary. In doing so we identify not only scales that the machine identifies as important, but also assess which subgrid physics processes the pipeline has learned automatically. In contrast to other works that have investigated the importance of feedback parameters in these simulations, we can make statements not only in regard to any given human-made summary statistic, but in regard to the total cosmological information.

To facilitate data efficiency, following the geometric deep learning paradigm~\citep{bronstein_21}, it is strongly advised to reflect important symmetries also in the machine learning model. Mathematical graphs provide for a natural way of achieving this. They consist of nodes connected by edges and can be constructed in such a way that translation and rotation symmetries of the underlying data are captured. Graph neural networks (GNNs)~\citep{battaglia_18,bronstein_21} operate on graphs and have already been explored for their capabilities in cosmological inference~\citep{anagnostidis_22,villanueva_22,shao_22,makinen_22,desanti_23,desanti_23b,massara_23,roncoli_23,shao_23}. An important feature of this setup is that it enables the input of variably sized data into the GNN as long as it can be structured into a graph. This architecture has been mostly explored with the moment network loss function~\citep{jeffrey_20}, which predicts the first two moments of the marginal parameter posterior. Refs.~\cite{desanti_23} and \cite{desanti_23b} in particular have shown that using positional information and velocity information in one dimension enables robust inference across feedback models. We build upon these works by coupling this graph architecture with a full SBI framework based on masked autoregressive flows~\citep{papamakarios_17}. The resulting architecture is similar to that of Ref.~\cite{nguyen_23} who have used such an approach to recover the dark matter profile of dwarf galaxies and to that of Ref.~\cite{ho_24}. Apart from a different use case here, our architecture also differs as we treat the number of learned summaries as an optimizable hyperparameter.

Once the architecture is trained, we are not only interested in the constraining power we achieve, but also in the Physics contained in the machine-learned summary. Concerning our interpretability efforts, we would like to mention similar works on interpretability for machine learning in the field of Cosmology by Refs.~\cite{lucie-smith_22,gong_24,lucie-smith_24,piras_24}. In these works additional constraints are set on the respective machine learning architectures to either disentangle the latent space or link the resulting latent variables with theoretical quantities a priori. We do not follow such an approach as to not prohibit the inference network from maximum performance. Instead, we investigate how much insight into the summaries we can get from purely analyzing them a posteriori. We leave the exploration of a priori more easily interpretable machine learning architectures as are used in these works up to further study.

This paper is structured as follows:
In Section~\ref{sec:simulations} we describe the simulations we use in our analysis. Section~\ref{sec:methodology} describes our inference pipeline and how we train and validate our models. We present our results in Section~\ref{sec:results}, before concluding in Section~\ref{sec:conclusions}.

\section{The CAMEL Simulations} \label{sec:simulations}

In order to find machine-learned summary statistics and perform mock inference we make use of the CAMELS suite of simulations~\citep{villaescusa_21}. For training statistical models over a range of cosmological and astrophysical parameters, we employ the latin-hypercube (LH) set, which consists of 3000 hydrodynamical simulations of side length $25\ \mathrm{Mpc}\,h^{-1}$. We note that the relatively small box size is the main limitation of these simulations and throughout this work (for cosmological parameter inference). While we are missing out on a lot of cosmological information from large scale modes, the goal of this work is to examine how much additional information can be gained from smaller scales. The simulations used in this work are grouped into three sets of 1000 simulations each, where each set employs a different subgrid physics model. The three models used are:

\begin{itemize}
    \item {\bf IllustrisTNG~\citep{weinberger_17,pillepich_18} model based on the Arepo~\citep{springel_10,weinberger_20} code.} Galactic winds from stellar feedback are implemented with temporarily hydrodynamically decoupled particles which are stochastically and isotropically ejected from star forming gas based on Ref.~\cite{springel_03}. A small part of the total energy is ejected thermally. The total ejected energy per unit star formation and the wind speed are modulated by the parameters $A_\mathrm{SN1}$ and $A_\mathrm{SN2}$ respectively. AGN feedback is modeled with a kinetic, thermal and radiative mode. While the radiative component is always active, the kinetic and thermal modes operate separately depending on the accretion rate. At high accretion rate the thermal mode deposits energy into a `feedback sphere'. In the low accretion kinetic mode, bursts of kinetic energy are injected at discrete times into a random direction in the `feedback sphere'. The amount of energy injected is modulated by $A_\mathrm{AGN1}$ and the minimum energy for a burst to occur is modulated by $A_\mathrm{AGN2}$.
    \item {\bf SIMBA~\citep{dave_19} model based on the Gizmo~\citep{hopkins_15} code.} Stellar feedback drives galactic winds kinetically via hydrodynamically decoupled wind particles. Star-forming gas elements are ejected according to a prescribed mass loading factor which is modulated by $A_\mathrm{SN1}$. The wind velocity is regulated by $A_\mathrm{SN2}$. Black hole feedback is modeled with two modes informed by observations~\citep{heckman_14}, one radiative `QSO' mode and one kinetic jet mode. Gas elements are ejected in the directions parallel and anti-parallel to the angular momentum vector of surrounding gas. The total momentum flux in both modes is regulated by $A_\mathrm{AGN1}$, and the maximum outflow velocity in the jet mode is modulated by $A_\mathrm{AGN2}$.
    \item {\bf Astrid~\citep{bird_22} model based on MP-Gadget~\citep{feng_18} with some additional modifications according to Refs.~\cite{ni_22,ni_23}.} Stellar feedback drives galactic winds where $A_\mathrm{SN1}$ modulates the energy ejection rate per unit star formation and $A_\mathrm{SN2}$ modulates the wind speed. AGN feedback is modeled with a kinetic and thermal feedback mode depending on the accretion rate. At low accretion rates, kinetic energy is deposited with the modulation factor $A_\mathrm{AGN1}$ and released in bursts once an energy threshold is met. The resulting kick to particles within the feedback sphere is in a random direction. Due to a lower kinetic feedback efficiency limit and more strict activation criterion kinetic feedback is overall milder than the IllustrisTNG model. At high accretion rates, energy is thermally injected in the feedback sphere, modulated by the parameter $A_\mathrm{AGN2}$. 
\end{itemize}

Each simulation is run with different values for cosmological and astrophysical parameters, which were sampled according to a latin-hypercube layout. The varied cosmological parameters are $\Omega_m$ and $\sigma_8$ and consequently these will be the parameters of interest in our inference task. Astrophysical effects are parametrized in the underlying feedback model with two parameters $A_\mathrm{AGN1}$ and $A_\mathrm{AGN2}$ regulating AGN feedback and two stellar feedback parameters $A_\mathrm{SN1}$ and $A_\mathrm{SN2}$ as described above. We note again, that the same astrophysical parameter does not necessarily correspond to the same physical process across feedback models. For example, $A_\mathrm{AGN2}$ regulates the ejection speed of AGN bursts in the IllustrisTNG simulations and the speed of kinetic jets in the SIMBA simulations. In the Astrid simulations on the other hand, the parameter describes the efficiency of thermal AGN feedback. Consequently, there is some caution warranted when comparing the impact of a given parameter across feedback implementations. The fiducial values of all parameters and the prior volume of the LH set are given in Table \ref{tab:params}. Each hydrodynamical simulation in the CAMELS suite has a corresponding n-body dark matter only run with identical initial conditions. These simulations are used to compute the power spectrum suppression caused by baryons, for which we compare power spectra from the hydrodynamical simulations and their n-body counterparts. For the inference task we use subhalo catalogs produced by the \texttt{SubFind} algorithm~\citep{springel_01} modified to take baryons into account~\citep{dolag_09}.

\begin{table}[tbp]
\centering
\begin{tabular}{|c|c|c|c|}
\hline
Parameter & Fiducial Value & LH Lower Bound & LH Upper Bound\\
\hline
$\Omega_m$ & 0.3 & 0.1 & 0.5 \\
\hline
$\sigma_8$ & 0.8 & 0.6 & 1.0 \\
\hline
$A_\mathrm{AGN1},\ A_\mathrm{SN1}$ & 1.0 & 0.25 & 4.0 \\
\hline
$A_\mathrm{AGN2}\ \mathrm{(except\ Astrid)},\ A_\mathrm{SN2}$ & 1.0 & 0.5 & 2.0 \\
\hline
$A_\mathrm{AGN2}$ (Astrid suite) & 1.0 & 0.25 & 4.0 \\
\hline
\end{tabular}
\caption{\label{tab:params}The fiducial values and LH set bounds on all varied parameters in the CAMEL simulations. The difference in the range of $A_\mathrm{AGN2}$ in the Astrid suite is because its function resembles that of $A_\mathrm{AGN1}$ in the other simulations, controlling AGN thermal feedback efficiency instead of gas ejection speed~\citep{ni_23}. Other $\Lambda$CDM parameters are fixed over all simulations used in this work to the following values: $\Omega_b=0.049$, $h=0.6711$, and $n_s=0.9624$.}
\end{table}

We note that this work uses the publicly available simulations and we could adapt our analysis to other simulations incorporating different baryonic feedback models or box-sizes. In particular, we leave it for future work to explore this algorithm on simulation pipelines directly developed for cosmological surveys.

\section{Simulation-Based Inference with Graph Neural Networks} \label{sec:methodology}

In this section we describe our compression and inference pipelines. First we give a brief overview of the standard SBI framework with masked autoregressive flows. In our pipeline this task is carried out by the \texttt{sbi}~\citep{tejero_20} library. We then describe the Graph neural network \texttt{CosmoGraphNet}~\citep{villanueva_22} which acts as the compression network. We end this section on details of the joint training process. For the impatient reader a graphical overview of our procedure can be found in Figure~\ref{fig:training_cartoon}.

\subsection{Neural Density Estimation for Simulation-Based Inference} \label{sec:sbi}

Simulation-based inference (SBI), or likelihood-free inference, turns the traditional Bayesian inference task into a problem of density estimation. The main advantage of the SBI framework is the omission of an explicit likelihood: The pipeline only needs a simulator and a probability density estimator. The role of the simulator is to sample parameters $\pmb \theta$ and produce an associated data sample $\pmb d$, i.e. drawing parameter-data pairs $\{\pmb\theta,\pmb d\}$ from the joint probability distribution function. This process takes over an explicit model of the likelihood, which is now implicitly shifted into the simulator. The density estimator, in principle, can either be a traditional one such as a kernel density estimator or a deep learning accelerated one, {\it i.e. a neural density estimator}. Traditional density estimators especially suffer from the curse of dimensionality, hence modern implementations of the SBI framework almost exclusively rely on such neural density estimators. Popular models include mixture density networks~\citep{bishop_94}, Real NVP~\citep{dinh_16}, or neural spline flows~\cite{durkan_19}. In this work, we make use of masked autoregressive flows (MAFs)~\cite{papamakarios_17} which we describe in the following.\\

A MAF estimates a density by learning a mapping back to the unit normal~\citep{papamakarios_17}. Let $\pmb x$ be an $N$ dimensional random variable distributed according to some probability density function $p(\pmb x)$ which is to be approximated. We wish to learn an invertible differentiable transformation $f$, such that
\begin{equation}
    \pmb x = f(\pmb u) \quad \quad \quad \quad \mathrm{where}\ \pmb u\sim \pi(\pmb u)~.
\end{equation}
$\pi(\pmb u)$ is some base density which we choose to be Gaussian and $\pmb u$ a random variable distributed according to $\pi$. The law of transformation of variables of probability distributions reads
\begin{equation} \label{equ:trafo_law}
    p(\pmb x) = \pi \left(f^{-1}(\pmb x)\right) \left|\det\left(\frac{\partial f^{-1}}{\partial \pmb x}\right)\right|~,
\end{equation}
from which we can identify two requirements we wish to impose on $f$: It needs to be easily invertible, and possess a tractable Jacobian. MAFs decompose the target density into a product of one dimensional conditional distributions
\begin{equation}
    p(\pmb{x}) = \prod_{i=1}^N p(x_i|\pmb{x}_{1:i-1})~,
\end{equation}
where each of them is modeled by a normal distribution. The notation $1:i-1$ indicates a slice over all indices from the first entry to and including the $(i-1)$th entry. To satisfy the requirements of invertibility and tractability emerging from Equation~\eqref{equ:trafo_law}, such a transformation can be parametrized with an affine block
\begin{equation} \label{equ:affine_block}
    u_i = (x_i-\mu_{\varphi,i}(\pmb x_{1:i-1}))/\sigma_{\varphi,i} (\pmb x_{1:i-1})~,
\end{equation}
where $i$ indexes the entries of the $N$ dimensional vectors and $\mu_{\varphi,i}$ and $\sigma_{\varphi,i}$ are functions learned by neural networks with learnable parameters $\varphi$. As a result of this parametrization, each entry $u_i$ is only dependent on the $i-1$ previous entries of $\pmb x$. This is called the autoregressive property and results in a convenient Jacobian as it is triangular. Its determinant is therefore just the product of the diagonal entries. Invertibility is also ensured as samples of $\pmb x$ can be generated via
\begin{equation}
    x_i = u_i \times \sigma_{\varphi,i} (\pmb u_{1:i-1}) + \mu_{\varphi,i}(\pmb u_{1:i-1})~.
\end{equation}
We note, that the neural networks themselves need not be invertible, as the functions $\mu_{\varphi,i}$ and $\sigma_{\varphi,i}$ are always evaluated in the forward direction. In practice these functions are learned with Masked Autoencoders for Distribution Estimation (MADEs)~\citep{germain_15}. MADEs are neural networks in which weights are masked in such a way that the autoregressive property of the product of conditionals is fulfilled. In practice, multiple of these transformations are stacked to give the network more flexibility to model complex distributions. Furthermore, to guarantee independence with respect to the factorization order, entries are shuffled in between transformations. The number of transformations is a hyperparameter. \\

In the use case of cosmological inference there are three probability density functions which are natural to be estimated within the above scheme~\citep{alsing_19}: the joint parameter-data probability density $p(\pmb\theta,\pmb d)$~\citep{alsing_18}, the posterior $p(\pmb\theta|\pmb d)$~\citep{papamakarios_16,lueckmann_17} and the likelihood $p(\pmb d|\pmb\theta)$~\citep{papamakarios_18,lueckmann_18}. We will only consider direct estimation of the posterior in this work as it reduces the dimensionality of the task and enables automatic compression of the data vector. In the literature this way of inference is known as neural posterior estimation (NPE)~\citep{greenberg_19}. As machine learning tasks generally perform better for lower dimensional problems, instead of the full data $\pmb d$ some compressed summary statistic $\pmb t$ is often considered~\citep{blum_12}. Going from Equation~\eqref{equ:affine_block} to estimating a conditional density can be straightforwardly implemented by letting $\pmb x = \pmb \theta$ and including the conditioning variable $\pmb t$ as an input to the neural networks $\mu_\varphi$ and $\sigma_\varphi$. This results in the conditional affine block
\begin{equation}
    u_i = (\theta_i-\mu_{\varphi,i}(\pmb \theta_{1:i-1}, \pmb t))/\sigma_{\varphi,i} (\pmb \theta_{1:i-1},\pmb t)~.
\end{equation}
We note that $\pmb t$ is merely a conditioning variable and the derivatives from Equation~\eqref{equ:trafo_law} remain with respect to $\pmb\theta$ only. As a result the Jacobian remains triangular. The final neural posterior estimator with $T$ transformations can therefore be written as~\citep{papamakarios_18}:
\begin{equation}
    q_\varphi(\pmb\theta|\pmb t) = \mathcal{N}(\pmb u^0|\pmb 0, \pmb 1 ) \times \prod_{t=1}^T \left|\det \left( \frac{\partial f_{\varphi,t}}{\partial \pmb u^t}\right) \right|^{-1}~,
\end{equation}
where the superscripted $\pmb u^t$ denotes the input variable to the $t$-th transformation. In order to optimize the underlying parameters $\varphi$ of the functions $\mu_\varphi$ and $\sigma_\varphi$, we wish to minimize the ``difference'' between the learned posterior estimate $q_\varphi(\pmb \theta|\pmb t)$ parametrized by neural network parameters $\varphi$ and the true posterior $p(\pmb \theta|\pmb t)$. A natural choice for this comparison in information geometry is the Kullback-Leibler divergence~\citep{kullback_51}. It is our aim to minimize its expectation value~\citep{jordan_99,rezende_15,papamakarios_17}
\begin{equation} \label{equ:kl_divergence}
    \begin{aligned}
    \mathbb{E}_{p(\pmb t)}[D_\mathrm{KL}(p(\pmb \theta|\pmb t)||q_\varphi(\pmb \theta|\pmb t))] &= \int d\pmb t p(\pmb t) \int d\pmb \theta p(\pmb \theta|\pmb t) \ln\left( \frac{p(\pmb \theta|\pmb t)}{q_\varphi (\pmb \theta | \pmb t)} \right) \\
    &= \int d\pmb\theta d\pmb t p(\pmb\theta,\pmb t) \ln \left( \frac{p(\pmb \theta|\pmb t)}{q_\varphi(\pmb \theta|\pmb t)} \right) \\
    &= -\mathbb{E}_{p(\pmb \theta, \pmb t)}[\ln q_\varphi(\pmb \theta|\pmb t)] + \mathrm{const.} \\
    &\approx -\frac{1}{N_{\mathrm{sim}}} \sum_n \ln q_\varphi(\pmb \theta_n|\pmb t_n) + \mathrm{const.}
    \end{aligned}
\end{equation}
In the last line the integral is approximated with its Monte Carlo estimate over the $N_\mathrm{sim}$ training examples indexed by $n$. As it is our goal to optimize the parameters $\varphi$, we bundle all terms independent of $\varphi$ into a constant. Optimizing the first term in the last line of Equation~\eqref{equ:kl_divergence} therefore also optimizes the expected Kullback-Leibler divergence up to some constant, even in the case where the true distribution $p(\pmb\theta|\pmb t)$ is {\it not} known, provided that the Monte Carlo estimate of the integral is faithful. \\

To conclude this discussion on simulation-based neural posterior estimation, we again stress the main capability of this approach: The SBI framework can estimate an {\it unknown} posterior, given a faithful approximation of the joint pdf in Equation~\eqref{equ:kl_divergence} with the help of a prior and a simulator. The posterior is then approximated by learning a transformation back to the unit normal. The entire approach is devoid of an explicit likelihood, as it is implicitly shifted into the simulator.

\subsection{Graph Neural Networks for Data Compression}

In order to apply the SBI scheme presented in the previous section, we need a (summary) data vector $\pmb t$. In this work, we will use galaxy catalogs from hydrodynamical simulations as the data $\pmb d$ and a compression network $h$ with learnable parameters $\lambda$ to extract the summaries $\pmb t = h_\lambda(\pmb d)$. We note that there are two steps preceeding the asymptotically optimal compression step $h_\lambda$: the construction of the galaxy catalog and the assembly of the galaxy graph. The full raw information contained in the hydrodynamic simulation would be the phase space of the $2\times 256^3$ particles with the addition of the tracked astrophysical quantities. As these are unobservable in reality, we begin compression at the galaxy level. This section will introduce how we extract a one-dimensional summary vector of length $N$ from galaxy catalogs of various lengths. 

A natural representation of a galaxy catalog is given in the form of a graph, which can encode various properties of the catalog (such as galaxy position, mass or velocity) in a structured fashion. An immediate advantage of this structure is, that galaxy catalogs of variable length can be encoded in such a way. We use \texttt{CosmoGraphNet}\footnote{https://github.com/PabloVD/CosmoGraphNet} as introduced by Ref.~\cite{villanueva_22}. This approach has already been used for the purposes of cosmological inference in Refs.~\cite{anagnostidis_22,shao_22,desanti_23,desanti_23b,massara_23,shao_23}. In contrast to these studies which have used \texttt{CosmoGraphNet} to predict the mean and standard deviation of the marginal posteriors, we employ it to infer the full two-dimensional posteriors and analyze the learned summary statistics. In the following we present the details of this architecture.\\

A graph is a structure consisting of nodes, edges and a global feature. Our graph is connected as follows. An edge $\pmb e_{ij}$ connects two nodes $i$ and $j$ if their separation is smaller than some linking length $r$ which is given in units of box length. While we encode positional information in the edges, each node can additionally contain features of its own $\pmb h_i$ (e.g. stellar mass, stellar metallicity etc.) which in our case is the peculiar velocity in the z-direction $\pmb h_i = h_i = v_z$. The graph is constructed such that it contains important symmetries of the galaxy catalogs: rotational and translational invariance. Following Ref.~\cite{villanueva_22} this can be achieved by setting the edges to
\begin{equation}
    \pmb e_{ij} = \left[ |\pmb d_{ij}|/r,\alpha_{ij},\beta_{ij} \right]~,
\end{equation}
where $\pmb d_{ij}$ is the relative distance between nodes $i$ and $j$. $\pmb p_i/|\pmb p_i|$ and $\pmb p_j/|\pmb p_j|$ are the normalized position vectors of the $i$th and $j$th galaxy respectively, in the euclidean coordinate system with origin $\pmb c = \sum_i^N \pmb p_i/N$. $\alpha_{ij}$ is their cosine:
\begin{equation}
    \alpha_{ij} = \frac{\pmb p_i}{|\pmb p_i|} \cdot \frac{\pmb p_j}{|\pmb p_j|}~.
\end{equation}
$\beta_{ij}$ is the cosine between the normalized vector $\pmb p_i/|\pmb p_i|$ and the corresponding normalized separation vector $\pmb d_{ij}/|\pmb d_{ij}|$:
\begin{equation}
    \beta_{ij} = \frac{\pmb p_i}{|\pmb p_i|} \cdot \frac{\pmb d_{ij}}{|\pmb d_{ij}|}~.
\end{equation}
The global feature $\pmb g$ is chosen as the logarithm of the number of galaxies in the graph $\pmb g = g = \log_{10}{n}$. Having defined the galaxy graph, we can now define a graph neural network (GNN) which operates on this graph structure.\\

The GNN can be broken down into blocks. Each block contains two functions parametrized by neural networks: an edge model $\phi$ and a node model $\psi$. The edge model of block $l+1$ takes an edge $\pmb e_{ij}^{(l)}$ and its adjacent nodes $\pmb h_i^{(l)}$ and $\pmb h_j^{(l)}$ of the previous block to produce an updated edge $\pmb e_{ij}^{(l+1)}$: 

\begin{equation}
    \pmb e^{(l+1)}_{ij} = \phi_{l+1}([\pmb h_i^{(l)},\pmb h_j^{(l)}, \pmb e_{ij}^{(l)}])~.
\end{equation}
The new edges are then used as input to the node model along with a node $\pmb h_i$ and the global feature $\pmb g$:
\begin{equation}
    \pmb h_i^{(l+1)} = \psi_{l+1}[\pmb h_i^{(l)},\bigoplus_{j\in \mathcal{N}_i} \pmb e^{(l+1)}_{ij}, \pmb g]~,
\end{equation}
where $\bigoplus$ is an operator aggregating all edges connected to the $i$th node in a permutation invariant way and where $\mathcal{N}_i$ denotes the neighborhood of the $i$th node, i.e. all connected edges. It has been shown that a combination of such operators tends to outperform the choice of a single one~\citep{corso_20}, which is why the \texttt{CosmoGraphNet} architecture concatenates the result of three possible aggregation operators: the maximum, the sum and the mean. The entire operation can consequently be expressed as
\begin{equation}
    \bigoplus_{j\in \mathcal{N}_i} \pmb{e}_{ij}^{(l+1)} = \left[\max_{j\in \mathcal{N}_i}\pmb{e}_{ij}^{(l+1)},\sum_{j\in \mathcal{N}_i}\pmb{e}_{ij}^{(l+1)}, \frac{\sum_{j\in \mathcal{N}_i}\pmb{e}_{ij}^{(l+1)}}{\sum_{j\in \mathcal{N}_i}}\right]~.
\end{equation}
The GNN consists of $L$ such blocks, the exact number of which is a hyperparameter. Every block except for the first makes use of residual, or skip connections, in which the input of each layer is added to its output. This has shown to smooth the resulting loss surface and therefore makes the training process more efficient~\citep{li_17,villanueva_22}. After the last block a final feed forward neural network $\xi$ produces a vector $\pmb t$ of length $N$ given all nodes of the final layer and the global feature, i.e.
\begin{equation}
    \pmb t = \xi ([\bigoplus_{i\in d^L} \pmb h^{(L)}_i, \pmb g])~.
\end{equation}

The resulting network we will call throughout this work the compression network $h_\lambda$, and the output we will call the summary statistic. On top of the standard hyperparameters of the \texttt{CosmoGraphNet} architecture, we also consider the dimension of the summary statistic $N$ as a additional hyperparameter. An exhaustive list of all hyperparameters is presented in the next section, where we provide details on the training procedure of this model. In general we will consider two quantities in the construction of our galaxy graph: positional information which enters via the edges as described above and the z-component of the velocity following Ref.~\cite{desanti_23}. We consider subhalos with a stellar mass above $1.95\times 10^8 \ M_\odot /h $ galaxies. The threshold was chosen according to the mean value used in Ref.~\cite{desanti_23} who additionally marginalize over this cutoff. 

\subsection{Putting it Together: Summary Extraction via Joint Training of Compression and Inference Networks}

\begin{figure}[tbp]
    \centering
    \includegraphics[width=1\textwidth]{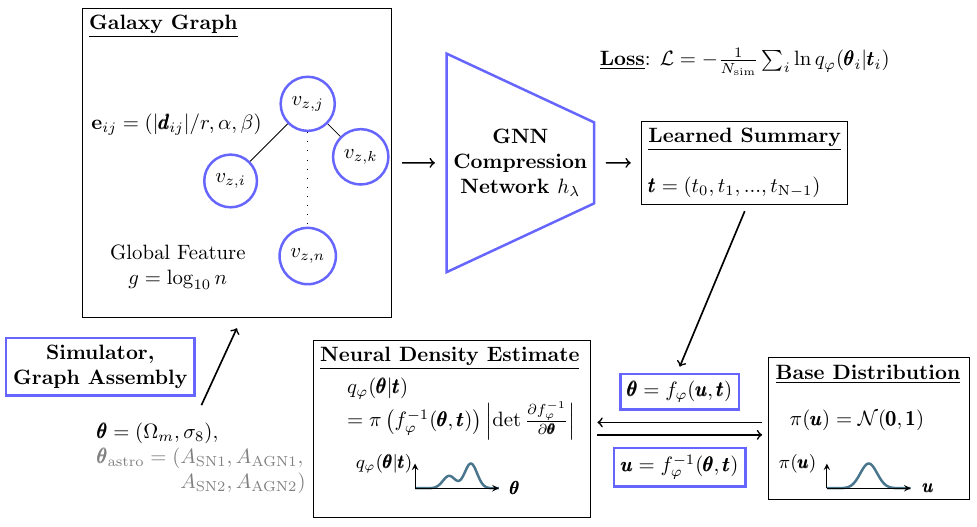}
    \caption{Schematic of our summary extractor. The galaxy graph $\pmb d$ consisting of positional, velocity, and number information is fed into a graph neural network which acts as the compression network, mapping the graph to a vector. The resulting summary vector is used for simulation-based neural posterior estimation with a masked autoregressive flow. The loss function is the expected Kullback-Leibler divergence the gradients of which are backpropagated through both networks. This setup results in a summary statistic which is optimized for the posed inference problem, maximizing the mutual information over the prior volume.}
    \label{fig:training_cartoon}
\end{figure}

In this work, we abstain from specifying a dedicated loss function for the compression network, as to not make any assumptions about the likelihood of the data or the resulting compressed data vector. Instead the initial affine block from the inference network is modified such that the functions $\mu_i$ and $\sigma_i$ themselves take neural networks as an input, as described in Ref.~\cite{radev_20}. The resulting first affine block is therefore:
\begin{equation}
    u_i = (\theta_i-\mu_{\varphi,i}(\pmb\theta_{1:i-1},h_\lambda(\pmb d)))/\sigma_{\varphi,i} (\pmb\theta_{1:i-1},h_\lambda(\pmb d))~,
\end{equation}
where $\mu_{\varphi,i}$, $\sigma_{\varphi,i}$ and $h_\lambda$ are functions parametrized by neural networks. $\mu_{\varphi,i}$ and $\sigma_{\varphi,i}$ are given by MADEs as explained in Section~\ref{sec:sbi}. The newly introduced function $h_\lambda$ now acts as a compression network, taking the original data $\pmb{d}$. We identify $h_\lambda$ with the GNN architecture presented in the previous section. This modified version of a normalizing flow can now be optimized as before, where in each backpropagation the gradients are propagated through the MADEs as well as the compression network. This way of compressing the data comes with an array of advantages compared to other methods:

\begin{itemize}
    \item Information optimality: The resulting summary statistic $\pmb t = h_\lambda(\pmb d)$ is the result of minimizing the expected KL-divergence via the inference network. A fully converged compression / inference joint network therefore maximizes the mutual information between the summary statistic and the cosmological parameters (cf.~Equation~19 and continued discussion in Ref.~\cite{radev_20}).
    \item Optimization over entire prior volume: Simulations drawn from the prior can directly be used to optimize the joint network. It is does not rely on either numerical derivatives at any point or the choice of one or multiple fiducial simulations at which the derivatives are evaluated, which do not exist in baryonic feedback modeling space. Neither does this method rely on explicit covariance estimates or assumptions on parameter independencies. 
    \item Compression and inference are coupled: Both networks are trained at the same time making this a computationally efficient operation.
    \item Interpretability: The dimensionality of the compressed data vector does not in general coincide with that of the parameter vector. As a result, the information on each of the parameters is not yet concentrated on a single number (or neuron respectively).
\end{itemize}

This approach of using solely the Kullback-Leibler divergence as a loss function for data compression has been previously explored by Ref.~\cite{cole_22} in the context of neural ratio estimation and Ref.~\cite{jeffrey_21, sharma_24} in the context of neural likelihood estimation and neural posterior estimation where the compression networks consisted of CNNs. In contrast to these works, we consider GNNs as the compression networks. Further we also perform inference with the same pipeline directly and provide interpretations of the learned summary statistics. We show a simple cartoon of our summary extractor in Figure~\ref{fig:training_cartoon}.

\subsection{Validation}

In order to check that the obtained posterior is properly calibrated we follow Ref.~\cite{talts_18}, who propose the simulation based calibration (sbc) test. sbc uses the fact, that the prior $p(\theta)$ can be expanded in the following way:
\begin{equation}
    p(\theta) = \int d\tilde{ t} d\tilde{\theta}\ p(\theta|\tilde{ t}) p(\tilde{ t}|\tilde{\theta}) p(\tilde{\theta} )~,
\end{equation}
i.e. can be expressed with the data averaged posterior. That is, if we take samples from the prior $p(\tilde{\theta})$, simulate for each of these samples, i.e. draw from the likelihood $p(\tilde{ t}|\tilde{\theta})$, and draw samples from the corresponding {\it true} posterior $p( \theta|\tilde{ t})$ for each of these simulations, we recover the prior distribution. As a consequence, our posterior estimate given by the SBI pipeline has to fulfill this same requirement if it is accurate. The sbc test therefore proceeds to compare the data averaged posterior with the prior via the so-called rank statistic. Given the prior sample $\tilde{ \theta}\sim p(\theta)$ and $M$ samples of the corresponding posterior estimate $\{\theta_1, ..., \theta_M\}\sim q_\varphi(\theta|t)$ the rank statistic $r$ is defined as~\citep{talts_18}
\begin{equation}
    r(\{\theta_1, ..., \theta_M\},\tilde\theta) \equiv \sum_{m=1}^M \mathbb{I}(\theta_m<\tilde{\theta})~,
\end{equation}
where $\mathbb{I}$ is the Heaviside step function, i.e.~$1$ if the condition is fulfilled and $0$ otherwise. Over many such samples, this rank statistic is uniformly distributed between $0$ and $M$ if the prior and the data averaged posterior estimate are similar. The test furthermore enables qualitative statements about wrongfully calibrated posteriors via the shape of the resulting rank statistic. Generally underconfident posteriors will result in a $\cap$-shaped (concave) rank statistic as the data averaged posterior is overdispersed relative to the prior. Vice-versa overconfident posteriors result in $\cup$-shaped (convex) distribution of ranks. Bias can be diagnosed if the rank statistic is a lopsided, non-symmetric distribution. For intuitive illustrations of the interpretation of test results we refer to Ref.~\cite{talts_18}. For visualization purposes it is further helpful to instead consider the empirical cumulative distribution functions (cdf) of the resulting histograms, making it easier to identify deviations. We follow this approach.

We note that this test does not guarantee the posterior estimate to be the true posterior. A simple example in which this test will be passed wrongfully, is if the posterior estimate is equal to the prior everywhere. This test is therefore only a necessary condition for correct inference, not a sufficient one. A further caveat of this test is the fact that only one dimensional posteriors can be calibrated this way. We therefore perform this test for each marginalized posterior.

This concludes the methodology of our approach. In the following, we present models trained within the scheme presented in this section.

\section{Results} \label{sec:results}

In this section we describe the performance of the inference process and we interpret the learned summary vector. We performed joint hyperparameter optimization for the usual hyperparameters of the GNN and masked autoregressive flow networks as well as the dimensionality of the summary statistic with the \texttt{optuna}\footnote{https://optuna.org/} package~\citep{akiba_19} using a Tree-Structured Parzen (TPE) Estimator~\citep{watanabe_23}. For more details on hyperparameter samplers we refer to Refs.~\cite{bergstra_11,bergstra_12}. A full listing of all hyperparameters considered and the corresponding boundaries and priors is given in Tab.~\ref{tab:hyperparams}. We allow for a warm up phase of ten trials, during which hyperparameters are drawn from their priors before the TPE strategy is applied. During training, we always retain 10\% of the entire dataset for validation, i.e.~these simulations are not used in the optimization of the model's parameters. In cases where we train on simulations of only one feedback model this results in 900 training simulations and 100 validation simulations, and in the case where we train over the entire dataset we consequently train on 2700 simulation and validate on 300. The validation set is used to monitor for early stopping: If the loss on the validation set does not improve for 20 epochs, training is stopped and the weights of the best performing model on the validation set are restored. \\

\begin{table}[tbp]
\centering
\begin{tabular}{|c|c|c|c|}
\hline
Hyperparameter & Lower & Upper & Flat Prior in\\
\hline
\hline
Learning Rate & $10^{-4}$ & $10^{-3}$ & Logarithm \\
Summary Dimension & 2 & 200 & Linear \\
Graph Linking Length [Box Length] & $5\times10^{-3}$ & $5\times10^{-1}$ & Logarithm \\
Batch Size & 5 & 50 & Categorical \\
\hline
Number of Layers & $1$ & $5$ & Linear\\
Hidden Channels & 64 & 256 & Categorical\\
\hline
Number of Transforms & 2 & 8 & Linear\\
Hidden Features & 32 & 512 & Categorical\\
\hline
\end{tabular}
\caption{\label{tab:hyperparams}All hyperparameters considered during optimization. The list is divided in general hyperparameters and those belonging to the GNN compression network, and the flow based inference network.}
\end{table}

\subsection{Performance}
\begin{figure}[tbp]
    \centering
    \begin{subfigure}{.49\textwidth}
    \includegraphics[width=\textwidth]{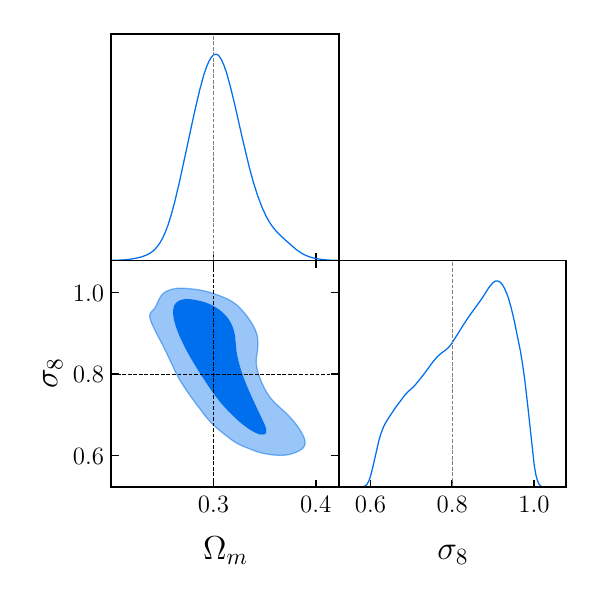}
    \caption{}
    \label{fig:full_posterior_ISA}
    \end{subfigure}
    \begin{subfigure}{.49\textwidth}
    \includegraphics[width=\textwidth]{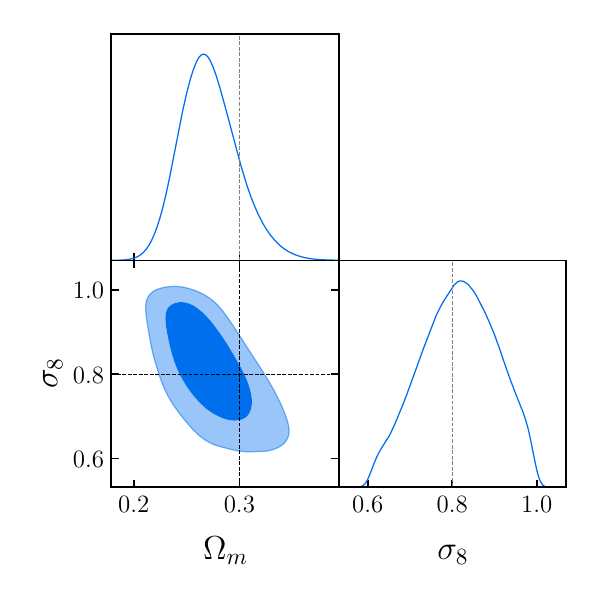}
    \caption{}
    \label{fig:full_posterior_Illustris}
    \end{subfigure}
    \caption{The trained models can correctly infer $\Omega_m$ on a fiducial cosmology box. We show posteriors for the best model trained on (a) all simulations and on the (b) IllustrisTNG subset. The cosmological parameters are recovered within the $1\sigma$ region. The $\sigma_8$ contour is not overcoming the prior, likely due to the small simulation box size. The models do however learn a correlation of the parameters in both cases.}
    \label{fig:full_posterior}
\end{figure}

\begin{figure}[tbp]
    \centering
    \begin{subfigure}{\textwidth}
    \includegraphics[width=\textwidth]{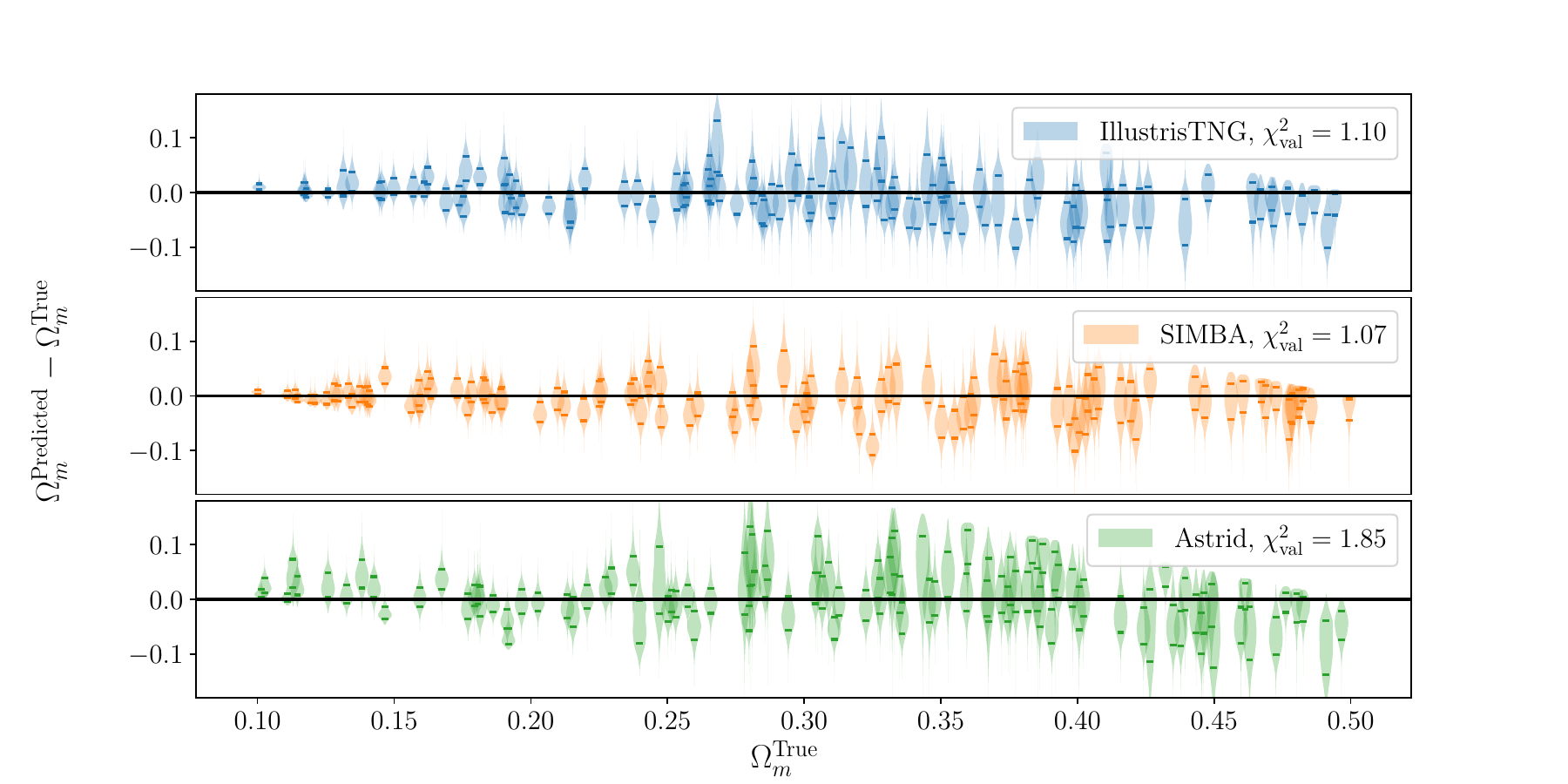}
    \caption{}
    \end{subfigure}
    \begin{subfigure}{\textwidth}
    \includegraphics[width=\textwidth]{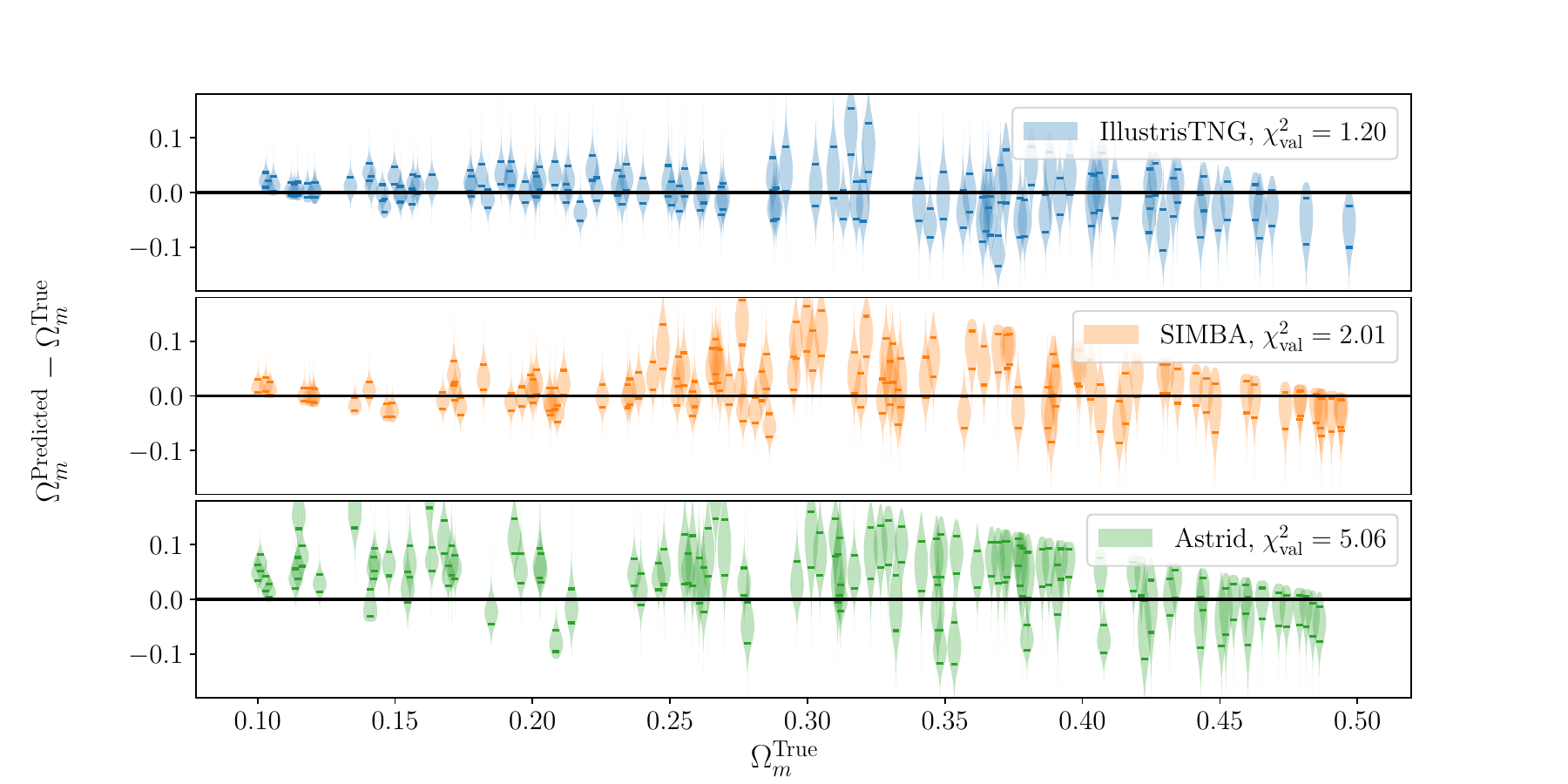}
    \caption{}
    \end{subfigure}
    \caption{$\Omega_m$ can be correctly inferred on the same simulations the model was trained on. When training on only the IllustrisTNG subset of the simulations, the model does not generalize well to the other simulations. We show marginal posteriors for $\Omega_m$ as predicted from the best performing model on the validation set. The violins are distributions of $100,000$ posterior samples, and the horizontal lines indicate the $16$th and $84$th percentile of these distributions. The model in (a) is trained on all simulations, (b) on the IllustrisTNG subset.}
    \label{fig:violin_om}
\end{figure}

\begin{figure}[tbp]
    \centering
    \begin{subfigure}{\textwidth}
    \includegraphics[width=\textwidth]{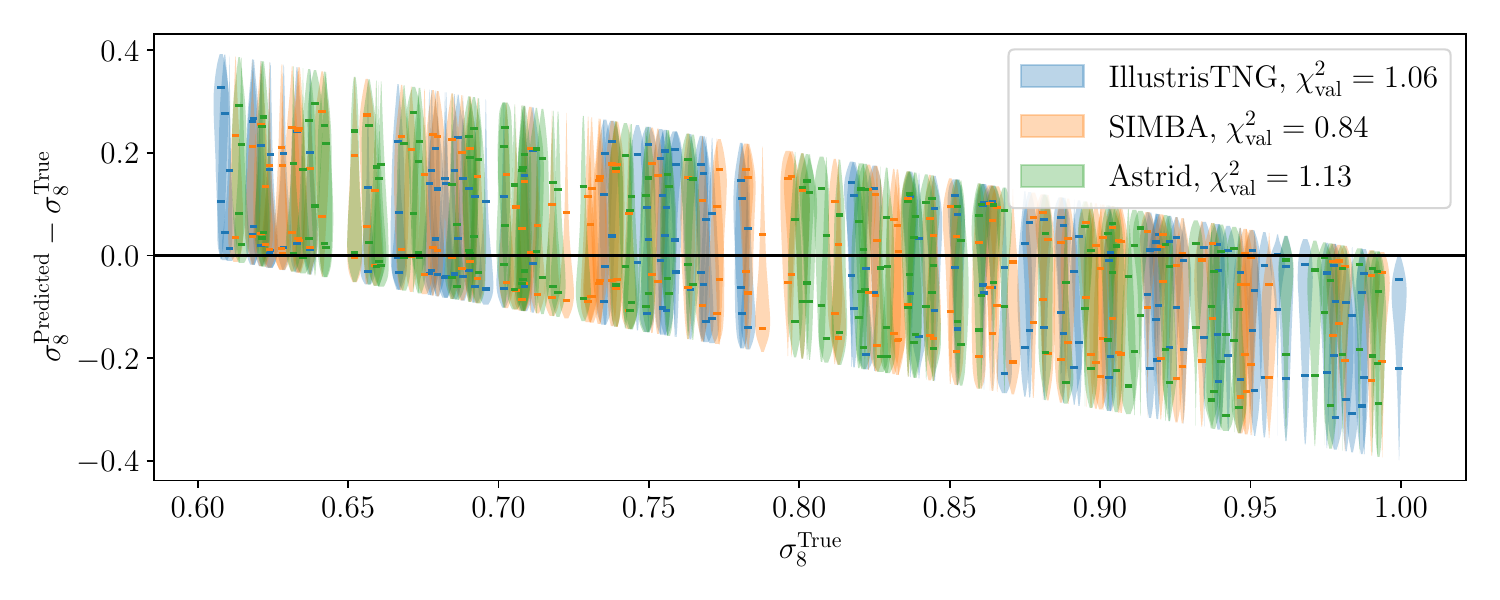}
    \caption{}
    \end{subfigure}
    \begin{subfigure}{\textwidth}
    \includegraphics[width=\textwidth]{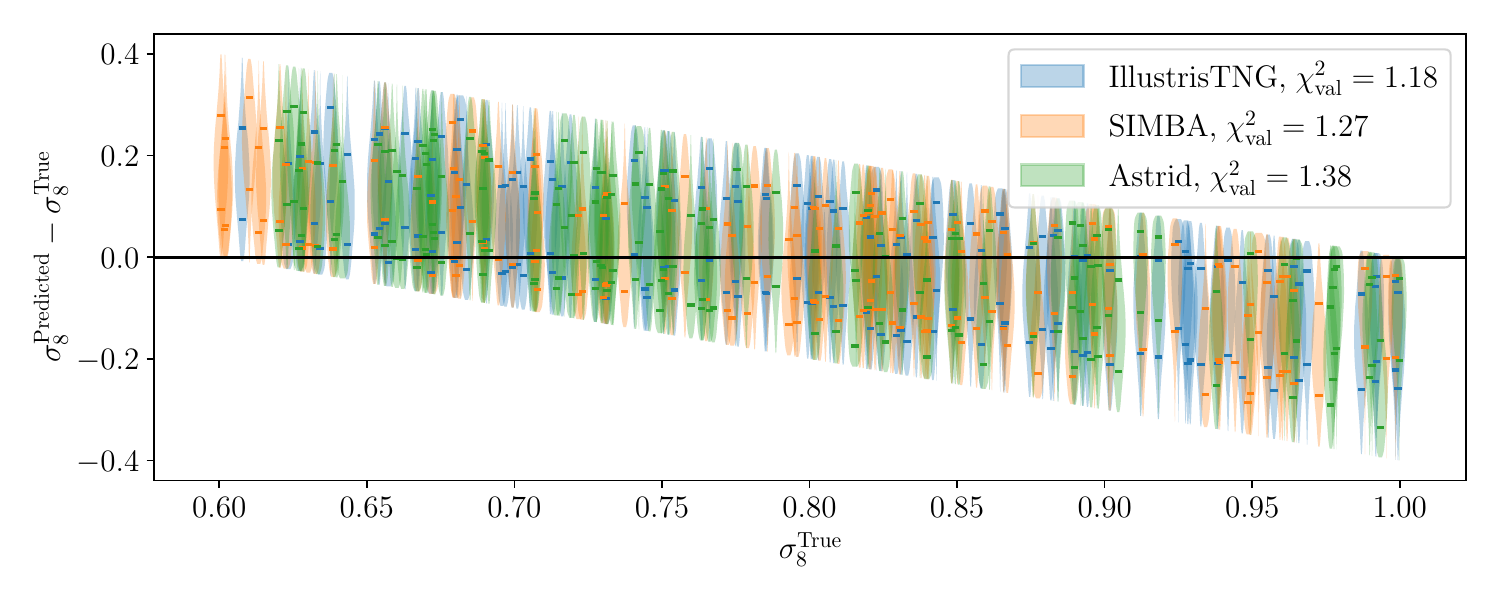}
    \caption{}
    \end{subfigure}
    \caption{Same as Fig.~\ref{fig:violin_om} but for $\sigma_8$. The model does not accurately constrain $\sigma_8$ and only learns the mean, likely due to the limited size of the simulation box and the resulting lack of large modes.}
    \label{fig:violin_sig8}
\end{figure}

\begin{figure}[tbp]
    \centering
    \begin{subfigure}{.49\textwidth}
    \includegraphics[width=\textwidth]{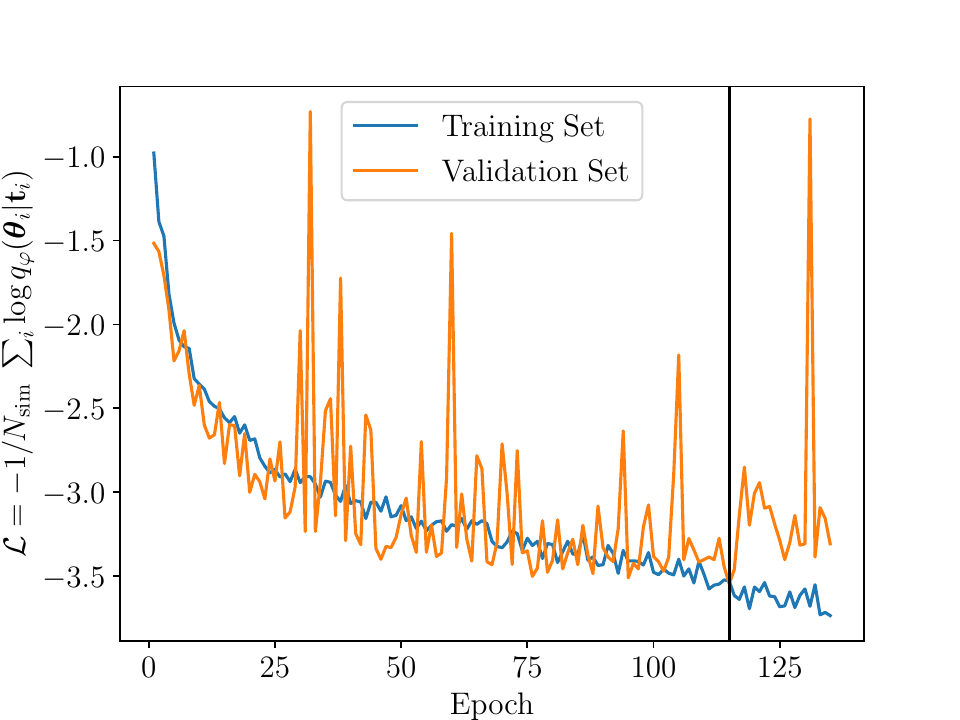}
    \caption{}
    \end{subfigure}
    \begin{subfigure}{.49\textwidth}
    \includegraphics[width=\textwidth]{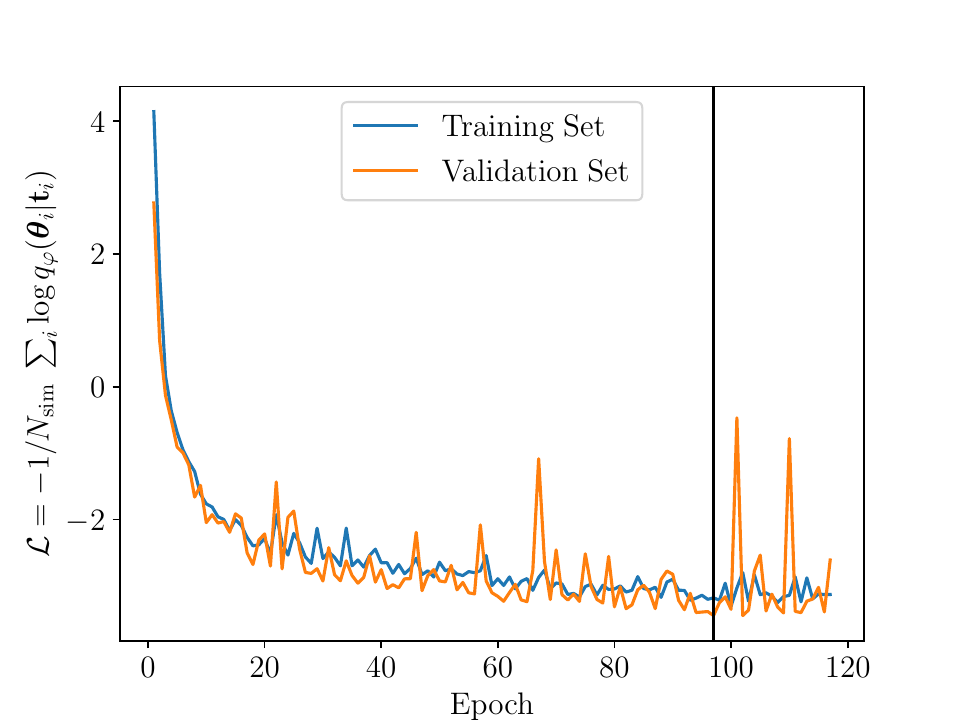}
    \caption{}
    \end{subfigure}
    \caption{The training history shows convergence of training for both of these models. Shown are (a) the best model trained on all simulations and (b) the best model trained on the IllustrisTNG subset. Early stopping sets in after 20 epochs without improvement and the weights of the best epoch are restored. The black vertical line indicates the epoch of lowest validation loss to which the model is restored.}
    \label{fig:training_hist}
\end{figure}

Following Ref.~\cite{desanti_23} we encode positions and velocities in one dimension into the galaxy graph to mimic the peculiar radial velocities in a galaxy survey. We show the posteriors of the best performing models at a fiducial cosmology IllustrisTNG box in Figure~\ref{fig:full_posterior}. The model shown in Figure~\ref{fig:full_posterior_ISA} was trained on all simulations while the model shown in Figure~\ref{fig:full_posterior_Illustris} was trained on only the IllustrisTNG subset. We note that while posteriors over the astrophysical nuisance parameters are not explicitly shown, they are implicitly marginalized over by varying them over the entire dataset. We show marginal posteriors over the entire validation set for the best models trained on all simulations and the IllustrisTNG subset in Figures \ref{fig:violin_om} and \ref{fig:violin_sig8}. These plots show the distribution of 100,000 marginal posterior samples at each validation example, with the true value subtracted from each sample. Horizontal lines indicate the $16$th and $84$th percentiles of each of these distributions. $\Omega_m$ can be robustly inferred over the entire validation set, whereas $\sigma_8$ shows a large degree of bias learning only the mean of the dataset. One reason for the pipeline's inability to correctly infer $\sigma_8$ may be the small simulation box size, which limits the presence of large scale modes. When training on the IllustrisTNG subset, the model is further unable to infer $\Omega_m$ on the other two simulation suites, as has been observed by Ref.~\cite{desanti_23}. The training history of these models is shown in Figure~\ref{fig:training_hist}, which confirms that training has converged.\\

\begin{figure}[tbp]
    \centering
    \begin{subfigure}{.8\textwidth}
    \includegraphics[width=\textwidth]{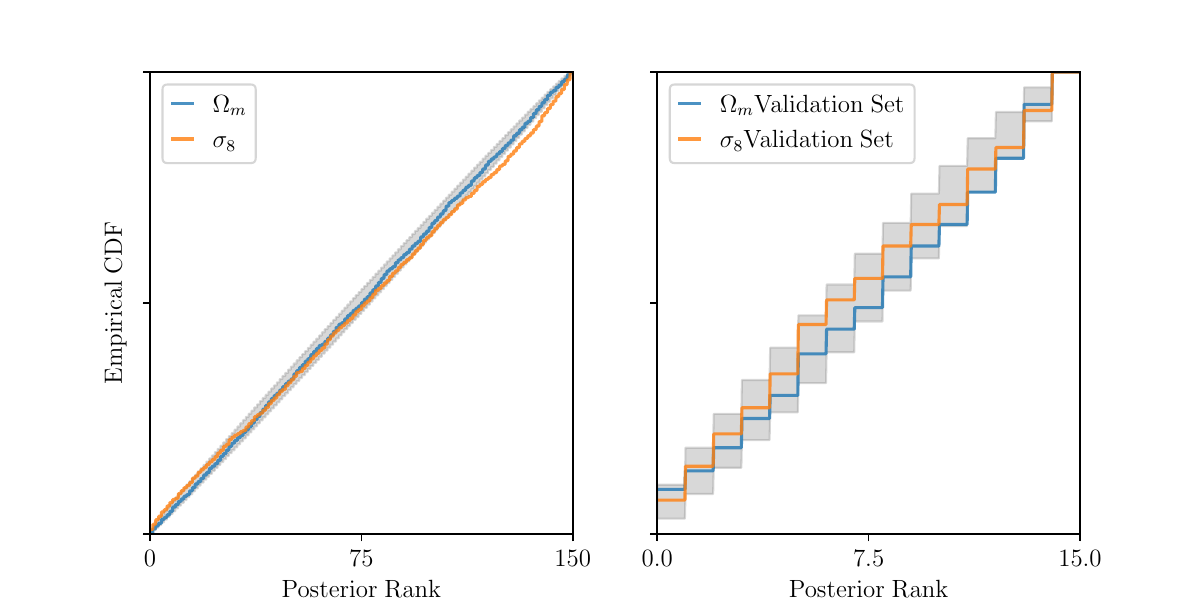}
    \caption{}
    \end{subfigure}
    \begin{subfigure}{.8\textwidth}
    \includegraphics[width=\textwidth]{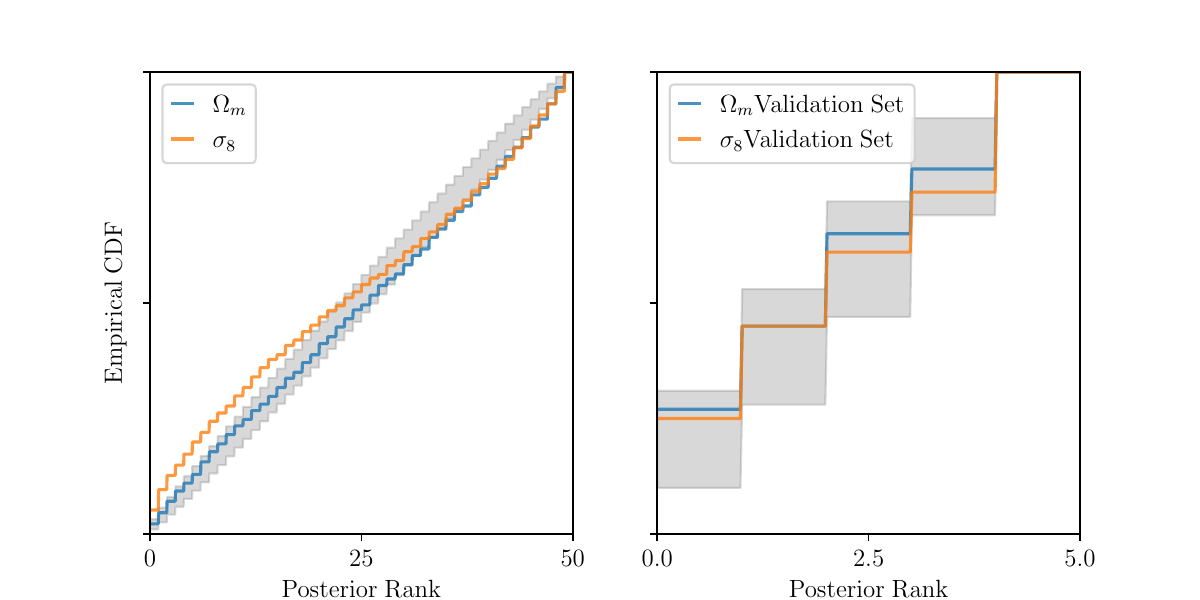}
    \caption{}
    \end{subfigure}
    \caption{The best model trained on all simulations passes the sbc test. We show the cumulative distribution function (cdf) of the rank histograms with the respective expected variance. Left panels show the cdf for the entire dataset and right panels show the cdf for only the validation set. In (a) we show the test for our best model trained over all simulations and in (b) our best model trained on only the IllustrisTNG subset.}
    \label{fig:sbc}
\end{figure}

In order to validate the produced posteriors, we perform sbc as described in the previous section. Figure~\ref{fig:sbc} shows the cdf of the distribution of ranks for the best performing model trained on all simulations and only the IllustrisTNG based subset. A correctly calibrated posterior will produce a uniform distribution and a linear cdf. The gray bands show the expected standard deviation from the cdf of a uniform distribution. We conclude from Figure~\ref{fig:sbc}, that the model trained on all simulations is well calibrated in $\Omega_m$ and slightly miscalibrated in $\sigma_8$. The model trained on only IllustrisTNG is mostly calibrated in $\Omega_m$ with slight deviations.

To further evaluate the performance we compute $\chi^2_\mathrm{red}$ scores defined as
\begin{equation}
    \chi^2_\mathrm{red} = \frac{1}{N_\mathrm{sim}}\sum^{N_\mathrm{sim}}_{i=1}\left(\frac{\theta_i-\mu_i}{\sigma_i}\right)^2~,
\end{equation}
where $\theta_i$ is the true parameter, $\mu_i$ is the posterior mean, and $\sigma_i$ is the standard deviation of the posterior. Values close to 1 indicate a good approximation. Moreover, we compute root mean square errors (RMSE):
\begin{equation}
    \mathrm{RMSE} = \sqrt{\frac{1}{N_\mathrm{sim}}\sum^{N_\mathrm{sim}}_{i=1}\left(\theta_i-\mu_i\right)^2}~,
\end{equation}
for which small values are desirable. We quote the exact values for the validation set in Table~\ref{tab:scores}. We also compare to values quoted by Ref.~\citep{desanti_23} when trained and tested on only one of the feedback models. Our performance is at least competitive with both scores slightly outperforming. A direct comparison should be cautioned, as the galaxy selection process is not identical compared to Ref.~\cite{desanti_23}. We can however state, that the increasing complexity of our model does not drastically reduce performance due to little training data. On the other hand, the increased model complexity gives us access to not only the mean and standard deviation of the posteriors, but the entire distributions. This also gives us access to e.g. maximum a posteriori (MAP) estimates.

\begin{table}[tbp]
\centering
\begin{tabular}{|c|c|c|c|c|c|} 
\hline
Tested on  & Parameter &  $\chi^2_\mathrm{red}$ & \cite{desanti_23} resp. $\chi^2_\mathrm{red}$ &  RMSE & \cite{desanti_23} resp. RMSE  \\
\hline
\multirow{2}{*}{IllustrisTNG} & $\Omega_m$ & 1.097 & 1.422  & 0.029 & 0.031 \\ 
\cline{2-6}
& $\sigma_8$ & 1.060 & N/A &  0.095 & N/A\\
\hline
\multirow{2}{*}{SIMBA} & $\Omega_m$ & 1.067  & 1.811 & 0.025 & 0.030 \\ 
\cline{2-6}
& $\sigma_8$ & 0.836 &N/A & 0.084 & N/A\\
\hline
\multirow{2}{*}{Astrid} & $\Omega_m$ & 1.851  & 1.647  & 0.039 & 0.043 \\ 
\cline{2-6}
& $\sigma_8$ & 1.130 & N/A &0.097 & N/A\\
\hline
\end{tabular}
\caption{\label{tab:scores}Comparison of our best model's performance on the validation set trained on all simulations to previous work by Ref.~\cite{desanti_23}, which was only trained on the respective simulation. Our $\chi^2_\mathrm{red}$ and RMSE values slightly outperform, however our training set is slightly different due to the fixed galaxy mass threshold. A direct comparison is therefore difficult.}
\end{table}

\subsection{Interpretability}

\begin{figure}[tbp]
    \centering
    \begin{subfigure}{.39\textwidth}
    \includegraphics[width=\textwidth]{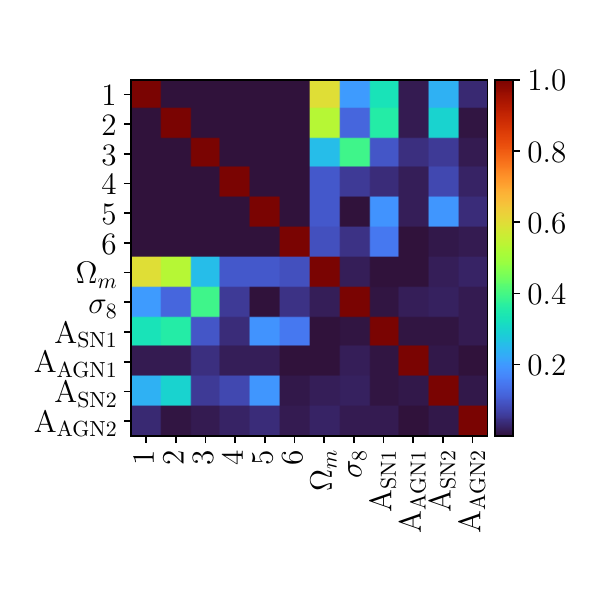}
    \caption{}
    \end{subfigure}
    \begin{subfigure}{.59\textwidth}
    \includegraphics[width=\textwidth]{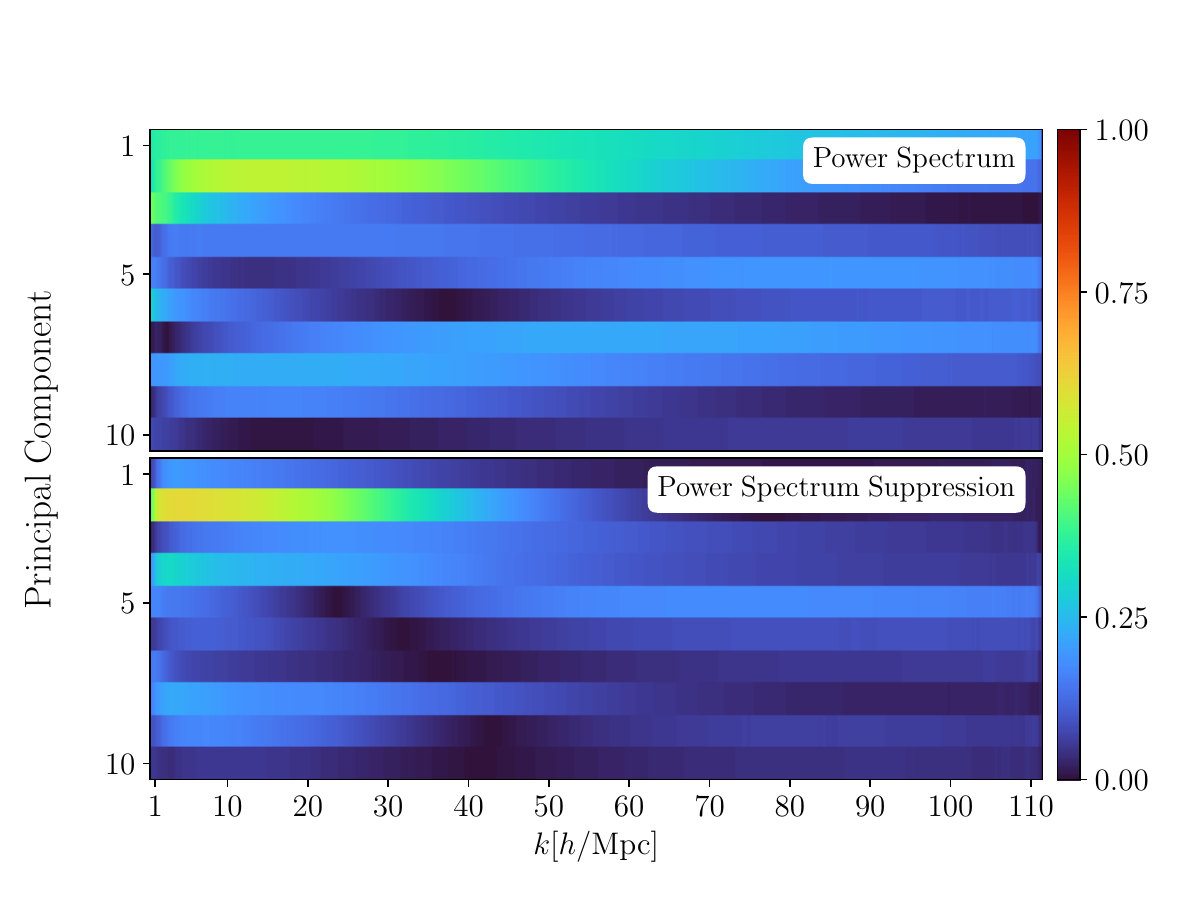}
    \caption{}
    \end{subfigure}
    \caption{Principal components of the machine-learned summary vector can be mapped back to known summaries and parameters. (a) shows the absolute correlation matrix of the first six principal components of the machine-learned summary and the simulation parameters. We can identify not only correlations with the cosmological parameters, but also with the astrophysical parameters which are unknown to the inference network. (b) shows absolute correlations of the first 10 principal components with the power spectrum, and power spectrum suppression caused by baryonic feedback.}
    \label{fig:pca_best_ISA}
\end{figure}

In order to interpret the machine-learned summary statistics, we perform further dimensionality reduction via principal component analysis. Principal components are then correlated to known parameters and summaries. In the case of the cosmological parameters, this is a crude linear approximation to the inference network, which has already learned the non-linear mapping from the summary statistics to the cosmological parameters. We note, however, that this is not the case for the astrophysical parameters, as they never enter the loss function used here. In order to quantify which scales are of importance in the inference problem, we also correlate the principal components with the matter power spectrum. We further do so for the power spectrum suppression induced by baryonic physics, which is defined as
\begin{equation}
    P_\mathrm{supp}(k) = \frac{P_\mathrm{hydro}(k)-P_\mathrm{DMO}(k)}{P_\mathrm{DMO}(k)}~,
\end{equation}
i.e.~the difference of the power spectra in hydrodynamic and n-body simulation, normalized by the latter. We show the joint absolute correlation matrix of the first 6 principal components with highest variance over the entire dataset and simulation parameters, as well as correlations with the power spectrum and the power spectrum suppression in Figure \ref{fig:pca_best_ISA} for the best model trained on all simulations. Interestingly, we find not only the expected correlation with $\Omega_m$, but also the feedback parameters $A_\mathrm{SN1}$ and $A_\mathrm{SN2}$. Even though these parameters are known neither to the compression nor the inference network, they are implicitly learned from the cosmological inference task. We also find correlations with $\sigma_8$ even though we have previously not been able to predict this parameter. One explanation for this is that it stems from the correlation between the cosmological parameters, which could be learned as shown in Figure~\ref{fig:full_posterior}. The principal components with the largest variance 1 and 2 contain information on scales $k\sim 1 h {\rm Mpc}^{-1}$ to $k\sim 30 h {\rm Mpc}^{-1}$. Upon comparing this to the correlation of the same principal components with the simulation parameters, we identify the suppression to be particularly important for the inference of $\Omega_m$. Principal component 3 which is correlated to $\sigma_8$, does not seem to correlate strongly with the power spectrum suppression. We can however find a correlation with the power spectrum at large scales.

\begin{figure}
    \centering
    \includegraphics[width=.6\textwidth]{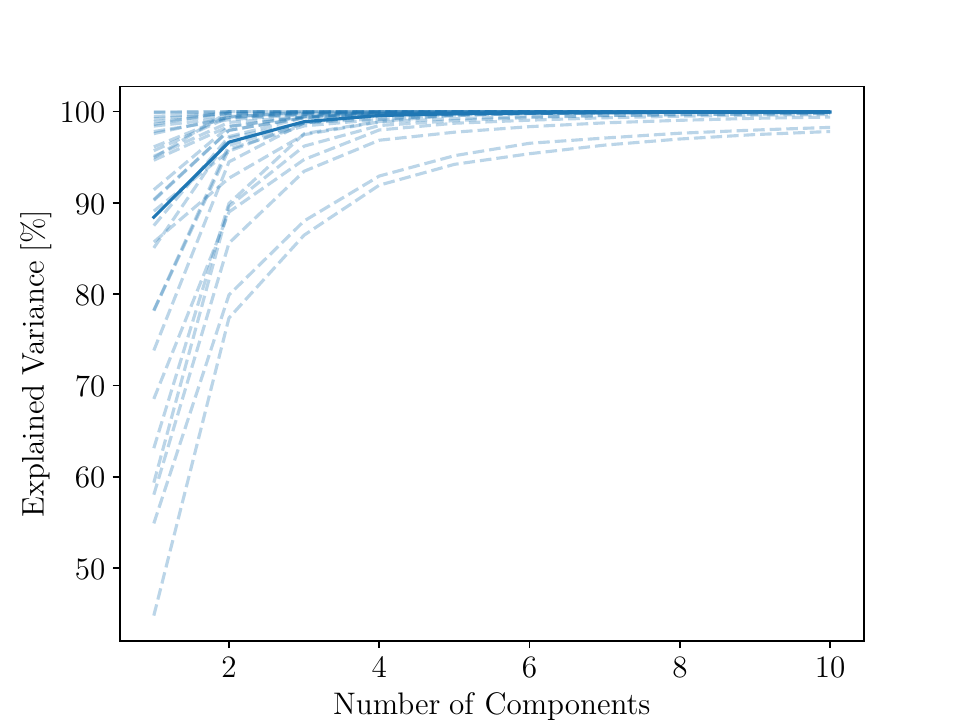}
    \caption{The cumulative variance is fully explained after considering only a few ($\mathcal{O}(5)$) principal components. Dashed lines show the 30 best models and the solid line shows the mean.} 
    \label{fig:pca_variance}
\end{figure}

How does the dimensionality of the summary vector influence the GNN's ability to compress the graph while retaining a maximum amount of information? 
To investigate this, we calculate the principal components of the summary vectors of the 30 best performing models and analyze the explained variance in Figure~\ref{fig:pca_variance}. We find across models with varying dimensionality of summary vectors that they can be reduced to $\mathcal{O}(5)$ principal components that encode a large amount of the cumulative variance.

\begin{figure}[tbp]
    \centering
    \includegraphics[width=.9\textwidth]{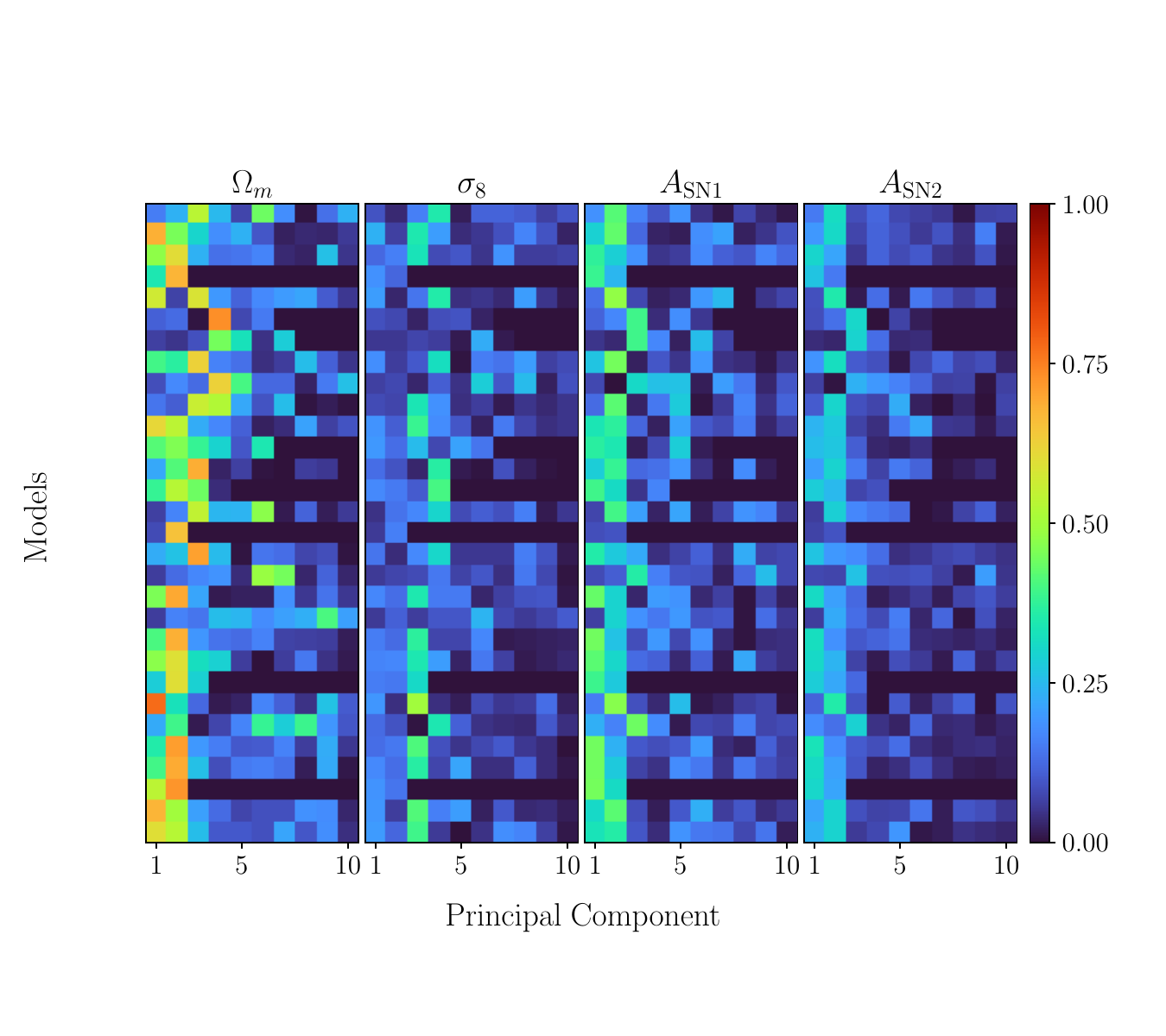}
    \caption{The correlations of principal components with parameters shows a similar pattern among many trained models. On the y-axis we show the 30 best models trained during hyperparameter optimization, the x-axis is each model's correlation of principal components with the corresponding parameter of each panel.}
    \label{fig:pca_many_models}
\end{figure}

In order to assess the robustness of the summary over many trained models, we compare the correlations of principal components over the 30 overall best performing models. We show the cross correlations with the cosmological and previously identified important astrophysical parameters in Figure~\ref{fig:pca_many_models}. A repeating pattern can be identified, as $\Omega_m$ is usually distributed across two or three different principal components. There is only one component correlating with $\sigma_8$, which is also orthogonal to the other correlations. $A_\mathrm{SN1}$ and $A_\mathrm{SN2}$ is mostly picked up by one or two principal components. \\

\begin{figure}[tbp]
    \centering
    \begin{subfigure}{.36\textwidth}
    \includegraphics[width=\textwidth]{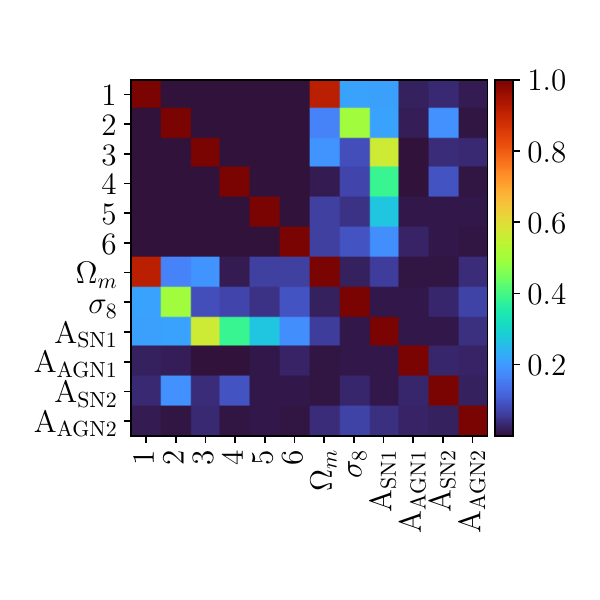}
    \caption{}
    \end{subfigure}
    \begin{subfigure}{.36\textwidth}
    \includegraphics[width=\textwidth]{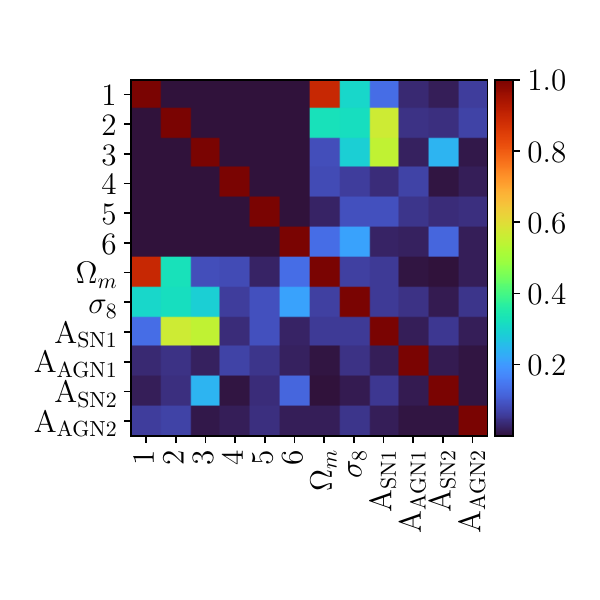}
    \caption{}
    \end{subfigure}
    \begin{subfigure}{.36\textwidth}
    \includegraphics[width=\textwidth]{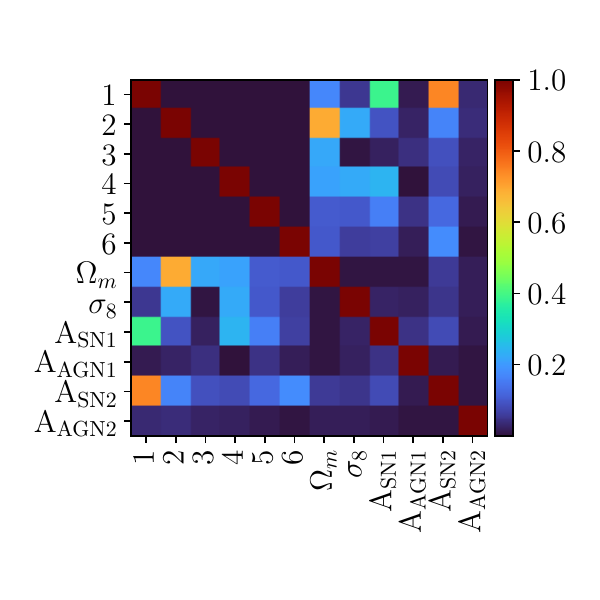}
    \caption{}
    \end{subfigure}
    \begin{subfigure}{.36\textwidth}
    \includegraphics[width=\textwidth]{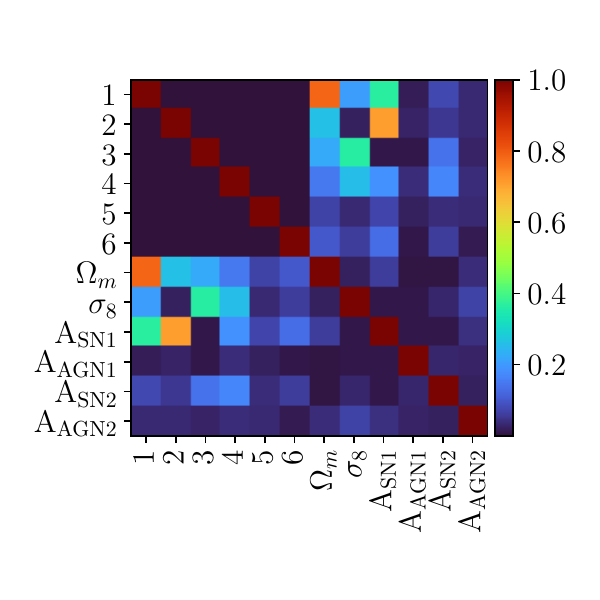}
    \caption{}
    \end{subfigure}
    \begin{subfigure}{.36\textwidth}
    \includegraphics[width=\textwidth]{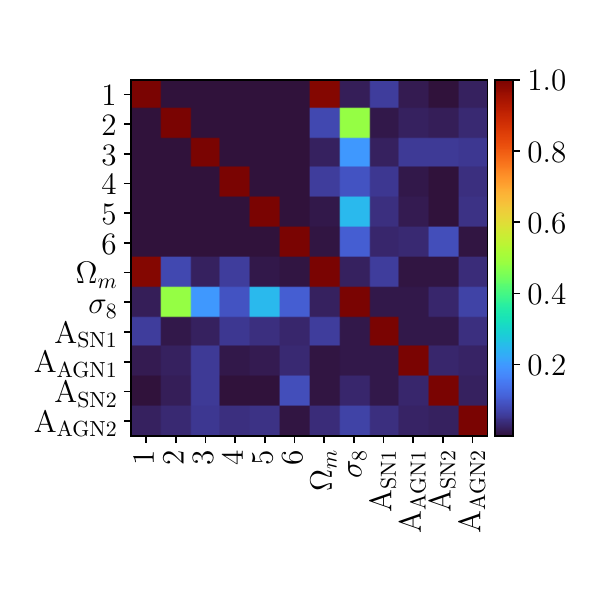}
    \caption{}
    \end{subfigure}
    \caption{Training data influences the learned summary vector. We show the best models from our hyperparameter scan trained on (a)IllustrisTNG, (b)SIMBA and (c)Astrid. (d) shows the best model trained on the IllustrisTNG set only on positional information, i.e. no velocity information. In (e) we show principal components of the best IllustrisTNG trained compression network applied to the corresponding n-body simulations.}
    \label{fig:pca_alternatives}
\end{figure}

While the summaries are robust to training multiple models, we find differences in the learned summaries depending on which underlying feedback prescription the model is trained on. This is shown in Figure~\ref{fig:pca_alternatives}. The panels (a), (b) and (c) show that when the model is trained on only one feedback prescription, the astrophysical parameters encoded in the principal component changes. While we have seen correlations with both $A_\mathrm{SN1}$ and $A_\mathrm{SN2}$ when training on all simulations, training on only IllustrisTNG or SIMBA removes a large amount of the correlation. This shows how different parameters are of importance for cosmological inference when the underlying feedback model is switched. Training on IllustrisTNG also yields a model that can decorrelate the stellar feedback parameter principal components from the $\Omega_m$ principal components. We investigate how this decorrelation is possible by retraining models on IllustrisTNG without the velocity information in panel (d). Once the velocity information is withheld, the best model can no longer decorrelate Astrophysics and Cosmology. This is in agreement with previous work that has shown the velocity field to be important for the robustness of models trained on only one feedback prescription~\citep{desanti_23}. We note, that we never find correlations of our summary statistic with either AGN parameter. In panel (e) we use the IllustrisTNG trained compression network on subhalo catalogs from the corresponding n-body gravity-only runs and compress them to a summary statistic. We then recompute the principal components and again correlate them with the simulation parameters. For the correlation with astrophysical parameters we choose the values of the hydrodynamic counterpart of each n-body simulation, even though they have no influence on the actual gravity-only n-body simulation. This way we can check that the compression network has not learned every astrophysical parameter from the latin-hypercube layout. We can observe that even though the compression network was trained on hydrodynamical simulations, it also extracts the cosmological information from n-body simulations. As expected, the correlations with the impactless astrophysical parameters is vanishing.  

\subsection{Simulation Comparison in Summary Space}

\begin{figure}[tbp]
    \centering
    \begin{subfigure}{.6\textwidth}
    \includegraphics[width=\textwidth]{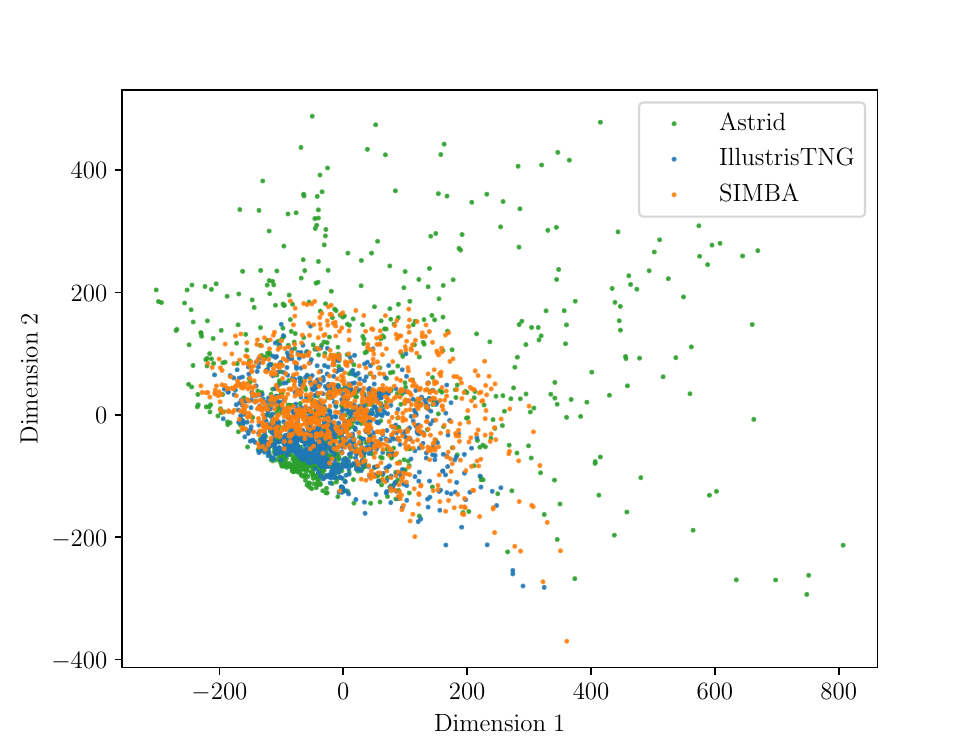}
    \caption{}
    \end{subfigure}
    \begin{subfigure}{\textwidth}
    \includegraphics[width=\textwidth]{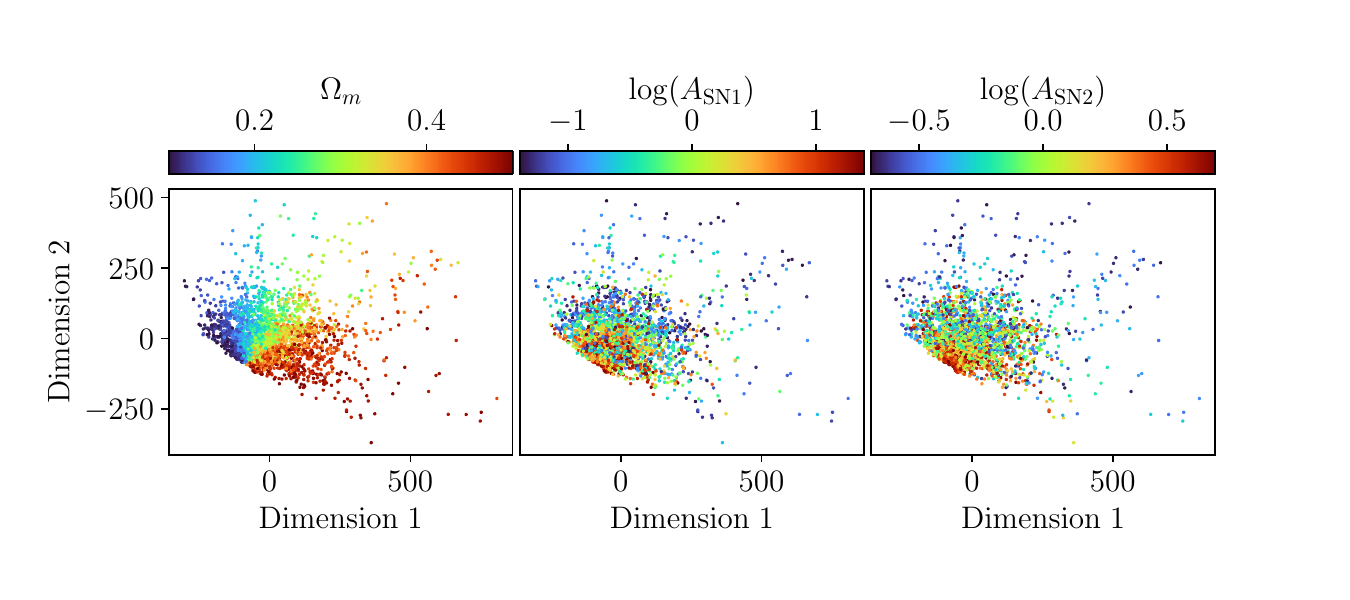}
    \caption{}
    \label{fig:pca_isomaps_params}
    \end{subfigure}
    \caption{Isomaps constructed from learned summaries of the best model trained on all simulations. (a) shows that the Astrid simulations occupy a larger volume in this summary space. As the cosmology priors are identical, this must stem from the parametrization of astrophysical effects. (b) shows how the model distributes cosmological and astrophysical summaries in summary space. Simulations with comparable $\Omega_m$ are close to each other along the polar angle. The supernovae parameters $A_\mathrm{SN1}$ and $A_\mathrm{SN2}$ are separated along the radial axis. We identify a forbidden region in the lower left part of the distribution, which is sharply cut off. We defer the study of this cutoff to further work.}
    \label{fig:pca_isomaps}
\end{figure}

\begin{figure}[tbp]
    \centering
    \includegraphics[width=.7\textwidth]{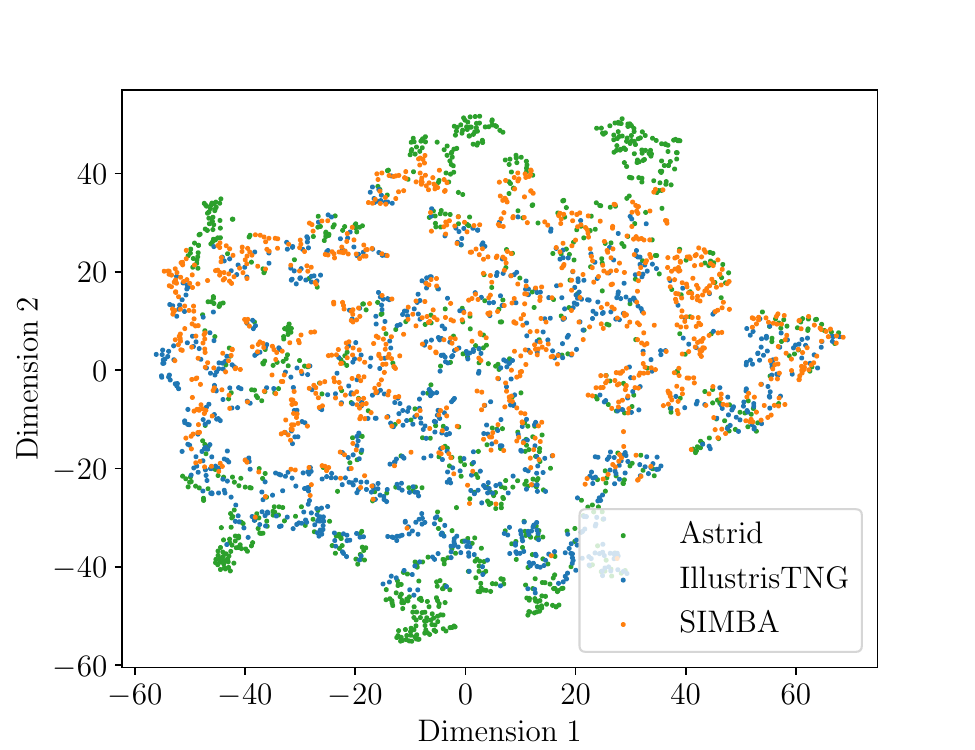}
    \caption{The Astrid simulations also occupy a larger volume in the t-sne embedding. We compare the two-dimensional t-sne summary space embeddings of all summary vectors over the entire dataset.}
    \label{fig:tsne}
\end{figure}

We pursue two different non-linear strategies to compare the distribution of simulations in summary space\footnote{We use the respective publicly available implementations from the \texttt{sklearn}~\citep{scikit-learn_11} library.}. First we construct Isomaps~\citep{tenenbaum_00}: all simulations points in summary space are put into a graph and $k$ nearest neighbors are connected. Eigenvectors are then computed for the resulting matrix of shortest paths between nodes. Coordinates of a d-dimensional subspace are then set to the top $d$ eigenvectors, which optimally preserves distances in between points. Figure \ref{fig:pca_isomaps} shows the structure of this embedding for the case $d=2$. To compare to a second method of visualization we also train t-distributed stochastic neighbor embeddings~\citep{vandermaaten_08} (t-sne) to compress the summary vectors. This method compresses high dimensional datapoints to a two-dimensional representation by representing similarities of datapoints with conditional probability distributions. Starting from Gaussian noise, samples in a two-dimensional space are then updated via gradient descent to mimic the same conditional probabilities as in the high dimensional summary space. This is done by minimizing the Kullback-Leibler divergence between the two pdfs. For more detail on this we refer to Ref.~\cite{vandermaaten_08}. Figure~\ref{fig:tsne} shows the distribution of all 3000 summary vectors with the compression network trained on all three simulation suites.

We observe the distributions produced by the Astrid simulations to be much broader than the other two feedback models in both compression schemes. This now gives us insight on many results obtained thus far. On one hand, we have observed that performance metrics have generally been worse on Astrid data. This can be explained by the fact that these simulations occupy a larger volume in summary space and can therefore not be generalized to by training on only IllustrisTNG or SIMBA. On the other hand, when Astrid was used for training Ref.~\cite{desanti_23} reported the best robustness to testing on the other two simulations. Again, we can explain this with the fact that the distribution of Astrid simulations in summary space covers the distributions of the other two. On the simulations side, we can therefore state, that as parametrized in the CAMEL simulations, the Astrid suite covers the largest part of the feedback model space. This was also mentioned by Ref.~\cite{ni_23}. 

We can therefore use these summaries to help the modeling community decide which feedback models to examine next, for example to sample previously unexplored regions of the model space. On the other hand, one could identify what parts of the summary space are close to observations and run simulations preferentially in those regions. While such statements are impossible to make on a galaxy catalog level without resorting to highly lossy summaries, they are easily diagnosed in the machine-learned summary space setting. Future simulation suites could now also be constructed not according to somewhat arbitrary parametrizations, but according to their distributions in summary space, potentially making future simulation suites more efficient for training machine learning inference pipelines. 

We further note that the model orders cosmological information in the Isomap embedding space along the polar angle as can be observed in Figure~\ref{fig:pca_isomaps_params}. The information on supernova feedback is mostly stored in the radial coordinate. While there is a large amount of scatter in this relation, the model seems to decorrelate baryonic effects from cosmological information along two separate coordinates. We can also identify a sharp cutoff of all distributions. We leave up to further study whether this is due to the chosen embedding or can be attributed to a physical meaning.

\subsection{Marginalizing over Baryonic Contributions}

\begin{figure}
    \centering
    \includegraphics[width=\textwidth]{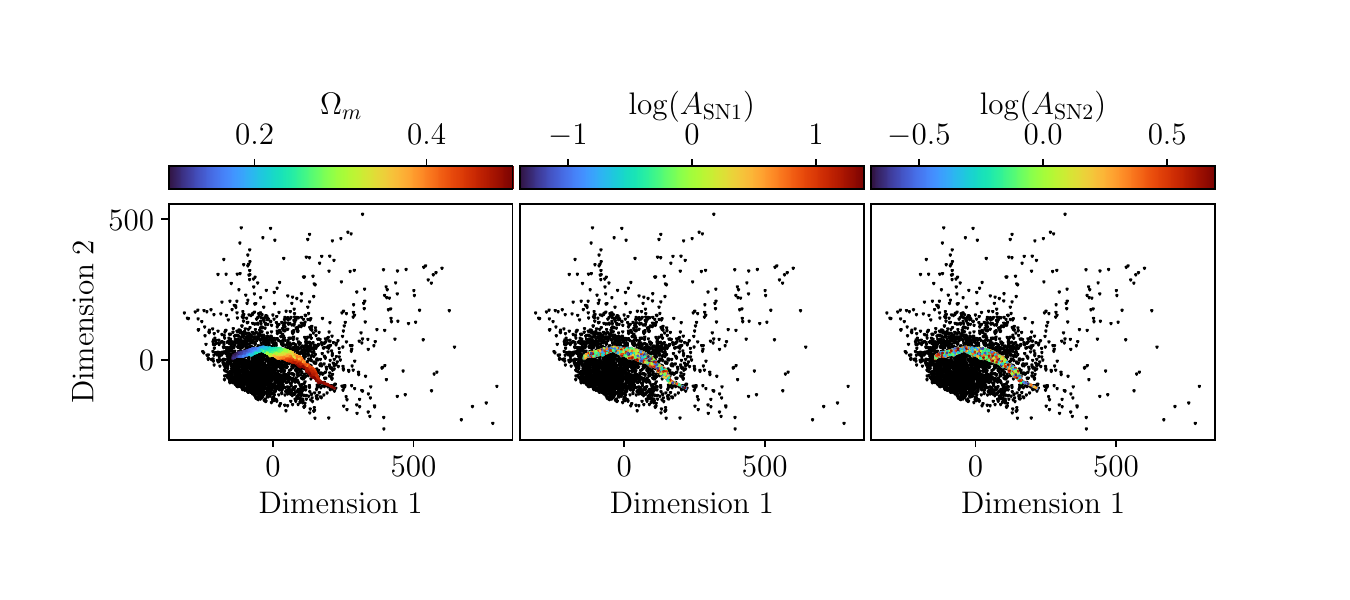}
    \caption{The joint Isomap representation of the emulated summaries and GNN summaries shows the former to be on a hypersurface in this space. We can therefore identify the region in summary space, in which marginalized summaries can be found.}
    \label{fig:regressor_isomap}
\end{figure}

\begin{figure}
    \centering
    \includegraphics[width=.4\textwidth]{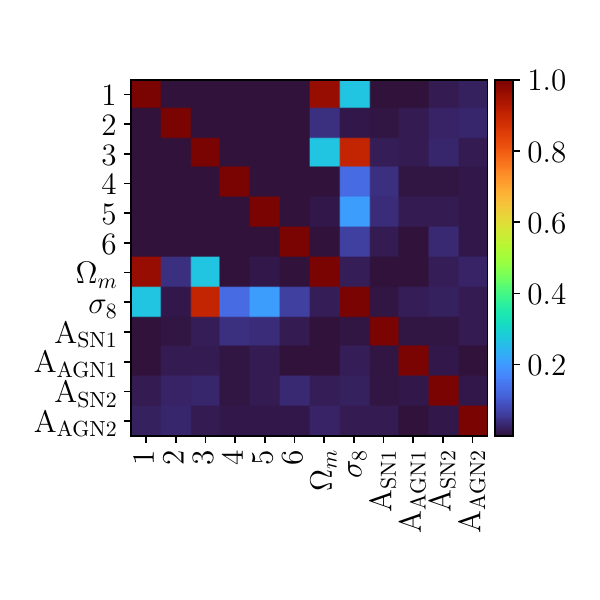}
    \caption{The correlation of the emulated principal components with the astrophysical parameters of the training simulations vanish, as they are marginalized over.}
    \label{fig:regressor_pcas}
\end{figure}

As we have shown in the previous sections, we find significant information on the effects of baryons in the machine-learned summary statistic. To continue the exploration of the learned summary space of the previous section, we can further ask the question: What volume of summary space is occupied by a baryon-robust summery, i.e. one where baryonic effects have been marginalized over within the scope of the feedback models explored here. We do so by training a fully connected feed-forward neural network based emulator to predict the summary, given cosmological parameters on a mean square error loss:
\begin{equation}
    \mathrm{MSE} = \frac{1}{N} \sum_i^N (\pmb t_{\mathrm{GNN},i} - \pmb t_{\mathrm{emulator},i})^2~.
\end{equation}
It is well known in the literature, that the MSE upon perfect convergence yields the mean of the posterior of the output~(e.g.~Ref.~\cite{murphy_22}). This can be shown easily as the neural network learns the minimum risk:
\begin{equation}
    \mathrm{argmin}_{\pmb{t}_\mathrm{emulator}} \int (\pmb t_{\mathrm{GNN}} - \pmb t_{\mathrm{emulator}})^2 p(\pmb t_\mathrm{GNN}|\pmb \theta) d\pmb t_\mathrm{GNN}~,
\end{equation}
which under perfect convergence has a vanishing first derivative
\begin{equation}
    \frac{\partial \int (\pmb t_{\mathrm{GNN}} - \pmb t_{\mathrm{emulator}})^2 p(\pmb t_\mathrm{GNN}|\pmb \theta) d\pmb t_\mathrm{GNN}}{\partial \pmb t_{\mathrm{emulator}}} = 0~.
\end{equation}
Via explicit calculation of the derivatives we obtain the optimal estimator
\begin{equation}
\begin{aligned}
    \pmb t_{\mathrm{emulator}} &= \int p(\pmb t_\mathrm{GNN}|\pmb \theta)\ \pmb t_\mathrm{GNN}\ d\pmb t_\mathrm{GNN}\\
    &= \int \int p(\pmb t_\mathrm{GNN}|\pmb\theta,\pmb \theta_\mathrm{astro})\ p(\pmb \theta_\mathrm{astro})\ \pmb t_\mathrm{GNN}\ d\pmb t_\mathrm{GNN}\ d\pmb \theta_\mathrm{astro}~,
\end{aligned}
\end{equation}
where we have introduced $\pmb \theta_\mathrm{astro}$ as the baryonic parameters. By training the emulator we therefore learn a baryon marginalized summary statistic and approximate an integral, which would otherwise be unfeasible due to the low simulation budget. In Figure~\ref{fig:regressor_isomap} we show the Isomap representation of the joint set of GNN and emulated summaries, with GNN summaries in black. We observe that the emulated summary is a connected hypersurface (up to possible projection effects) in the learned summary space. We can therefore attribute a subvolume of the summary space to galaxy catalog representations which are marginalized over the baryonic models sampled by the CAMELS simulations. As a consistency check, we also confirm the absence of baryonic effects in the summary by showing the same correlation matrices as in the previous section. As the emulator is not prompted with baryonic parameters, we use for the computation of the correlation coefficients the values of the corresponding hydrodynamic simulations. Figure~\ref{fig:regressor_pcas} shows that the correlation coefficients no longer appear. We note, however, that an absence of correlations in this test does {\it not} guarantee independence as it only picks up on linear correlations. As a consistency check we are never the less encouraged by the fact that linear correlations vanish and refer to the theoretical argument above. We view this exercise as one of the first steps into understanding the structure of the learned summary space and with it the space of feedback models.


\section{Conclusions}\label{sec:conclusions}

We presented a novel approach to extract optimally informative summary statistics from the large-scale structure. Summary statistics are created as a byproduct when optimizing a cosmological inference problem. In order to capture underlying symmetries of the galaxy catalogs in simulations, we construct galaxy graphs. A GNN then serves as an embedding network to a MAF based neural posterior estimator. By varying the dimensionality of the summary space during hyperparameter optimization, we enable the model to find a high dimensional embedding of the cosmological information which eases inference. We show that even though summary vectors are $\mathcal{O}(10-100)$, the inherent linear information is explained with only a few principal components. While our approach to computing summary statistics is shown to be robust to training many models, the learned summaries vary if the underlying feedback model is switched. By training models on different subgrid physics prescriptions, we can compare them via the resulting summary statistics.

We perform interpretation of the summary vectors, by correlating its principal components to input parameters, the power spectrum and the power spectrum suppression caused by baryonic physics. While we lose asymptotic informative optimality by compressing via PCA, we only do so for the interpretation step. We find that different astrophysical parameters are implicitly learned, even though they are not explicitly known to the model as they never enter the loss function. Whereas the importance of single parameters varies with feedback implementation, we do not find correlations with the AGN parameters. Examining correlations with the power spectra, we can identify scales of relevance to the inference problem.

As a next step we analyze the distribution of summary vectors in summary space, when the model is trained over all simulations. We non-linearly compress the summaries to two dimensions with both Isomaps and t-sne. In both cases we identify the Astrid simulations to occupy a larger volume in this compressed space. With this observation we can explain previously found results which showed that models trained on other simulations do not generalize to Astrid but models trained on the Astrid simulations generalize to the other feedback models. As Astrid occupies a larger volume for the same cosmological priors, we can explain this with a wider range of feedback model space that is sampled within these simulations. We further observe that the model does indeed decorrelate part of the baryonic noise from cosmological information.

To further understand the structure of the learned summary representation, we train a neural network emulator to predict the summaries, given the cosmological parameters. We show from well known theoretical arguments and correlations that this leads to a summary statistic which is marginalized over the baryonic models sampled in the CAMELS simulations. We compare the emulated summary with the GNN produced summary and identify a hypersurface which encloses all emulated summaries. This is one of the first steps to understand the entirety of the produced galaxy catalogs in these simulations.

We also find that our inference results are at least competitive with existing ML inference work~\cite{desanti_23} as we observe similar RMSE scores and slightly reduced $\chi^2_\mathrm{red}$ results. It is worth to re-iterate that our method provides us with more statistical information such as an MAP estimate.

We leave many possibilities for the extension of this work to future efforts. The most straightforward next step would be to include astrophysical observables in the graph and compare if we obtain similar results. The scope of our approach could also be increased by considering an even larger range of feedback models as well as applications to larger boxes. Moreover, we are interested in a deeper understanding of galaxy catalogs from their representation in summary space.

\acknowledgments

We thank Oliver Friedrich, David Gebauer, Zhengyangguang Gong, Daniel Gr\"un, Anik Halder, Jed Homer, Luisa Lucie-Smith, Lucas Makinen, Tilman Plehn, Fabian Schmidt, Benedikt Schosser, Jesse Thaler and Beatriz Tucci for insightful discussions. We acknowledge support via the KISS consortium (05D23WM1) funded by the German Federal Ministry of Education and Research BMBF in the ErUM-Data action plan. We are also grateful for support from the Cambridge-LMU strategic partnership. This work has been partially supported by STFC consolidated grants ST/T000694/1 and ST/X000664/1.

\paragraph{Software Acknowledgments} In addition to the software packages mentioned in the main body of this work we acknowledge the \texttt{matplotlib}~\citep{matplotlib_07}, \texttt{getdist}~\citep{getdist_19}, \texttt{pytorch}~\citep{pytorch_19} and \texttt{numpy}~\citep{numpy_20} libraries.

\bibliographystyle{unsrt}
\bibliography{bibliography}

\begin{thebibliography}{100}

\bibitem{sptsz_15}
L.~E. {Bleem}, B.~{Stalder}, T.~{de Haan}, K.~A. {Aird}, S.~W. {Allen}, D.~E.
  {Applegate}, M.~L.~N. {Ashby}, M.~{Bautz}, M.~{Bayliss}, B.~A. {Benson},
  S.~{Bocquet}, M.~{Brodwin}, J.~E. {Carlstrom}, C.~L. {Chang}, I.~{Chiu},
  H.~M. {Cho}, A.~{Clocchiatti}, T.~M. {Crawford}, A.~T. {Crites}, S.~{Desai},
  J.~P. {Dietrich}, M.~A. {Dobbs}, R.~J. {Foley}, W.~R. {Forman}, E.~M.
  {George}, M.~D. {Gladders}, A.~H. {Gonzalez}, N.~W. {Halverson}, C.~{Hennig},
  H.~{Hoekstra}, G.~P. {Holder}, W.~L. {Holzapfel}, J.~D. {Hrubes}, C.~{Jones},
  R.~{Keisler}, L.~{Knox}, A.~T. {Lee}, E.~M. {Leitch}, J.~{Liu}, M.~{Lueker},
  D.~{Luong-Van}, A.~{Mantz}, D.~P. {Marrone}, M.~{McDonald}, J.~J. {McMahon},
  S.~S. {Meyer}, L.~{Mocanu}, J.~J. {Mohr}, S.~S. {Murray}, S.~{Padin},
  C.~{Pryke}, C.~L. {Reichardt}, A.~{Rest}, J.~{Ruel}, J.~E. {Ruhl}, B.~R.
  {Saliwanchik}, A.~{Saro}, J.~T. {Sayre}, K.~K. {Schaffer}, T.~{Schrabback},
  E.~{Shirokoff}, J.~{Song}, H.~G. {Spieler}, S.~A. {Stanford},
  Z.~{Staniszewski}, A.~A. {Stark}, K.~T. {Story}, C.~W. {Stubbs},
  K.~{Vanderlinde}, J.~D. {Vieira}, A.~{Vikhlinin}, R.~{Williamson}, O.~{Zahn},
  and A.~{Zenteno}.
\newblock {Galaxy Clusters Discovered via the Sunyaev-Zel'dovich Effect in the
  2500-Square-Degree SPT-SZ Survey}.
\newblock {\em \apjs}, 216(2):27, February 2015.

\bibitem{hetdex_08}
G.~J. {Hill}, K.~{Gebhardt}, E.~{Komatsu}, N.~{Drory}, P.~J. {MacQueen},
  J.~{Adams}, G.~A. {Blanc}, R.~{Koehler}, M.~{Rafal}, M.~M. {Roth}, A.~{Kelz},
  C.~{Gronwall}, R.~{Ciardullo}, and D.~P. {Schneider}.
\newblock {The Hobby-Eberly Telescope Dark Energy Experiment (HETDEX):
  Description and Early Pilot Survey Results}.
\newblock In T.~{Kodama}, T.~{Yamada}, and K.~{Aoki}, editors, {\em Panoramic
  Views of Galaxy Formation and Evolution}, volume 399 of {\em Astronomical
  Society of the Pacific Conference Series}, page 115, October 2008.

\bibitem{kids_13}
Jelte T.~A. {de Jong}, Gijs~A. {Verdoes Kleijn}, Konrad~H. {Kuijken}, and
  Edwin~A. {Valentijn}.
\newblock {The Kilo-Degree Survey}.
\newblock {\em Experimental Astronomy}, 35(1-2):25--44, January 2013.

\bibitem{hsc_15}
Satoshi {Miyazaki}, Masamune {Oguri}, Takashi {Hamana}, Masayuki {Tanaka},
  Lance {Miller}, Yousuke {Utsumi}, Yutaka {Komiyama}, Hisanori {Furusawa},
  Junya {Sakurai}, Satoshi {Kawanomoto}, Fumiaki {Nakata}, Fumihiro {Uraguchi},
  Michitaro {Koike}, Daigo {Tomono}, Robert {Lupton}, James~E. {Gunn}, Hiroshi
  {Karoji}, Hiroaki {Aihara}, Hitoshi {Murayama}, and Masahiro {Takada}.
\newblock {Properties of Weak Lensing Clusters Detected on Hyper
  Suprime-Cam{\textquoteright}s 2.3 deg$^{2}$ field}.
\newblock {\em \apj}, 807(1):22, July 2015.

\bibitem{des_16}
{Dark Energy Survey Collaboration}, T.~{Abbott}, F.~B. {Abdalla},
  J.~{Aleksi{\'c}}, S.~{Allam}, A.~{Amara}, D.~{Bacon}, E.~{Balbinot},
  M.~{Banerji}, K.~{Bechtol}, A.~{Benoit-L{\'e}vy}, G.~M. {Bernstein},
  E.~{Bertin}, J.~{Blazek}, C.~{Bonnett}, S.~{Bridle}, D.~{Brooks}, R.~J.
  {Brunner}, E.~{Buckley-Geer}, D.~L. {Burke}, G.~B. {Caminha}, D.~{Capozzi},
  J.~{Carlsen}, A.~{Carnero-Rosell}, M.~{Carollo}, M.~{Carrasco-Kind},
  J.~{Carretero}, F.~J. {Castander}, L.~{Clerkin}, T.~{Collett},
  C.~{Conselice}, M.~{Crocce}, C.~E. {Cunha}, C.~B. {D'Andrea}, L.~N. {da
  Costa}, T.~M. {Davis}, S.~{Desai}, H.~T. {Diehl}, J.~P. {Dietrich},
  S.~{Dodelson}, P.~{Doel}, A.~{Drlica-Wagner}, J.~{Estrada}, J.~{Etherington},
  A.~E. {Evrard}, J.~{Fabbri}, D.~A. {Finley}, B.~{Flaugher}, R.~J. {Foley},
  P.~{Fosalba}, J.~{Frieman}, J.~{Garc{\'\i}a-Bellido}, E.~{Gaztanaga}, D.~W.
  {Gerdes}, T.~{Giannantonio}, D.~A. {Goldstein}, D.~{Gruen}, R.~A. {Gruendl},
  P.~{Guarnieri}, G.~{Gutierrez}, W.~{Hartley}, K.~{Honscheid}, B.~{Jain},
  D.~J. {James}, T.~{Jeltema}, S.~{Jouvel}, R.~{Kessler}, A.~{King}, D.~{Kirk},
  R.~{Kron}, K.~{Kuehn}, N.~{Kuropatkin}, O.~{Lahav}, T.~S. {Li}, M.~{Lima},
  H.~{Lin}, M.~A.~G. {Maia}, M.~{Makler}, M.~{Manera}, C.~{Maraston}, J.~L.
  {Marshall}, P.~{Martini}, R.~G. {McMahon}, P.~{Melchior}, A.~{Merson}, C.~J.
  {Miller}, R.~{Miquel}, J.~J. {Mohr}, X.~{Morice-Atkinson}, K.~{Naidoo},
  E.~{Neilsen}, R.~C. {Nichol}, B.~{Nord}, R.~{Ogando}, F.~{Ostrovski},
  A.~{Palmese}, A.~{Papadopoulos}, H.~V. {Peiris}, J.~{Peoples}, W.~J.
  {Percival}, A.~A. {Plazas}, S.~L. {Reed}, A.~{Refregier}, A.~K. {Romer},
  A.~{Roodman}, A.~{Ross}, E.~{Rozo}, E.~S. {Rykoff}, I.~{Sadeh}, M.~{Sako},
  C.~{S{\'a}nchez}, E.~{Sanchez}, B.~{Santiago}, V.~{Scarpine}, M.~{Schubnell},
  I.~{Sevilla-Noarbe}, E.~{Sheldon}, M.~{Smith}, R.~C. {Smith},
  M.~{Soares-Santos}, F.~{Sobreira}, M.~{Soumagnac}, E.~{Suchyta},
  M.~{Sullivan}, M.~{Swanson}, G.~{Tarle}, J.~{Thaler}, D.~{Thomas}, R.~C.
  {Thomas}, D.~{Tucker}, J.~D. {Vieira}, V.~{Vikram}, A.~R. {Walker}, R.~H.
  {Wechsler}, J.~{Weller}, W.~{Wester}, L.~{Whiteway}, H.~{Wilcox}, B.~{Yanny},
  Y.~{Zhang}, and J.~{Zuntz}.
\newblock {The Dark Energy Survey: more than dark energy - an overview}.
\newblock {\em \mnras}, 460(2):1270--1299, August 2016.

\bibitem{boss_13}
Kyle~S. {Dawson}, David~J. {Schlegel}, Christopher~P. {Ahn}, Scott~F.
  {Anderson}, {\'E}ric {Aubourg}, Stephen {Bailey}, Robert~H. {Barkhouser},
  Julian~E. {Bautista}, Alessandra {Beifiori}, Andreas~A. {Berlind}, Vaishali
  {Bhardwaj}, Dmitry {Bizyaev}, Cullen~H. {Blake}, Michael~R. {Blanton},
  Michael {Blomqvist}, Adam~S. {Bolton}, Arnaud {Borde}, Jo~{Bovy}, W.~N.
  {Brandt}, Howard {Brewington}, Jon {Brinkmann}, Peter~J. {Brown}, Joel~R.
  {Brownstein}, Kevin {Bundy}, N.~G. {Busca}, William {Carithers}, Aurelio~R.
  {Carnero}, Michael~A. {Carr}, Yanmei {Chen}, Johan {Comparat}, Natalia
  {Connolly}, Frances {Cope}, Rupert A.~C. {Croft}, Antonio~J. {Cuesta},
  Luiz~N. {da Costa}, James R.~A. {Davenport}, Timoth{\'e}e {Delubac}, Roland
  {de Putter}, Saurav {Dhital}, Anne {Ealet}, Garrett~L. {Ebelke}, Daniel~J.
  {Eisenstein}, S.~{Escoffier}, Xiaohui {Fan}, N.~{Filiz Ak}, Hayley {Finley},
  Andreu {Font-Ribera}, R.~{G{\'e}nova-Santos}, James~E. {Gunn}, Hong {Guo},
  Daryl {Haggard}, Patrick~B. {Hall}, Jean-Christophe {Hamilton}, Ben {Harris},
  David~W. {Harris}, Shirley {Ho}, David~W. {Hogg}, Diana {Holder}, Klaus
  {Honscheid}, Joe {Huehnerhoff}, Beatrice {Jordan}, Wendell~P. {Jordan},
  Guinevere {Kauffmann}, Eyal~A. {Kazin}, David {Kirkby}, Mark~A. {Klaene},
  Jean-Paul {Kneib}, Jean-Marc {Le Goff}, Khee-Gan {Lee}, Daniel~C. {Long},
  Craig~P. {Loomis}, Britt {Lundgren}, Robert~H. {Lupton}, Marcio A.~G. {Maia},
  Martin {Makler}, Elena {Malanushenko}, Viktor {Malanushenko}, Rachel
  {Mandelbaum}, Marc {Manera}, Claudia {Maraston}, Daniel {Margala}, Karen~L.
  {Masters}, Cameron~K. {McBride}, Patrick {McDonald}, Ian~D. {McGreer},
  Richard~G. {McMahon}, Olga {Mena}, Jordi {Miralda-Escud{\'e}}, Antonio~D.
  {Montero-Dorta}, Francesco {Montesano}, Demitri {Muna}, Adam~D. {Myers},
  Tracy {Naugle}, Robert~C. {Nichol}, Pasquier {Noterdaeme}, Sebasti{\'a}n~E.
  {Nuza}, Matthew~D. {Olmstead}, Audrey {Oravetz}, Daniel~J. {Oravetz}, Russell
  {Owen}, Nikhil {Padmanabhan}, Nathalie {Palanque-Delabrouille}, Kaike {Pan},
  John~K. {Parejko}, Isabelle {P{\^a}ris}, Will~J. {Percival}, Ismael
  {P{\'e}rez-Fournon}, Ignasi {P{\'e}rez-R{\`a}fols}, Patrick {Petitjean},
  Robert {Pfaffenberger}, Janine {Pforr}, Matthew~M. {Pieri}, Francisco
  {Prada}, Adrian~M. {Price-Whelan}, M.~Jordan {Raddick}, Rafael {Rebolo},
  James {Rich}, Gordon~T. {Richards}, Constance~M. {Rockosi}, Natalie~A. {Roe},
  Ashley~J. {Ross}, Nicholas~P. {Ross}, Graziano {Rossi}, J.~A.
  {Rubi{\~n}o-Martin}, Lado {Samushia}, Ariel~G. {S{\'a}nchez}, Conor {Sayres},
  Sarah~J. {Schmidt}, Donald~P. {Schneider}, C.~G. {Sc{\'o}ccola}, Hee-Jong
  {Seo}, Alaina {Shelden}, Erin {Sheldon}, Yue {Shen}, Yiping {Shu}, An{\v{z}}e
  {Slosar}, Stephen~A. {Smee}, Stephanie~A. {Snedden}, Fritz {Stauffer}, Oliver
  {Steele}, Michael~A. {Strauss}, Alina {Streblyanska}, Nao {Suzuki}, Molly
  E.~C. {Swanson}, Tomer {Tal}, Masayuki {Tanaka}, Daniel {Thomas}, Jeremy~L.
  {Tinker}, Rita {Tojeiro}, Christy~A. {Tremonti}, M.~{Vargas Maga{\~n}a},
  Licia {Verde}, Matteo {Viel}, David~A. {Wake}, Mike {Watson}, Benjamin~A.
  {Weaver}, David~H. {Weinberg}, Benjamin~J. {Weiner}, Andrew~A. {West}, Martin
  {White}, W.~M. {Wood-Vasey}, Christophe {Yeche}, Idit {Zehavi}, Gong-Bo
  {Zhao}, and Zheng {Zheng}.
\newblock {The Baryon Oscillation Spectroscopic Survey of SDSS-III}.
\newblock {\em \aj}, 145(1):10, January 2013.

\bibitem{eboss_16}
Gong-Bo {Zhao}, Yuting {Wang}, Ashley~J. {Ross}, Sarah {Shandera}, Will~J.
  {Percival}, Kyle~S. {Dawson}, Jean-Paul {Kneib}, Adam~D. {Myers}, Joel~R.
  {Brownstein}, Johan {Comparat}, Timoth{\'e}e {Delubac}, Pengyuan {Gao},
  Alireza {Hojjati}, Kazuya {Koyama}, Cameron~K. {McBride}, Andr{\'e}s {Meza},
  Jeffrey~A. {Newman}, Nathalie {Palanque-Delabrouille}, Levon {Pogosian},
  Francisco {Prada}, Graziano {Rossi}, Donald~P. {Schneider}, Hee-Jong {Seo},
  Charling {Tao}, Dandan {Wang}, Christophe {Y{\`e}che}, Hanyu {Zhang},
  Yuecheng {Zhang}, Xu~{Zhou}, Fangzhou {Zhu}, and Hu~{Zou}.
\newblock {The extended Baryon Oscillation Spectroscopic Survey: a cosmological
  forecast}.
\newblock {\em \mnras}, 457(3):2377--2390, April 2016.

\bibitem{erosita_21}
P.~{Predehl}, R.~{Andritschke}, V.~{Arefiev}, V.~{Babyshkin}, O.~{Batanov},
  W.~{Becker}, H.~{B{\"o}hringer}, A.~{Bogomolov}, T.~{Boller}, K.~{Borm},
  W.~{Bornemann}, H.~{Br{\"a}uninger}, M.~{Br{\"u}ggen}, H.~{Brunner},
  M.~{Brusa}, E.~{Bulbul}, M.~{Buntov}, V.~{Burwitz}, W.~{Burkert}, N.~{Clerc},
  E.~{Churazov}, D.~{Coutinho}, T.~{Dauser}, K.~{Dennerl}, V.~{Doroshenko},
  J.~{Eder}, V.~{Emberger}, T.~{Eraerds}, A.~{Finoguenov}, M.~{Freyberg},
  P.~{Friedrich}, S.~{Friedrich}, M.~{F{\"u}rmetz}, A.~{Georgakakis},
  M.~{Gilfanov}, S.~{Granato}, C.~{Grossberger}, A.~{Gueguen}, P.~{Gureev},
  F.~{Haberl}, O.~{H{\"a}lker}, G.~{Hartner}, G.~{Hasinger}, H.~{Huber},
  L.~{Ji}, A.~v. {Kienlin}, W.~{Kink}, F.~{Korotkov}, I.~{Kreykenbohm},
  G.~{Lamer}, I.~{Lomakin}, I.~{Lapshov}, T.~{Liu}, C.~{Maitra},
  N.~{Meidinger}, B.~{Menz}, A.~{Merloni}, T.~{Mernik}, B.~{Mican}, J.~{Mohr},
  S.~{M{\"u}ller}, K.~{Nandra}, V.~{Nazarov}, F.~{Pacaud}, M.~{Pavlinsky},
  E.~{Perinati}, E.~{Pfeffermann}, D.~{Pietschner}, M.~E. {Ramos-Ceja},
  A.~{Rau}, J.~{Reiffers}, T.~H. {Reiprich}, J.~{Robrade}, M.~{Salvato},
  J.~{Sanders}, A.~{Santangelo}, M.~{Sasaki}, H.~{Scheuerle}, C.~{Schmid},
  J.~{Schmitt}, A.~{Schwope}, A.~{Shirshakov}, M.~{Steinmetz}, I.~{Stewart},
  L.~{Str{\"u}der}, R.~{Sunyaev}, C.~{Tenzer}, L.~{Tiedemann},
  J.~{Tr{\"u}mper}, V.~{Voron}, P.~{Weber}, J.~{Wilms}, and V.~{Yaroshenko}.
\newblock {The eROSITA X-ray telescope on SRG}.
\newblock {\em \aap}, 647:A1, March 2021.

\bibitem{desi_13}
Michael {Levi}, Chris {Bebek}, Timothy {Beers}, Robert {Blum}, Robert {Cahn},
  Daniel {Eisenstein}, Brenna {Flaugher}, Klaus {Honscheid}, Richard {Kron},
  Ofer {Lahav}, Patrick {McDonald}, Natalie {Roe}, David {Schlegel}, and
  {representing the DESI collaboration}.
\newblock {The DESI Experiment, a whitepaper for Snowmass 2013}.
\newblock {\em arXiv e-prints}, page arXiv:1308.0847, August 2013.

\bibitem{lsst_09}
{LSST Science Collaboration}, Paul~A. {Abell}, Julius {Allison}, Scott~F.
  {Anderson}, John~R. {Andrew}, J.~Roger~P. {Angel}, Lee {Armus}, David
  {Arnett}, S.~J. {Asztalos}, Tim~S. {Axelrod}, Stephen {Bailey}, D.~R.
  {Ballantyne}, Justin~R. {Bankert}, Wayne~A. {Barkhouse}, Jeffrey~D. {Barr},
  L.~Felipe {Barrientos}, Aaron~J. {Barth}, James~G. {Bartlett}, Andrew~C.
  {Becker}, Jacek {Becla}, Timothy~C. {Beers}, Joseph~P. {Bernstein}, Rahul
  {Biswas}, Michael~R. {Blanton}, Joshua~S. {Bloom}, John~J. {Bochanski}, Pat
  {Boeshaar}, Kirk~D. {Borne}, Marusa {Bradac}, W.~N. {Brandt}, Carrie~R.
  {Bridge}, Michael~E. {Brown}, Robert~J. {Brunner}, James~S. {Bullock},
  Adam~J. {Burgasser}, James~H. {Burge}, David~L. {Burke}, Phillip~A.
  {Cargile}, Srinivasan {Chandrasekharan}, George {Chartas}, Steven~R.
  {Chesley}, You-Hua {Chu}, David {Cinabro}, Mark~W. {Claire}, Charles~F.
  {Claver}, Douglas {Clowe}, A.~J. {Connolly}, Kem~H. {Cook}, Jeff {Cooke},
  Asantha {Cooray}, Kevin~R. {Covey}, Christopher~S. {Culliton}, Roelof {de
  Jong}, Willem~H. {de Vries}, Victor~P. {Debattista}, Francisco {Delgado},
  Ian~P. {Dell'Antonio}, Saurav {Dhital}, Rosanne {Di Stefano}, Mark
  {Dickinson}, Benjamin {Dilday}, S.~G. {Djorgovski}, Gregory {Dobler}, Ciro
  {Donalek}, Gregory {Dubois-Felsmann}, Josef {Durech}, Ardis {Eliasdottir},
  Michael {Eracleous}, Laurent {Eyer}, Emilio~E. {Falco}, Xiaohui {Fan},
  Christopher~D. {Fassnacht}, Harry~C. {Ferguson}, Yanga~R. {Fernandez},
  Brian~D. {Fields}, Douglas {Finkbeiner}, Eduardo~E. {Figueroa}, Derek~B.
  {Fox}, Harold {Francke}, James~S. {Frank}, Josh {Frieman}, Sebastien
  {Fromenteau}, Muhammad {Furqan}, Gaspar {Galaz}, A.~{Gal-Yam}, Peter
  {Garnavich}, Eric {Gawiser}, John {Geary}, Perry {Gee}, Robert~R. {Gibson},
  Kirk {Gilmore}, Emily~A. {Grace}, Richard~F. {Green}, William~J. {Gressler},
  Carl~J. {Grillmair}, Salman {Habib}, J.~S. {Haggerty}, Mario {Hamuy}, Alan~W.
  {Harris}, Suzanne~L. {Hawley}, Alan~F. {Heavens}, Leslie {Hebb}, Todd~J.
  {Henry}, Edward {Hileman}, Eric~J. {Hilton}, Keri {Hoadley}, J.~B. {Holberg},
  Matt~J. {Holman}, Steve~B. {Howell}, Leopoldo {Infante}, Zeljko {Ivezic},
  Suzanne~H. {Jacoby}, Bhuvnesh {Jain}, {R}, {Jedicke}, M.~James {Jee},
  J.~{Garrett Jernigan}, Saurabh~W. {Jha}, Kathryn~V. {Johnston}, R.~Lynne
  {Jones}, Mario {Juric}, Mikko {Kaasalainen}, {Styliani}, {Kafka}, Steven~M.
  {Kahn}, Nathan~A. {Kaib}, Jason {Kalirai}, Jeff {Kantor}, Mansi~M.
  {Kasliwal}, Charles~R. {Keeton}, Richard {Kessler}, Zoran {Knezevic}, Adam
  {Kowalski}, Victor~L. {Krabbendam}, K.~Simon {Krughoff}, Shrinivas
  {Kulkarni}, Stephen {Kuhlman}, Mark {Lacy}, Sebastien {Lepine}, Ming {Liang},
  Amy {Lien}, Paulina {Lira}, Knox~S. {Long}, Suzanne {Lorenz}, Jennifer~M.
  {Lotz}, R.~H. {Lupton}, Julie {Lutz}, Lucas~M. {Macri}, Ashish~A. {Mahabal},
  Rachel {Mandelbaum}, Phil {Marshall}, Morgan {May}, Peregrine~M. {McGehee},
  Brian~T. {Meadows}, Alan {Meert}, Andrea {Milani}, Christopher~J. {Miller},
  Michelle {Miller}, David {Mills}, Dante {Minniti}, David {Monet}, Anjum~S.
  {Mukadam}, Ehud {Nakar}, Douglas~R. {Neill}, Jeffrey~A. {Newman}, Sergei
  {Nikolaev}, Martin {Nordby}, Paul {O'Connor}, Masamune {Oguri}, John
  {Oliver}, Scot~S. {Olivier}, Julia~K. {Olsen}, Knut {Olsen}, Edward~W.
  {Olszewski}, Hakeem {Oluseyi}, Nelson~D. {Padilla}, Alex {Parker}, Joshua
  {Pepper}, John~R. {Peterson}, Catherine {Petry}, Philip~A. {Pinto}, James~L.
  {Pizagno}, Bogdan {Popescu}, Andrej {Prsa}, Veljko {Radcka}, M.~Jordan
  {Raddick}, Andrew {Rasmussen}, Arne {Rau}, Jeonghee {Rho}, James~E. {Rhoads},
  Gordon~T. {Richards}, Stephen~T. {Ridgway}, Brant~E. {Robertson}, Rok
  {Roskar}, Abhijit {Saha}, Ata {Sarajedini}, Evan {Scannapieco}, Terry
  {Schalk}, Rafe {Schindler}, Samuel {Schmidt}, Sarah {Schmidt}, Donald~P.
  {Schneider}, German {Schumacher}, Ryan {Scranton}, Jacques {Sebag}, Lynn~G.
  {Seppala}, Ohad {Shemmer}, Joshua~D. {Simon}, M.~{Sivertz}, Howard~A.
  {Smith}, J.~{Allyn Smith}, Nathan {Smith}, Anna~H. {Spitz}, Adam {Stanford},
  Keivan~G. {Stassun}, Jay {Strader}, Michael~A. {Strauss}, Christopher~W.
  {Stubbs}, Donald~W. {Sweeney}, Alex {Szalay}, Paula {Szkody}, Masahiro
  {Takada}, Paul {Thorman}, David~E. {Trilling}, Virginia {Trimble}, Anthony
  {Tyson}, Richard {Van Berg}, Daniel {Vanden Berk}, Jake {VanderPlas}, Licia
  {Verde}, Bojan {Vrsnak}, Lucianne~M. {Walkowicz}, Benjamin~D. {Wandelt},
  Sheng {Wang}, Yun {Wang}, Michael {Warner}, Risa~H. {Wechsler}, Andrew~A.
  {West}, Oliver {Wiecha}, Benjamin~F. {Williams}, Beth {Willman}, David
  {Wittman}, Sidney~C. {Wolff}, W.~Michael {Wood-Vasey}, Przemek {Wozniak},
  Patrick {Young}, Andrew {Zentner}, and Hu~{Zhan}.
\newblock {LSST Science Book, Version 2.0}.
\newblock {\em arXiv e-prints}, page arXiv:0912.0201, December 2009.

\bibitem{euclid_11}
R.~{Laureijs}, J.~{Amiaux}, S.~{Arduini}, J.~L. {Augu{\`e}res},
  J.~{Brinchmann}, R.~{Cole}, M.~{Cropper}, C.~{Dabin}, L.~{Duvet}, A.~{Ealet},
  B.~{Garilli}, P.~{Gondoin}, L.~{Guzzo}, J.~{Hoar}, H.~{Hoekstra},
  R.~{Holmes}, T.~{Kitching}, T.~{Maciaszek}, Y.~{Mellier}, F.~{Pasian},
  W.~{Percival}, J.~{Rhodes}, G.~{Saavedra Criado}, M.~{Sauvage},
  R.~{Scaramella}, L.~{Valenziano}, S.~{Warren}, R.~{Bender}, F.~{Castander},
  A.~{Cimatti}, O.~{Le F{\`e}vre}, H.~{Kurki-Suonio}, M.~{Levi}, P.~{Lilje},
  G.~{Meylan}, R.~{Nichol}, K.~{Pedersen}, V.~{Popa}, R.~{Rebolo Lopez}, H.~W.
  {Rix}, H.~{Rottgering}, W.~{Zeilinger}, F.~{Grupp}, P.~{Hudelot},
  R.~{Massey}, M.~{Meneghetti}, L.~{Miller}, S.~{Paltani},
  S.~{Paulin-Henriksson}, S.~{Pires}, C.~{Saxton}, T.~{Schrabback},
  G.~{Seidel}, J.~{Walsh}, N.~{Aghanim}, L.~{Amendola}, J.~{Bartlett},
  C.~{Baccigalupi}, J.~P. {Beaulieu}, K.~{Benabed}, J.~G. {Cuby}, D.~{Elbaz},
  P.~{Fosalba}, G.~{Gavazzi}, A.~{Helmi}, I.~{Hook}, M.~{Irwin}, J.~P. {Kneib},
  M.~{Kunz}, F.~{Mannucci}, L.~{Moscardini}, C.~{Tao}, R.~{Teyssier},
  J.~{Weller}, G.~{Zamorani}, M.~R. {Zapatero Osorio}, O.~{Boulade}, J.~J.
  {Foumond}, A.~{Di Giorgio}, P.~{Guttridge}, A.~{James}, M.~{Kemp},
  J.~{Martignac}, A.~{Spencer}, D.~{Walton}, T.~{Bl{\"u}mchen}, C.~{Bonoli},
  F.~{Bortoletto}, C.~{Cerna}, L.~{Corcione}, C.~{Fabron}, K.~{Jahnke},
  S.~{Ligori}, F.~{Madrid}, L.~{Martin}, G.~{Morgante}, T.~{Pamplona},
  E.~{Prieto}, M.~{Riva}, R.~{Toledo}, M.~{Trifoglio}, F.~{Zerbi},
  F.~{Abdalla}, M.~{Douspis}, C.~{Grenet}, S.~{Borgani}, R.~{Bouwens},
  F.~{Courbin}, J.~M. {Delouis}, P.~{Dubath}, A.~{Fontana}, M.~{Frailis},
  A.~{Grazian}, J.~{Koppenh{\"o}fer}, O.~{Mansutti}, M.~{Melchior},
  M.~{Mignoli}, J.~{Mohr}, C.~{Neissner}, K.~{Noddle}, M.~{Poncet},
  M.~{Scodeggio}, S.~{Serrano}, N.~{Shane}, J.~L. {Starck}, C.~{Surace},
  A.~{Taylor}, G.~{Verdoes-Kleijn}, C.~{Vuerli}, O.~R. {Williams},
  A.~{Zacchei}, B.~{Altieri}, I.~{Escudero Sanz}, R.~{Kohley},
  T.~{Oosterbroek}, P.~{Astier}, D.~{Bacon}, S.~{Bardelli}, C.~{Baugh},
  F.~{Bellagamba}, C.~{Benoist}, D.~{Bianchi}, A.~{Biviano}, E.~{Branchini},
  C.~{Carbone}, V.~{Cardone}, D.~{Clements}, S.~{Colombi}, C.~{Conselice},
  G.~{Cresci}, N.~{Deacon}, J.~{Dunlop}, C.~{Fedeli}, F.~{Fontanot},
  P.~{Franzetti}, C.~{Giocoli}, J.~{Garcia-Bellido}, J.~{Gow}, A.~{Heavens},
  P.~{Hewett}, C.~{Heymans}, A.~{Holland}, Z.~{Huang}, O.~{Ilbert},
  B.~{Joachimi}, E.~{Jennins}, E.~{Kerins}, A.~{Kiessling}, D.~{Kirk},
  R.~{Kotak}, O.~{Krause}, O.~{Lahav}, F.~{van Leeuwen}, J.~{Lesgourgues},
  M.~{Lombardi}, M.~{Magliocchetti}, K.~{Maguire}, E.~{Majerotto}, R.~{Maoli},
  F.~{Marulli}, S.~{Maurogordato}, H.~{McCracken}, R.~{McLure},
  A.~{Melchiorri}, A.~{Merson}, M.~{Moresco}, M.~{Nonino}, P.~{Norberg},
  J.~{Peacock}, R.~{Pello}, M.~{Penny}, V.~{Pettorino}, C.~{Di Porto},
  L.~{Pozzetti}, C.~{Quercellini}, M.~{Radovich}, A.~{Rassat}, N.~{Roche},
  S.~{Ronayette}, E.~{Rossetti}, B.~{Sartoris}, P.~{Schneider}, E.~{Semboloni},
  S.~{Serjeant}, F.~{Simpson}, C.~{Skordis}, G.~{Smadja}, S.~{Smartt},
  P.~{Spano}, S.~{Spiro}, M.~{Sullivan}, A.~{Tilquin}, R.~{Trotta}, L.~{Verde},
  Y.~{Wang}, G.~{Williger}, G.~{Zhao}, J.~{Zoubian}, and E.~{Zucca}.
\newblock {Euclid Definition Study Report}.
\newblock {\em arXiv e-prints}, page arXiv:1110.3193, October 2011.

\bibitem{4most_12}
Roelof~S. {de Jong}, Olga {Bellido-Tirado}, Cristina {Chiappini}, {\'E}ric
  {Depagne}, Roger {Haynes}, Diana {Johl}, Olivier {Schnurr}, Axel {Schwope},
  Jakob {Walcher}, Frank {Dionies}, Dionne {Haynes}, Andreas {Kelz},
  Francisco~S. {Kitaura}, Georg {Lamer}, Ivan {Minchev}, Volker {M{\"u}ller},
  Sebasti{\'a}n.~E. {Nuza}, Jean-Christophe {Olaya}, Tilmann {Piffl}, Emil
  {Popow}, Matthias {Steinmetz}, Ugur {Ural}, Mary {Williams}, Roland
  {Winkler}, Lutz {Wisotzki}, Wolfgang~R. {Ansorge}, Manda {Banerji}, Eduardo
  {Gonzalez Solares}, Mike {Irwin}, Robert~C. {Kennicutt}, Dave {King},
  Richard~G. {McMahon}, Sergey {Koposov}, Ian~R. {Parry}, David {Sun},
  Nicholas~A. {Walton}, Gert {Finger}, Olaf {Iwert}, Mirko {Krumpe}, Jean-Louis
  {Lizon}, Mainieri {Vincenzo}, Jean-Philippe {Amans}, Piercarlo {Bonifacio},
  Mathieu {Cohen}, Patrick {Francois}, Pascal {Jagourel}, Shan~B. {Mignot},
  Fr{\'e}d{\'e}ric {Royer}, Paola {Sartoretti}, Ralf {Bender}, Frank {Grupp},
  Hans-Joachim {Hess}, Florian {Lang-Bardl}, Bernard {Muschielok}, Hans
  {B{\"o}hringer}, Thomas {Boller}, Angela {Bongiorno}, Marcella {Brusa}, Tom
  {Dwelly}, Andrea {Merloni}, Kirpal {Nandra}, Mara {Salvato}, Johannes~H.
  {Pragt}, Ram{\'o}n {Navarro}, Gerrit {Gerlofsma}, Ronald {Roelfsema},
  Gavin~B. {Dalton}, Kevin~F. {Middleton}, Ian~A. {Tosh}, Corrado {Boeche},
  Elisabetta {Caffau}, Norbert {Christlieb}, Eva~K. {Grebel}, Camilla {Hansen},
  Andreas {Koch}, Hans-G. {Ludwig}, Andreas {Quirrenbach}, Luca {Sbordone},
  Walter {Seifert}, Guido {Thimm}, Trifon {Trifonov}, Amina {Helmi}, Scott~C.
  {Trager}, Sofia {Feltzing}, Andreas {Korn}, and Wilfried {Boland}.
\newblock {4MOST: 4-metre multi-object spectroscopic telescope}.
\newblock In Ian~S. {McLean}, Suzanne~K. {Ramsay}, and Hideki {Takami},
  editors, {\em Ground-based and Airborne Instrumentation for Astronomy IV},
  volume 8446 of {\em Society of Photo-Optical Instrumentation Engineers (SPIE)
  Conference Series}, page 84460T, September 2012.

\bibitem{pfs_14}
Masahiro {Takada}, Richard~S. {Ellis}, Masashi {Chiba}, Jenny~E. {Greene},
  Hiroaki {Aihara}, Nobuo {Arimoto}, Kevin {Bundy}, Judith {Cohen}, Olivier
  {Dor{\'e}}, Genevieve {Graves}, James~E. {Gunn}, Timothy {Heckman},
  Christopher~M. {Hirata}, Paul {Ho}, Jean-Paul {Kneib}, Olivier {Le
  F{\`e}vre}, Lihwai {Lin}, Surhud {More}, Hitoshi {Murayama}, Tohru {Nagao},
  Masami {Ouchi}, Michael {Seiffert}, John~D. {Silverman}, Laerte {Sodr{\'e}},
  David~N. {Spergel}, Michael~A. {Strauss}, Hajime {Sugai}, Yasushi {Suto},
  Hideki {Takami}, and Rosemary {Wyse}.
\newblock {Extragalactic science, cosmology, and Galactic archaeology with the
  Subaru Prime Focus Spectrograph}.
\newblock {\em \pasj}, 66(1):R1, February 2014.

\bibitem{ska_15}
Roy {Maartens}, Filipe~B. {Abdalla}, Matt {Jarvis}, and Mario~G. {Santos}.
\newblock {Cosmology with the SKA -- overview}.
\newblock {\em arXiv e-prints}, page arXiv:1501.04076, January 2015.

\bibitem{roman_15}
D.~{Spergel}, N.~{Gehrels}, C.~{Baltay}, D.~{Bennett}, J.~{Breckinridge},
  M.~{Donahue}, A.~{Dressler}, B.~S. {Gaudi}, T.~{Greene}, O.~{Guyon},
  C.~{Hirata}, J.~{Kalirai}, N.~J. {Kasdin}, B.~{Macintosh}, W.~{Moos},
  S.~{Perlmutter}, M.~{Postman}, B.~{Rauscher}, J.~{Rhodes}, Y.~{Wang},
  D.~{Weinberg}, D.~{Benford}, M.~{Hudson}, W.~S. {Jeong}, Y.~{Mellier},
  W.~{Traub}, T.~{Yamada}, P.~{Capak}, J.~{Colbert}, D.~{Masters}, M.~{Penny},
  D.~{Savransky}, D.~{Stern}, N.~{Zimmerman}, R.~{Barry}, L.~{Bartusek},
  K.~{Carpenter}, E.~{Cheng}, D.~{Content}, F.~{Dekens}, R.~{Demers},
  K.~{Grady}, C.~{Jackson}, G.~{Kuan}, J.~{Kruk}, M.~{Melton}, B.~{Nemati},
  B.~{Parvin}, I.~{Poberezhskiy}, C.~{Peddie}, J.~{Ruffa}, J.~K. {Wallace},
  A.~{Whipple}, E.~{Wollack}, and F.~{Zhao}.
\newblock {Wide-Field InfrarRed Survey Telescope-Astrophysics Focused Telescope
  Assets WFIRST-AFTA 2015 Report}.
\newblock {\em arXiv e-prints}, page arXiv:1503.03757, March 2015.

\bibitem{rudd_08}
Douglas~H. {Rudd}, Andrew~R. {Zentner}, and Andrey~V. {Kravtsov}.
\newblock {Effects of Baryons and Dissipation on the Matter Power Spectrum}.
\newblock {\em \apj}, 672(1):19--32, January 2008.

\bibitem{chisari_19}
Nora~Elisa {Chisari}, Alexander~J. {Mead}, Shahab {Joudaki}, Pedro~G.
  {Ferreira}, Aurel {Schneider}, Joseph {Mohr}, Tilman {Tr{\"o}ster}, David
  {Alonso}, Ian~G. {McCarthy}, Sergio {Martin-Alvarez}, Julien {Devriendt},
  Adrianne {Slyz}, and Marcel~P. {van Daalen}.
\newblock {Modelling baryonic feedback for survey cosmology}.
\newblock {\em The Open Journal of Astrophysics}, 2(1):4, June 2019.

\bibitem{vandaalen_20}
Marcel~P. {van Daalen}, Ian~G. {McCarthy}, and Joop {Schaye}.
\newblock {Exploring the effects of galaxy formation on matter clustering
  through a library of simulation power spectra}.
\newblock {\em \mnras}, 491(2):2424--2446, January 2020.

\bibitem{villaescusa_21}
Francisco {Villaescusa-Navarro}, Daniel {Angl{\'e}s-Alc{\'a}zar}, Shy {Genel},
  David~N. {Spergel}, Rachel~S. {Somerville}, Romeel {Dave}, Annalisa
  {Pillepich}, Lars {Hernquist}, Dylan {Nelson}, Paul {Torrey}, Desika
  {Narayanan}, Yin {Li}, Oliver {Philcox}, Valentina {La Torre}, Ana {Maria
  Delgado}, Shirley {Ho}, Sultan {Hassan}, Blakesley {Burkhart}, Digvijay
  {Wadekar}, Nicholas {Battaglia}, Gabriella {Contardo}, and Greg~L. {Bryan}.
\newblock {The CAMELS Project: Cosmology and Astrophysics with Machine-learning
  Simulations}.
\newblock {\em \apj}, 915(1):71, July 2021.

\bibitem{moser_22}
Emily {Moser}, Nicholas {Battaglia}, Daisuke {Nagai}, Erwin {Lau},
  Luis~Fernando {Machado Poletti Valle}, Francisco {Villaescusa-Navarro},
  Stefania {Amodeo}, Daniel {Angl{\'e}s-Alc{\'a}zar}, Greg~L. {Bryan}, Romeel
  {Dave}, Lars {Hernquist}, and Mark {Vogelsberger}.
\newblock {The Circumgalactic Medium from the CAMELS Simulations: Forecasting
  Constraints on Feedback Processes from Future Sunyaev-Zeldovich
  Observations}.
\newblock {\em \apj}, 933(2):133, July 2022.

\bibitem{wadekar_23}
Digvijay {Wadekar}, Leander {Thiele}, J.~Colin {Hill}, Shivam {Pandey},
  Francisco {Villaescusa-Navarro}, David~N. {Spergel}, Miles {Cranmer}, Daisuke
  {Nagai}, Daniel {Angl{\'e}s-Alc{\'a}zar}, Shirley {Ho}, and Lars {Hernquist}.
\newblock {The SZ flux-mass (Y-M) relation at low-halo masses: improvements
  with symbolic regression and strong constraints on baryonic feedback}.
\newblock {\em \mnras}, 522(2):2628--2643, June 2023.

\bibitem{pandey_23}
Shivam {Pandey}, Kai {Lehman}, Eric~J. {Baxter}, Yueying {Ni}, Daniel
  {Angl{\'e}s-Alc{\'a}zar}, Shy {Genel}, Francisco {Villaescusa-Navarro},
  Ana~Maria {Delgado}, and Tiziana {di Matteo}.
\newblock {Inferring the impact of feedback on the matter distribution using
  the Sunyaev Zel'dovich effect: insights from CAMELS simulations and ACT + DES
  data}.
\newblock {\em \mnras}, 525(2):1779--1794, October 2023.

\bibitem{delgado_23}
Ana~Maria {Delgado}, Daniel {Angl{\'e}s-Alc{\'a}zar}, Leander {Thiele}, Shivam
  {Pandey}, Kai {Lehman}, Rachel~S. {Somerville}, Michelle {Ntampaka}, Shy
  {Genel}, Francisco {Villaescusa-Navarro}, and Lars {Hernquist}.
\newblock {Predicting the impact of feedback on matter clustering with machine
  learning in CAMELS}.
\newblock {\em \mnras}, 526(4):5306--5325, December 2023.

\bibitem{to_24}
Chun-Hao {To}, Shivam {Pandey}, Elisabeth {Krause}, Nihar {Dalal}, Dhayaa
  {Anbajagane}, and David~H. {Weinberg}.
\newblock {Deciphering baryonic feedback with galaxy clusters}.
\newblock {\em arXiv e-prints}, page arXiv:2402.00110, January 2024.

\bibitem{bigwood_24}
L.~{Bigwood}, A.~{Amon}, A.~{Schneider}, J.~{Salcido}, I.~G. {McCarthy},
  C.~{Preston}, D.~{Sanchez}, D.~{Sijacki}, E.~{Schaan}, S.~{Ferraro},
  N.~{Battaglia}, A.~{Chen}, S.~{Dodelson}, A.~{Roodman}, A.~{Pieres},
  A.~{Ferte}, A.~{Alarcon}, A.~{Drlica-Wagner}, A.~{Choi}, A.~{Navarro-Alsina},
  A.~{Campos}, A.~J. {Ross}, A.~{Carnero Rosell}, B.~{Yin}, B.~{Yanny},
  C.~{Sanchez}, C.~{Chang}, C.~{Davis}, C.~{Doux}, D.~{Gruen}, E.~S. {Rykoff},
  E.~M. {Huff}, E.~{Sheldon}, F.~{Tarsitano}, F.~{Andrade-Oliveira}, G.~M.
  {Bernstein}, G.~{Giannini}, H.~T. {Diehl}, H.~{Huang}, I.~{Harrison},
  I.~{Sevilla-Noarbe}, I.~{Tutusaus}, J.~{Elvin-Poole}, J.~{McCullough},
  J.~{Zuntz}, J.~{Blazek}, J.~{DeRose}, J.~{Cordero}, J.~{Prat}, J.~{Myles},
  K.~{Eckert}, K.~{Bechtol}, K.~{Herner}, L.~F. {Secco}, M.~{Gatti},
  M.~{Raveri}, M.~{Carrasco Kind}, M.~R. {Becker}, M.~A. {Troxel}, M.~{Jarvis},
  N.~{MacCrann}, O.~{Friedrich}, O.~{Alves}, P.~F. {Leget}, R.~{Chen}, R.~P.
  {Rollins}, R.~H. {Wechsler}, R.~A. {Gruendl}, R.~{Cawthon}, S.~{Allam}, S.~L.
  {Bridle}, S.~{Pandey}, S.~{Everett}, T.~{Shin}, W.~G. {Hartley}, X.~{Fang},
  Y.~{Zhang}, M.~{Aguena}, J.~{Annis}, D.~{Bacon}, E.~{Bertin}, S.~{Bocquet},
  D.~{Brooks}, J.~{Carretero}, F.~J. {Castander}, L.~N. {da Costa}, M.~E.~S.
  {Pereira}, J.~{De Vicente}, S.~{Desai}, P.~{Doel}, I.~{Ferrero},
  B.~{Flaugher}, J.~{Frieman}, J.~{Garcia-Bellido}, E.~{Gaztanaga},
  G.~{Gutierrez}, S.~R. {Hinton}, D.~L. {Hollowood}, K.~{Honscheid},
  D.~{Huterer}, D.~J. {James}, K.~{Kuehn}, O.~{Lahav}, S.~{Lee}, J.~L.
  {Marshall}, J.~{Mena-Fernandez}, R.~{Miquel}, J.~{Muir}, M.~{Paterno}, A.~A.
  {Plazas Malagon}, A.~{Porredon}, A.~K. {Romer}, S.~{Samuroff}, E.~{Sanchez},
  D.~{Sanchez Cid}, M.~{Smith}, M.~{Soares-Santos}, E.~{Suchyta}, M.~E.~C.
  {Swanson}, G.~{Tarle}, C.~{To}, N.~{Weaverdyck}, J.~{Weller}, P.~{Wiseman},
  and M.~{Yamamoto}.
\newblock {Weak lensing combined with the kinetic Sunyaev Zel'dovich effect: A
  study of baryonic feedback}.
\newblock {\em arXiv e-prints}, page arXiv:2404.06098, April 2024.

\bibitem{mccarthy_24}
Ian~G. {McCarthy}, Alexandra {Amon}, Joop {Schaye}, Emmanuel {Schaan}, Raul~E.
  {Angulo}, Jaime {Salcido}, Matthieu {Schaller}, Leah {Bigwood}, Willem
  {Elbers}, Roi {Kugel}, John~C. {Helly}, Victor~J. {Forouhar Moreno},
  Carlos~S. {Frenk}, Robert~J. {McGibbon}, Lurdes {Ondaro-Mallea}, and
  Marcel~P. {van Daalen}.
\newblock {FLAMINGO: combining kinetic SZ effect and galaxy-galaxy lensing
  measurements to gauge the impact of feedback on large-scale structure}.
\newblock {\em arXiv e-prints}, page arXiv:2410.19905, October 2024.

\bibitem{bahar_24}
Y.~E. {Bahar}, E.~{Bulbul}, V.~{Ghirardini}, J.~S. {Sanders}, X.~{Zhang},
  A.~{Liu}, N.~{Clerc}, E.~{Artis}, F.~{Balzer}, V.~{Biffi}, S.~{Bose},
  J.~{Comparat}, K.~{Dolag}, C.~{Garrel}, B.~{Hadzhiyska},
  C.~{Hern{\'a}ndez-Aguayo}, L.~{Hernquist}, M.~{Kluge}, S.~{Krippendorf},
  A.~{Merloni}, K.~{Nandra}, R.~{Pakmor}, P.~{Popesso}, M.~{Ramos-Ceja},
  R.~{Seppi}, V.~{Springel}, J.~{Weller}, and S.~{Zelmer}.
\newblock {The SRG/eROSITA All-Sky Survey: Constraints on AGN Feedback in
  Galaxy Groups}.
\newblock {\em arXiv e-prints}, page arXiv:2401.17276, January 2024.

\bibitem{theis_24}
Alexander {Theis}, Steffen {Hagstotz}, Robert {Reischke}, and Jochen {Weller}.
\newblock {Galaxy dispersion measured by Fast Radio Bursts as a probe of
  baryonic feedback models}.
\newblock {\em arXiv e-prints}, page arXiv:2403.08611, March 2024.

\bibitem{medlock_24}
Isabel {Medlock}, Daisuke {Nagai}, Priyanka {Singh}, Benjamin {Oppenheimer},
  Daniel {Angl{\'e}s Alc{\'a}zar}, and Francisco {Villaescusa-Navarro}.
\newblock {Probing the Circum-Galactic Medium with Fast Radio Bursts: Insights
  from the CAMELS Simulations}.
\newblock {\em arXiv e-prints}, page arXiv:2403.02313, March 2024.

\bibitem{chiang_14}
Chi-Ting {Chiang}, Christian {Wagner}, Fabian {Schmidt}, and Eiichiro
  {Komatsu}.
\newblock {Position-dependent power spectrum of the large-scale structure: a
  novel method to measure the squeezed-limit bispectrum}.
\newblock {\em \jcap}, 2014(5):048, May 2014.

\bibitem{chiang_15}
Chi-Ting {Chiang}, Christian {Wagner}, Ariel~G. {S{\'a}nchez}, Fabian
  {Schmidt}, and Eiichiro {Komatsu}.
\newblock {Position-dependent correlation function from the SDSS-III Baryon
  Oscillation Spectroscopic Survey Data Release 10 CMASS sample}.
\newblock {\em \jcap}, 2015(9):028--028, September 2015.

\bibitem{halder_21}
Anik {Halder}, Oliver {Friedrich}, Stella {Seitz}, and Tamas~N. {Varga}.
\newblock {The integrated three-point correlation function of cosmic shear}.
\newblock {\em \mnras}, 506(2):2780--2803, September 2021.

\bibitem{halder_23}
Anik {Halder}, Zhengyangguang {Gong}, Alexandre {Barreira}, Oliver {Friedrich},
  Stella {Seitz}, and Daniel {Gruen}.
\newblock {Beyond 3{\texttimes}2-point cosmology: the integrated shear and
  galaxy 3-point correlation functions}.
\newblock {\em \jcap}, 2023(10):028, October 2023.

\bibitem{friedrich_18}
O.~{Friedrich}, D.~{Gruen}, J.~{DeRose}, D.~{Kirk}, E.~{Krause},
  T.~{McClintock}, E.~S. {Rykoff}, S.~{Seitz}, R.~H. {Wechsler}, G.~M.
  {Bernstein}, J.~{Blazek}, C.~{Chang}, S.~{Hilbert}, B.~{Jain}, A.~{Kovacs},
  O.~{Lahav}, F.~B. {Abdalla}, S.~{Allam}, J.~{Annis}, K.~{Bechtol},
  A.~{Benoit-L{\'e}vy}, E.~{Bertin}, D.~{Brooks}, A.~{Carnero Rosell},
  M.~{Carrasco Kind}, J.~{Carretero}, C.~E. {Cunha}, C.~B. {D'Andrea}, L.~N.
  {da Costa}, C.~{Davis}, S.~{Desai}, H.~T. {Diehl}, J.~P. {Dietrich},
  A.~{Drlica-Wagner}, T.~F. {Eifler}, P.~{Fosalba}, J.~{Frieman},
  J.~{Garc{\'\i}a-Bellido}, E.~{Gaztanaga}, D.~W. {Gerdes}, T.~{Giannantonio},
  R.~A. {Gruendl}, J.~{Gschwend}, G.~{Gutierrez}, K.~{Honscheid}, D.~J.
  {James}, M.~{Jarvis}, K.~{Kuehn}, N.~{Kuropatkin}, M.~{Lima}, M.~{March},
  J.~L. {Marshall}, P.~{Melchior}, F.~{Menanteau}, R.~{Miquel}, J.~J. {Mohr},
  B.~{Nord}, A.~A. {Plazas}, E.~{Sanchez}, V.~{Scarpine}, R.~{Schindler},
  M.~{Schubnell}, I.~{Sevilla-Noarbe}, E.~{Sheldon}, M.~{Smith},
  M.~{Soares-Santos}, F.~{Sobreira}, E.~{Suchyta}, M.~E.~C. {Swanson},
  G.~{Tarle}, D.~{Thomas}, M.~A. {Troxel}, V.~{Vikram}, J.~{Weller}, and {DES
  Collaboration}.
\newblock {Density split statistics: Joint model of counts and lensing in
  cells}.
\newblock {\em \prd}, 98(2):023508, July 2018.

\bibitem{gruen_18}
D.~{Gruen}, O.~{Friedrich}, E.~{Krause}, J.~{DeRose}, R.~{Cawthon}, C.~{Davis},
  J.~{Elvin-Poole}, E.~S. {Rykoff}, R.~H. {Wechsler}, A.~{Alarcon}, G.~M.
  {Bernstein}, J.~{Blazek}, C.~{Chang}, J.~{Clampitt}, M.~{Crocce}, J.~{De
  Vicente}, M.~{Gatti}, M.~S.~S. {Gill}, W.~G. {Hartley}, S.~{Hilbert},
  B.~{Hoyle}, B.~{Jain}, M.~{Jarvis}, O.~{Lahav}, N.~{MacCrann},
  T.~{McClintock}, J.~{Prat}, R.~P. {Rollins}, A.~J. {Ross}, E.~{Rozo},
  S.~{Samuroff}, C.~{S{\'a}nchez}, E.~{Sheldon}, M.~A. {Troxel}, J.~{Zuntz},
  T.~M.~C. {Abbott}, F.~B. {Abdalla}, S.~{Allam}, J.~{Annis}, K.~{Bechtol},
  A.~{Benoit-L{\'e}vy}, E.~{Bertin}, S.~L. {Bridle}, D.~{Brooks},
  E.~{Buckley-Geer}, A.~{Carnero Rosell}, M.~{Carrasco Kind}, J.~{Carretero},
  C.~E. {Cunha}, C.~B. {D'Andrea}, L.~N. {da Costa}, S.~{Desai}, H.~T. {Diehl},
  J.~P. {Dietrich}, P.~{Doel}, A.~{Drlica-Wagner}, E.~{Fernandez},
  B.~{Flaugher}, P.~{Fosalba}, J.~{Frieman}, J.~{Garc{\'\i}a-Bellido},
  E.~{Gaztanaga}, T.~{Giannantonio}, R.~A. {Gruendl}, J.~{Gschwend},
  G.~{Gutierrez}, K.~{Honscheid}, D.~J. {James}, T.~{Jeltema}, K.~{Kuehn},
  N.~{Kuropatkin}, M.~{Lima}, M.~{March}, J.~L. {Marshall}, P.~{Martini},
  P.~{Melchior}, F.~{Menanteau}, R.~{Miquel}, J.~J. {Mohr}, A.~A. {Plazas},
  A.~{Roodman}, E.~{Sanchez}, V.~{Scarpine}, M.~{Schubnell},
  I.~{Sevilla-Noarbe}, M.~{Smith}, R.~C. {Smith}, M.~{Soares-Santos},
  F.~{Sobreira}, M.~E.~C. {Swanson}, G.~{Tarle}, D.~{Thomas}, V.~{Vikram},
  A.~R. {Walker}, J.~{Weller}, Y.~{Zhang}, and {DES Collaboration}.
\newblock {Density split statistics: Cosmological constraints from counts and
  lensing in cells in DES Y1 and SDSS data}.
\newblock {\em \prd}, 98(2):023507, July 2018.

\bibitem{nguyen_24}
Nhat-Minh {Nguyen}, Fabian {Schmidt}, Beatriz {Tucci}, Martin {Reinecke}, and
  Andrija {Kosti{\'c}}.
\newblock {How much information can be extracted from galaxy clustering at the
  field level?}
\newblock {\em arXiv e-prints}, page arXiv:2403.03220, March 2024.

\bibitem{alsing_19}
Justin {Alsing}, Tom {Charnock}, Stephen {Feeney}, and Benjamin {Wandelt}.
\newblock {Fast likelihood-free cosmology with neural density estimators and
  active learning}.
\newblock {\em \mnras}, 488(3):4440--4458, September 2019.

\bibitem{cranmer_20}
Kyle {Cranmer}, Johann {Brehmer}, and Gilles {Louppe}.
\newblock {The frontier of simulation-based inference}.
\newblock {\em Proceedings of the National Academy of Science},
  117(48):30055--30062, December 2020.

\bibitem{papamakarios_17}
George {Papamakarios}, Theo {Pavlakou}, and Iain {Murray}.
\newblock {Masked Autoregressive Flow for Density Estimation}.
\newblock {\em arXiv e-prints}, page arXiv:1705.07057, May 2017.

\bibitem{rubin_84}
Donald~B. Rubin.
\newblock {Bayesianly Justifiable and Relevant Frequency Calculations for the
  Applied Statistician}.
\newblock {\em The Annals of Statistics}, 12(4):1151 -- 1172, 1984.

\bibitem{beaumont_02}
Mark~A Beaumont, Wenyang Zhang, and David~J Balding.
\newblock Approximate bayesian computation in population genetics.
\newblock {\em Genetics}, 162(4):2025--2035, December 2002.

\bibitem{akaret_15}
Jo{\"e}l {Akeret}, Alexandre {Refregier}, Adam {Amara}, Sebastian {Seehars},
  and Caspar {Hasner}.
\newblock {Approximate Bayesian computation for forward modeling in cosmology}.
\newblock {\em \jcap}, 2015(8):043--043, August 2015.

\bibitem{papamakarios_19}
George {Papamakarios}, Eric {Nalisnick}, Danilo {Jimenez Rezende}, Shakir
  {Mohamed}, and Balaji {Lakshminarayanan}.
\newblock {Normalizing Flows for Probabilistic Modeling and Inference}.
\newblock {\em arXiv e-prints}, page arXiv:1912.02762, December 2019.

\bibitem{legin_21}
Ronan {Legin}, Yashar {Hezaveh}, Laurence {Perreault Levasseur}, and Benjamin
  {Wandelt}.
\newblock {Simulation-Based Inference of Strong Gravitational Lensing
  Parameters}.
\newblock In {\em Machine Learning and the Physical Sciences Workshop},
  page~95, January 2021.

\bibitem{anau_22}
Noemi {Anau Montel} and Christoph {Weniger}.
\newblock {Detection is truncation: studying source populations with truncated
  marginal neural ratio estimation}.
\newblock {\em arXiv e-prints}, page arXiv:2211.04291, November 2022.

\bibitem{hahn_22b}
ChangHoon {Hahn} and Peter {Melchior}.
\newblock {Accelerated Bayesian SED Modeling Using Amortized Neural Posterior
  Estimation}.
\newblock {\em \apj}, 938(1):11, October 2022.

\bibitem{khullar_22}
Gourav {Khullar}, Brian {Nord}, Aleksandra {{\'C}iprijanovi{\'c}}, Jason {Poh},
  and Fei {Xu}.
\newblock {DIGS: deep inference of galaxy spectra with neural posterior
  estimation}.
\newblock {\em Machine Learning: Science and Technology}, 3(4):04LT04, December
  2022.

\bibitem{alvey_23}
James {Alvey}, Uddipta {Bhardwaj}, Samaya {Nissanke}, and Christoph {Weniger}.
\newblock {What to do when things get crowded? Scalable joint analysis of
  overlapping gravitational wave signals}.
\newblock {\em arXiv e-prints}, page arXiv:2308.06318, August 2023.

\bibitem{alvey_23b}
James {Alvey}, Mathis {Gerdes}, and Christoph {Weniger}.
\newblock {Albatross: a scalable simulation-based inference pipeline for
  analysing stellar streams in the Milky Way}.
\newblock {\em \mnras}, 525(3):3662--3681, November 2023.

\bibitem{bhardwaj_23}
Uddipta {Bhardwaj}, James {Alvey}, Benjamin~Kurt {Miller}, Samaya {Nissanke},
  and Christoph {Weniger}.
\newblock {Sequential simulation-based inference for gravitational wave
  signals}.
\newblock {\em \prd}, 108(4):042004, August 2023.

\bibitem{crisostomi_23}
Marco {Crisostomi}, Kallol {Dey}, Enrico {Barausse}, and Roberto {Trotta}.
\newblock {Neural posterior estimation with guaranteed exact coverage: The
  ringdown of GW150914}.
\newblock {\em \prd}, 108(4):044029, August 2023.

\bibitem{darc_23}
P.~{Darc}, Clecio~R. {Bom}, Bernardo M.~O. {Fraga}, and Charlie~D.
  {Kilpatrick}.
\newblock {Kilonova Spectral Inverse Modelling with Simulation-Based Inference:
  An Amortized Neural Posterior Estimation Analysis}.
\newblock {\em arXiv e-prints}, page arXiv:2311.09471, November 2023.

\bibitem{dax_23}
Maximilian {Dax}, Stephen~R. {Green}, Jonathan {Gair}, Michael {P{\"u}rrer},
  Jonas {Wildberger}, Jakob~H. {Macke}, Alessandra {Buonanno}, and Bernhard
  {Sch{\"o}lkopf}.
\newblock {Neural Importance Sampling for Rapid and Reliable Gravitational-Wave
  Inference}.
\newblock {\em \prl}, 130(17):171403, April 2023.

\bibitem{dupourque_23}
S.~{Dupourqu{\'e}}, N.~{Clerc}, E.~{Pointecouteau}, D.~{Eckert}, S.~{Ettori},
  and F.~{Vazza}.
\newblock {Investigating the turbulent hot gas in X-COP galaxy clusters}.
\newblock {\em \aap}, 673:A91, May 2023.

\bibitem{gebhard_23}
Timothy~D. {Gebhard}, Jonas {Wildberger}, Maximilian {Dax}, Daniel
  {Angerhausen}, Sascha~P. {Quanz}, and Bernhard {Sch{\"o}lkopf}.
\newblock {Inferring Atmospheric Properties of Exoplanets with Flow Matching
  and Neural Importance Sampling}.
\newblock {\em arXiv e-prints}, page arXiv:2312.08295, December 2023.

\bibitem{graber_23}
Vanessa {Graber}, Michele {Ronchi}, Celsa {Pardo-Araujo}, and Nanda {Rea}.
\newblock {Isolated pulsar population synthesis with simulation-based
  inference}.
\newblock {\em arXiv e-prints}, page arXiv:2312.14848, December 2023.

\bibitem{hahn_23c}
ChangHoon {Hahn}, Connor {Bottrell}, and Khee-Gan {Lee}.
\newblock {${\rm H{\scriptsize ALO}F{\scriptsize LOW}}$ I: Neural Inference of
  Halo Mass from Galaxy Photometry and Morphology}.
\newblock {\em arXiv e-prints}, page arXiv:2310.04503, October 2023.

\bibitem{prelogovic_23}
David {Prelogovi{\'c}} and Andrei {Mesinger}.
\newblock {Exploring the likelihood of the 21-cm power spectrum with
  simulation-based inference}.
\newblock {\em \mnras}, 524(3):4239--4255, September 2023.

\bibitem{vasist_23}
Malavika {Vasist}, Fran{\c{c}}ois {Rozet}, Olivier {Absil}, Paul
  {Molli{\`e}re}, Evert {Nasedkin}, and Gilles {Louppe}.
\newblock {Neural posterior estimation for exoplanetary atmospheric retrieval}.
\newblock {\em \aap}, 672:A147, April 2023.

\bibitem{wang_23}
Bingjie {Wang}, Joel {Leja}, V.~Ashley {Villar}, and Joshua~S. {Speagle}.
\newblock {SBI$^{++}$: Flexible, Ultra-fast Likelihood-free Inference
  Customized for Astronomical Applications}.
\newblock {\em \apjl}, 952(1):L10, July 2023.

\bibitem{zhao_23}
Xiaosheng {Zhao}, Yi~{Mao}, Shifan {Zuo}, and Benjamin~D. {Wandelt}.
\newblock {Simulation-based Inference of Reionization Parameters from 3D
  Tomographic 21 cm Light-cone Images -- II: Application of Solid Harmonic
  Wavelet Scattering Transform}.
\newblock {\em arXiv e-prints}, page arXiv:2310.17602, October 2023.

\bibitem{alvey_24}
James {Alvey}, Uddipta {Bhardwaj}, Valerie {Domcke}, Mauro {Pieroni}, and
  Christoph {Weniger}.
\newblock {Simulation-based inference for stochastic gravitational wave
  background data analysis}.
\newblock {\em \prd}, 109(8):083008, April 2024.

\bibitem{christy_24}
Katharena {Christy}, Eric~J. {Baxter}, and Jason {Kumar}.
\newblock {Applying Simulation-Based Inference to Spectral and Spatial
  Information from the Galactic Center Gamma-Ray Excess}.
\newblock {\em arXiv e-prints}, page arXiv:2402.04549, February 2024.

\bibitem{coogan_24}
Adam {Coogan}, Noemi {Anau Montel}, Konstantin {Karchev}, Meiert~W. {Grootes},
  Francesco {Nattino}, and Christoph {Weniger}.
\newblock {The effect of the perturber population on subhalo measurements in
  strong gravitational lenses}.
\newblock {\em \mnras}, 527(1):66--78, January 2024.

\bibitem{moser_24}
Beatrice {Moser}, Tomasz {Kacprzak}, Silvan {Fischbacher}, Alexandre
  {Refregier}, Dominic {Grimm}, and Luca {Tortorelli}.
\newblock {Simulation-based inference of deep fields: galaxy population model
  and redshift distributions}.
\newblock {\em arXiv e-prints}, page arXiv:2401.06846, January 2024.

\bibitem{sun_24}
Tian-Yang {Sun}, Chun-Yu {Xiong}, Shang-Jie {Jin}, Yu-Xin {Wang}, Jing-Fei
  {Zhang}, and Xin {Zhang}.
\newblock {Efficient parameter inference for gravitational wave signals in the
  presence of transient noises using temporal and time-spectral fusion
  normalizing flow}.
\newblock {\em Chinese Physics C}, 48(4):045108, April 2024.

\bibitem{xiong_24}
Chun-Yu {Xiong}, Tian-Yang {Sun}, Jing-Fei {Zhang}, and Xin {Zhang}.
\newblock {Robust inference of gravitational wave source parameters in the
  presence of noise transients using normalizing flows}.
\newblock {\em arXiv e-prints}, page arXiv:2405.09475, May 2024.

\bibitem{dimitriou_22}
Androniki {Dimitriou}, Christoph {Weniger}, and Camila~A. {Correa}.
\newblock {Towards reconstructing the halo clustering and halo mass function of
  N-body simulations using neural ratio estimation}.
\newblock {\em arXiv e-prints}, page arXiv:2206.11312, June 2022.

\bibitem{hahn_22}
ChangHoon {Hahn}, Michael {Eickenberg}, Shirley {Ho}, Jiamin {Hou}, Pablo
  {Lemos}, Elena {Massara}, Chirag {Modi}, Azadeh {Moradinezhad Dizgah}, Bruno
  {R{\'e}galdo-Saint Blancard}, and Muntazir~M. {Abidi}.
\newblock {${\rm S{\scriptsize IM}BIG}$: A Forward Modeling Approach To
  Analyzing Galaxy Clustering}.
\newblock {\em arXiv e-prints}, page arXiv:2211.00723, November 2022.

\bibitem{reza_22}
Moonzarin {Reza}, Yuanyuan {Zhang}, Brian {Nord}, Jason {Poh}, Aleksandra
  {Ciprijanovic}, and Louis {Strigari}.
\newblock {Estimating Cosmological Constraints from Galaxy Cluster Abundance
  using Simulation-Based Inference}.
\newblock In {\em Machine Learning for Astrophysics}, page~20, July 2022.

\bibitem{zhang_22}
Gemma {Zhang}, Siddharth {Mishra-Sharma}, and Cora {Dvorkin}.
\newblock {Inferring subhalo effective density slopes from strong lensing
  observations with neural likelihood-ratio estimation}.
\newblock {\em \mnras}, 517(3):4317--4326, December 2022.

\bibitem{zhao_22}
Xiaosheng {Zhao}, Yi~{Mao}, and Benjamin~D. {Wandelt}.
\newblock {Implicit Likelihood Inference of Reionization Parameters from the 21
  cm Power Spectrum}.
\newblock {\em \apj}, 933(2):236, July 2022.

\bibitem{anau_23}
Noemi {Anau Montel}, Adam {Coogan}, Camila {Correa}, Konstantin {Karchev}, and
  Christoph {Weniger}.
\newblock {Estimating the warm dark matter mass from strong lensing images with
  truncated marginal neural ratio estimation}.
\newblock {\em \mnras}, 518(2):2746--2760, January 2023.

\bibitem{hahn_23a}
ChangHoon {Hahn}, Pablo {Lemos}, Liam {Parker}, Bruno {R{\'e}galdo-Saint
  Blancard}, Michael {Eickenberg}, Shirley {Ho}, Jiamin {Hou}, Elena {Massara},
  Chirag {Modi}, Azadeh {Moradinezhad Dizgah}, and David {Spergel}.
\newblock {${\rm S{\scriptsize IM}BIG}$: The First Cosmological Constraints
  from Non-Gaussian and Non-Linear Galaxy Clustering}.
\newblock {\em arXiv e-prints}, page arXiv:2310.15246, October 2023.

\bibitem{hahn_23b}
ChangHoon {Hahn}, Michael {Eickenberg}, Shirley {Ho}, Jiamin {Hou}, Pablo
  {Lemos}, Elena {Massara}, Chirag {Modi}, Azadeh {Moradinezhad Dizgah}, Liam
  {Parker}, and Bruno {R{\'e}galdo-Saint Blancard}.
\newblock {${\rm S{\scriptsize IM}BIG}$: The First Cosmological Constraints
  from the Non-Linear Galaxy Bispectrum}.
\newblock {\em arXiv e-prints}, page arXiv:2310.15243, October 2023.

\bibitem{blancard_23}
Bruno {R{\'e}galdo-Saint Blancard}, ChangHoon {Hahn}, Shirley {Ho}, Jiamin
  {Hou}, Pablo {Lemos}, Elena {Massara}, Chirag {Modi}, Azadeh {Moradinezhad
  Dizgah}, Liam {Parker}, Yuling {Yao}, and Michael {Eickenberg}.
\newblock {${\rm S{\scriptsize IM}BIG}$: Galaxy Clustering Analysis with the
  Wavelet Scattering Transform}.
\newblock {\em arXiv e-prints}, page arXiv:2310.15250, October 2023.

\bibitem{gagnon_23}
Samuel {Gagnon-Hartman}, John {Ruan}, and Daryl {Haggard}.
\newblock {Debiasing standard siren inference of the Hubble constant with
  marginal neural ratio estimation}.
\newblock {\em \mnras}, 520(1):1--13, March 2023.

\bibitem{jo_23}
Yongseok {Jo}, Shy {Genel}, Benjamin {Wandelt}, Rachel~S. {Somerville},
  Francisco {Villaescusa-Navarro}, Greg~L. {Bryan}, Daniel
  {Angl{\'e}s-Alc{\'a}zar}, Daniel {Foreman-Mackey}, Dylan {Nelson}, and
  Ji-hoon {Kim}.
\newblock {Calibrating Cosmological Simulations with Implicit Likelihood
  Inference Using Galaxy Growth Observables}.
\newblock {\em \apj}, 944(1):67, February 2023.

\bibitem{karchev_23}
Konstantin {Karchev}, Roberto {Trotta}, and Christoph {Weniger}.
\newblock {SICRET: Supernova Ia Cosmology with truncated marginal neural Ratio
  EsTimation}.
\newblock {\em \mnras}, 520(1):1056--1072, March 2023.

\bibitem{lemos_23}
Pablo {Lemos}, Liam~H. {Parker}, ChangHoon {Hahn}, Shirley {Ho}, Michael
  {Eickenberg}, Jiamin {Hou}, Elena {Massara}, Chirag {Modi}, Azadeh
  {Moradinezhad Dizgah}, Bruno {R{\'e}galdo-Saint Blancard}, and David
  {Spergel}.
\newblock {SimBIG: Field-level Simulation-based Inference of Large-scale
  Structure}.
\newblock In {\em Machine Learning for Astrophysics}, page~18, July 2023.

\bibitem{list_23}
Florian {List}, Noemi {Anau Montel}, and Christoph {Weniger}.
\newblock {Bayesian Simulation-based Inference for Cosmological Initial
  Conditions}.
\newblock {\em arXiv e-prints}, page arXiv:2310.19910, October 2023.

\bibitem{modi_23}
Chirag {Modi}, Shivam {Pandey}, Matthew {Ho}, ChangHoon {Hahn}, Bruno
  {R'egaldo-Saint Blancard}, and Benjamin {Wandelt}.
\newblock {Sensitivity Analysis of Simulation-Based Inference for Galaxy
  Clustering}.
\newblock {\em arXiv e-prints}, page arXiv:2309.15071, September 2023.

\bibitem{modi_23b}
Chirag {Modi} and Oliver H.~E. {Philcox}.
\newblock {Hybrid SBI or How I Learned to Stop Worrying and Learn the
  Likelihood}.
\newblock {\em arXiv e-prints}, page arXiv:2309.10270, September 2023.

\bibitem{saxena_23}
Anchal {Saxena}, Alex {Cole}, Simon {Gazagnes}, P.~Daniel {Meerburg}, Christoph
  {Weniger}, and Samuel~J. {Witte}.
\newblock {Constraining the X-ray heating and reionization using 21-cm power
  spectra with Marginal Neural Ratio Estimation}.
\newblock {\em \mnras}, 525(4):6097--6111, November 2023.

\bibitem{zheng_23}
Yanyan {Zheng}, Nikolaos {Kouvatsos}, Jacob {Golomb}, Marco {Cavagli{\`a}},
  Arianna~I. {Renzini}, and Mairi {Sakellariadou}.
\newblock {Angular Power Spectrum of Gravitational-Wave Transient Sources as a
  Probe of the Large-Scale Structure}.
\newblock {\em \prl}, 131(17):171403, October 2023.

\bibitem{baxter_24}
Eric~J. {Baxter} and Shivam {Pandey}.
\newblock {Inferring galaxy cluster masses from cosmic microwave background
  lensing with neural simulation based inference}.
\newblock {\em arXiv e-prints}, page arXiv:2401.08910, January 2024.

\bibitem{hou_24}
Jiamin {Hou}, Azadeh {Moradinezhad Dizgah}, ChangHoon {Hahn}, Michael
  {Eickenberg}, Shirley {Ho}, Pablo {Lemos}, Elena {Massara}, Chirag {Modi},
  Liam {Parker}, and Bruno {R{\'e}galdo-Saint Blancard}.
\newblock {${\rm S{\scriptsize IM}BIG}$: Cosmological Constraints from the
  Redshift-Space Galaxy Skew Spectra}.
\newblock {\em arXiv e-prints}, page arXiv:2401.15074, January 2024.

\bibitem{massara_24}
Elena {Massara}, ChangHoon {Hahn}, Michael {Eickenberg}, Shirley {Ho}, Jiamin
  {Hou}, Pablo {Lemos}, Chirag {Modi}, Azadeh {Moradinezhad Dizgah}, Liam
  {Parker}, and Bruno {R{\'e}galdo-Saint Blancard}.
\newblock {{\sc SimBIG}: Cosmological Constraints using Simulation-Based
  Inference of Galaxy Clustering with Marked Power Spectra}.
\newblock {\em arXiv e-prints}, page arXiv:2404.04228, April 2024.

\bibitem{tucci_23}
Beatriz {Tucci} and Fabian {Schmidt}.
\newblock {EFTofLSS meets simulation-based inference: $\sigma_8$ from biased
  tracers}.
\newblock {\em arXiv e-prints}, page arXiv:2310.03741, October 2023.

\bibitem{yongseok_23}
Yongseok {Jo}, Shy {Genel}, Benjamin {Wandelt}, Rachel~S. {Somerville},
  Francisco {Villaescusa-Navarro}, Greg~L. {Bryan}, Daniel
  {Angl{\'e}s-Alc{\'a}zar}, Daniel {Foreman-Mackey}, Dylan {Nelson}, and
  Ji-hoon {Kim}.
\newblock {Calibrating Cosmological Simulations with Implicit Likelihood
  Inference Using Galaxy Growth Observables}.
\newblock {\em \apj}, 944(1):67, February 2023.

\bibitem{davies_24}
Frederick~B. {Davies}, Sarah E.~I. {Bosman}, Prakash {Gaikwad}, Fahad {Nasir},
  Joseph~F. {Hennawi}, George~D. {Becker}, Martin~G. {Haehnelt}, Valentina
  {D'Odorico}, Manuela {Bischetti}, Anna-Christina {Eilers}, Laura~C.
  {Keating}, Girish {Kulkarni}, Samuel {Lai}, Chiara {Mazzucchelli}, Yuxiang
  {Qin}, Sindhu {Satyavolu}, Feige {Wang}, Jinyi {Yang}, and Yongda {Zhu}.
\newblock {Constraints on the Evolution of the Ionizing Background and Ionizing
  Photon Mean Free Path at the End of Reionization}.
\newblock {\em \apj}, 965(2):134, April 2024.

\bibitem{gatti_24}
M.~{Gatti}, N.~{Jeffrey}, L.~{Whiteway}, J.~{Williamson}, B.~{Jain},
  V.~{Ajani}, D.~{Anbajagane}, G.~{Giannini}, C.~{Zhou}, A.~{Porredon},
  J.~{Prat}, M.~{Yamamoto}, J.~{Blazek}, T.~{Kacprzak}, S.~{Samuroff},
  A.~{Alarcon}, A.~{Amon}, K.~{Bechtol}, M.~{Becker}, G.~{Bernstein},
  A.~{Campos}, C.~{Chang}, R.~{Chen}, A.~{Choi}, C.~{Davis}, J.~{Derose}, H.~T.
  {Diehl}, S.~{Dodelson}, C.~{Doux}, K.~{Eckert}, J.~{Elvin-Poole},
  S.~{Everett}, A.~{Ferte}, D.~{Gruen}, R.~{Gruendl}, I.~{Harrison}, W.~G.
  {Hartley}, K.~{Herner}, E.~M. {Huff}, M.~{Jarvis}, N.~{Kuropatkin}, P.~F.
  {Leget}, N.~{MacCrann}, J.~{McCullough}, J.~{Myles}, A.~{Navarro-Alsina},
  S.~{Pandey}, M.~{Raveri}, R.~P. {Rollins}, A.~{Roodman}, C.~{Sanchez}, L.~F.
  {Secco}, I.~{Sevilla-Noarbe}, E.~{Sheldon}, T.~{Shin}, M.~{Troxel},
  I.~{Tutusaus}, T.~N. {Varga}, B.~{Yanny}, B.~{Yin}, Y.~{Zhang}, J.~{Zuntz},
  M.~{Aguena}, O.~{Alves}, J.~{Annis}, D.~{Brooks}, J.~{Carretero}, F.~J.
  {Castander}, R.~{Cawthon}, M.~{Costanzi}, L.~N. {da Costa}, M.~E.~S.
  {Pereira}, A.~E. {Evrard}, B.~{Flaugher}, P.~{Fosalba}, J.~{Frieman},
  J.~{Garc{\'\i}a-Bellido}, D.~W. {Gerdes}, D.~{Gruen}, R.~A. {Gruendl},
  J.~{Gschwend}, G.~{Gutierrez}, D.~L. {Hollowood}, K.~{Honscheid}, D.~J.
  {James}, K.~{Kuehn}, O.~{Lahav}, S.~{Lee}, J.~L. {Marshall},
  J.~{Mena-Fern{\'a}ndez}, F.~{Menanteau}, R.~{Miquel}, R.~L.~C. {Ogando},
  M.~E.~S. {Pereira}, A.~{Pieres}, A.~A. {Plazas Malag{\'o}n}, E.~{Sanchez},
  M.~{Smith}, E.~{Suchyta}, M.~E.~C. {Swanson}, G.~{Tarle}, N.~{Weaverdyck},
  J.~{Weller}, P.~{Wiseman}, and {DES Collaboration}.
\newblock {Dark Energy Survey Year 3 results: Simulation-based cosmological
  inference with wavelet harmonics, scattering transforms, and moments of weak
  lensing mass maps. Validation on simulations}.
\newblock {\em \prd}, 109(6):063534, March 2024.

\bibitem{gatti_24b}
M.~{Gatti}, G.~{Campailla}, N.~{Jeffrey}, L.~{Whiteway}, A.~{Porredon},
  J.~{Prat}, J.~{Williamson}, M.~{Raveri}, B.~{Jain}, V.~{Ajani},
  G.~{Giannini}, M.~{Yamamoto}, C.~{Zhou}, J.~{Blazek}, D.~{Anbajagane},
  S.~{Samuroff}, T.~{Kacprzak}, A.~{Alarcon}, A.~{Amon}, K.~{Bechtol},
  M.~{Becker}, G.~{Bernstein}, A.~{Campos}, C.~{Chang}, R.~{Chen}, A.~{Choi},
  C.~{Davis}, J.~{Derose}, H.~T. {Diehl}, S.~{Dodelson}, C.~{Doux},
  K.~{Eckert}, J.~{Elvin-Poole}, S.~{Everett}, A.~{Ferte}, D.~{Gruen},
  R.~{Gruendl}, I.~{Harrison}, W.~G. {Hartley}, K.~{Herner}, E.~M. {Huff},
  M.~{Jarvis}, N.~{Kuropatkin}, P.~F. {Leget}, N.~{MacCrann}, J.~{McCullough},
  J.~{Myles}, A.~{Navarro-Alsina}, S.~{Pandey}, R.~P. {Rollins}, A.~{Roodman},
  C.~{Sanchez}, L.~F. {Secco}, I.~{Sevilla-Noarbe}, E.~{Sheldon}, T.~{Shin},
  M.~{Troxel}, I.~{Tutusaus}, T.~N. {Varga}, B.~{Yanny}, B.~{Yin}, Y.~{Zhang},
  J.~{Zuntz}, T.~M.~C. {Abbott}, M.~{Aguena}, S.~S. {Allam}, O.~{Alves},
  F.~{Andrade-Oliveira}, D.~{Bacon}, S.~{Bocquet}, D.~{Brooks}, A.~{Carnero
  Rosell}, J.~{Carretero}, L.~N. {da Costa}, M.~E.~S. {Pereira}, J.~{De
  Vicente}, I.~{Ferrero}, J.~{Frieman}, J.~{Garc{\'\i}a-Bellido},
  E.~{Gaztanaga}, G.~{Gutierrez}, S.~R. {Hinton}, D.~L. {Hollowood},
  K.~{Honscheid}, D.~J. {James}, K.~{Kuehn}, O.~{Lahav}, S.~{Lee}, J.~L.
  {Marshall}, J.~{Mena-Fern{\'a}ndez}, R.~{Miquel}, A.~{Pieres}, A.~A. {Plazas
  Malag{\'o}n}, E.~{Sanchez}, D.~{Sanchez Cid}, M.~{Schubnell}, M.~{Smith},
  E.~{Suchyta}, G.~{Tarle}, N.~{Weaverdyck}, J.~{Weller}, and P.~{Wiseman}.
\newblock {Dark Energy Survey Year 3 results: simulation-based cosmological
  inference with wavelet harmonics, scattering transforms, and moments of weak
  lensing mass maps II. Cosmological results}.
\newblock {\em arXiv e-prints}, page arXiv:2405.10881, May 2024.

\bibitem{jeffrey_24}
N.~{Jeffrey}, L.~{Whiteway}, M.~{Gatti}, J.~{Williamson}, J.~{Alsing},
  A.~{Porredon}, J.~{Prat}, C.~{Doux}, B.~{Jain}, C.~{Chang}, T.~Y. {Cheng},
  T.~{Kacprzak}, P.~{Lemos}, A.~{Alarcon}, A.~{Amon}, K.~{Bechtol}, M.~R.
  {Becker}, G.~M. {Bernstein}, A.~{Campos}, A.~{Carnero Rosell}, R.~{Chen},
  A.~{Choi}, J.~{DeRose}, A.~{Drlica-Wagner}, K.~{Eckert}, S.~{Everett},
  A.~{Fert{\'e}}, D.~{Gruen}, R.~A. {Gruendl}, K.~{Herner}, M.~{Jarvis},
  J.~{McCullough}, J.~{Myles}, A.~{Navarro-Alsina}, S.~{Pandey}, M.~{Raveri},
  R.~P. {Rollins}, E.~S. {Rykoff}, C.~{S{\'a}nchez}, L.~F. {Secco},
  I.~{Sevilla-Noarbe}, E.~{Sheldon}, T.~{Shin}, M.~A. {Troxel}, I.~{Tutusaus},
  T.~N. {Varga}, B.~{Yanny}, B.~{Yin}, J.~{Zuntz}, M.~{Aguena}, S.~S. {Allam},
  O.~{Alves}, D.~{Bacon}, S.~{Bocquet}, D.~{Brooks}, L.~N. {da Costa}, T.~M.
  {Davis}, J.~{De Vicente}, S.~{Desai}, H.~T. {Diehl}, I.~{Ferrero},
  J.~{Frieman}, J.~{Garc{\'\i}a-Bellido}, E.~{Gaztanaga}, G.~{Giannini},
  G.~{Gutierrez}, S.~R. {Hinton}, D.~L. {Hollowood}, K.~{Honscheid},
  D.~{Huterer}, D.~J. {James}, O.~{Lahav}, S.~{Lee}, J.~L. {Marshall},
  J.~{Mena-Fern{\'a}ndez}, R.~{Miquel}, A.~{Pieres}, A.~A. {Plazas
  Malag{\'o}n}, A.~{Roodman}, M.~{Sako}, E.~{Sanchez}, D.~{Sanchez Cid},
  M.~{Smith}, E.~{Suchyta}, M.~E.~C. {Swanson}, G.~{Tarle}, D.~L. {Tucker},
  N.~{Weaverdyck}, J.~{Weller}, P.~{Wiseman}, and M.~{Yamamoto}.
\newblock {Dark Energy Survey Year 3 results: likelihood-free, simulation-based
  $w$CDM inference with neural compression of weak-lensing map statistics}.
\newblock {\em arXiv e-prints}, page arXiv:2403.02314, March 2024.

\bibitem{schosser_24}
Benedikt {Schosser}, Caroline {Heneka}, and Tilman {Plehn}.
\newblock {Optimal, fast, and robust inference of reionization-era cosmology
  with the 21cmPIE-INN}.
\newblock {\em arXiv e-prints}, page arXiv:2401.04174, January 2024.

\bibitem{wietersheim_24}
Maximilian {von Wietersheim-Kramsta}, Kiyam {Lin}, Nicolas {Tessore}, Benjamin
  {Joachimi}, Arthur {Loureiro}, Robert {Reischke}, and Angus~H. {Wright}.
\newblock {KiDS-SBI: Simulation-Based Inference Analysis of KiDS-1000 Cosmic
  Shear}.
\newblock {\em arXiv e-prints}, page arXiv:2404.15402, April 2024.

\bibitem{blum_12}
M.~G.~B. {Blum}, M.~A. {Nunes}, D.~{Prangle}, and S.~A. {Sisson}.
\newblock {A Comparative Review of Dimension Reduction Methods in Approximate
  Bayesian Computation}.
\newblock {\em arXiv e-prints}, page arXiv:1202.3819, February 2012.

\bibitem{blum_08}
M.~G.~B. {Blum} and O.~{Francois}.
\newblock {Non-linear regression models for Approximate Bayesian Computation}.
\newblock {\em arXiv e-prints}, page arXiv:0809.4178, September 2008.

\bibitem{fearnhead_10}
Paul {Fearnhead} and Dennis {Prangle}.
\newblock {Constructing Summary Statistics for Approximate Bayesian
  Computation: Semi-automatic ABC}.
\newblock {\em arXiv e-prints}, page arXiv:1004.1112, April 2010.

\bibitem{jiang_15}
Bai {Jiang}, Tung-yu {Wu}, Charles {Zheng}, and Wing~H. {Wong}.
\newblock {Learning Summary Statistic for Approximate Bayesian Computation via
  Deep Neural Network}.
\newblock {\em arXiv e-prints}, page arXiv:1510.02175, October 2015.

\bibitem{murphy_22}
Kevin~P. Murphy.
\newblock {\em Probabilistic Machine Learning: An introduction}.
\newblock MIT Press, 2022.

\bibitem{charnock_18}
Tom {Charnock}, Guilhem {Lavaux}, and Benjamin~D. {Wandelt}.
\newblock {Automatic physical inference with information maximizing neural
  networks}.
\newblock {\em \prd}, 97(8):083004, April 2018.

\bibitem{heavens_00}
Alan~F. {Heavens}, Raul {Jimenez}, and Ofer {Lahav}.
\newblock {Massive lossless data compression and multiple parameter estimation
  from galaxy spectra}.
\newblock {\em \mnras}, 317(4):965--972, October 2000.

\bibitem{prelogovic_24}
David {Prelogovi{\'c}} and Andrei {Mesinger}.
\newblock {How informative are summaries of the cosmic 21-cm signal?}
\newblock {\em arXiv e-prints}, page arXiv:2401.12277, January 2024.

\bibitem{makinen_24}
T.~Lucas {Makinen}, Alan {Heavens}, Natalia {Porqueres}, Tom {Charnock}, Axel
  {Lapel}, and Benjamin~D. {Wandelt}.
\newblock {Hybrid summary statistics: neural weak lensing inference beyond the
  power spectrum}.
\newblock {\em arXiv e-prints}, page arXiv:2407.18909, July 2024.

\bibitem{jeffrey_21}
Niall {Jeffrey}, Justin {Alsing}, and Fran{\c{c}}ois {Lanusse}.
\newblock {Likelihood-free inference with neural compression of DES SV weak
  lensing map statistics}.
\newblock {\em \mnras}, 501(1):954--969, February 2021.

\bibitem{radev_20}
Stefan~T. {Radev}, Ulf~K. {Mertens}, Andreas {Voss}, Lynton {Ardizzone}, and
  Ullrich {K{\"o}the}.
\newblock {BayesFlow: Learning complex stochastic models with invertible neural
  networks}.
\newblock {\em arXiv e-prints}, page arXiv:2003.06281, March 2020.

\bibitem{sharma_24}
Divij {Sharma}, Biwei {Dai}, and Uros {Seljak}.
\newblock {A comparative study of cosmological constraints from weak lensing
  using Convolutional Neural Networks}.
\newblock {\em arXiv e-prints}, page arXiv:2403.03490, March 2024.

\bibitem{lanzieri_24}
Denise {Lanzieri}, Justine {Zeghal}, T.~Lucas {Makinen}, Alexandre {Boucaud},
  Jean-Luc {Starck}, and Fran{\c{c}}ois {Lanusse}.
\newblock {Optimal Neural Summarisation for Full-Field Weak Lensing
  Cosmological Implicit Inference}.
\newblock {\em arXiv e-prints}, page arXiv:2407.10877, July 2024.

\bibitem{makinen_24b}
T.~Lucas {Makinen}, Ce~{Sui}, Benjamin~D. {Wandelt}, Natalia {Porqueres}, and
  Alan {Heavens}.
\newblock {Hybrid Summary Statistics}.
\newblock {\em arXiv e-prints}, page arXiv:2410.07548, October 2024.

\bibitem{bronstein_21}
Michael~M. {Bronstein}, Joan {Bruna}, Taco {Cohen}, and Petar
  {Veli{\v{c}}kovi{\'c}}.
\newblock {Geometric Deep Learning: Grids, Groups, Graphs, Geodesics, and
  Gauges}.
\newblock {\em arXiv e-prints}, page arXiv:2104.13478, April 2021.

\bibitem{battaglia_18}
Peter~W. {Battaglia}, Jessica~B. {Hamrick}, Victor {Bapst}, Alvaro
  {Sanchez-Gonzalez}, Vinicius {Zambaldi}, Mateusz {Malinowski}, Andrea
  {Tacchetti}, David {Raposo}, Adam {Santoro}, Ryan {Faulkner}, Caglar
  {Gulcehre}, Francis {Song}, Andrew {Ballard}, Justin {Gilmer}, George {Dahl},
  Ashish {Vaswani}, Kelsey {Allen}, Charles {Nash}, Victoria {Langston}, Chris
  {Dyer}, Nicolas {Heess}, Daan {Wierstra}, Pushmeet {Kohli}, Matt {Botvinick},
  Oriol {Vinyals}, Yujia {Li}, and Razvan {Pascanu}.
\newblock {Relational inductive biases, deep learning, and graph networks}.
\newblock {\em arXiv e-prints}, page arXiv:1806.01261, June 2018.

\bibitem{anagnostidis_22}
Sotiris {Anagnostidis}, Arne {Thomsen}, Tomasz {Kacprzak}, Tilman
  {Tr{\"o}ster}, Luca {Biggio}, Alexandre {Refregier}, and Thomas {Hofmann}.
\newblock {Cosmology from Galaxy Redshift Surveys with PointNet}.
\newblock {\em arXiv e-prints}, page arXiv:2211.12346, November 2022.

\bibitem{villanueva_22}
Pablo {Villanueva-Domingo} and Francisco {Villaescusa-Navarro}.
\newblock {Learning Cosmology and Clustering with Cosmic Graphs}.
\newblock {\em \apj}, 937(2):115, October 2022.

\bibitem{shao_22}
Helen {Shao}, Francisco {Villaescusa-Navarro}, Pablo {Villanueva-Domingo},
  Romain {Teyssier}, Lehman~H. {Garrison}, Marco {Gatti}, Derek {Inman},
  Yueying {Ni}, Ulrich~P. {Steinwandel}, Mihir {Kulkarni}, Eli {Visbal},
  Greg~L. {Bryan}, Daniel {Angles-Alcazar}, Tiago {Castro}, Elena
  {Hernandez-Martinez}, and Klaus {Dolag}.
\newblock {Robust field-level inference with dark matter halos}.
\newblock {\em arXiv e-prints}, page arXiv:2209.06843, September 2022.

\bibitem{makinen_22}
T.~Lucas {Makinen}, Tom {Charnock}, Pablo {Lemos}, Natalia {Porqueres}, Alan~F.
  {Heavens}, and Benjamin~D. {Wandelt}.
\newblock {The Cosmic Graph: Optimal Information Extraction from Large-Scale
  Structure using Catalogues}.
\newblock {\em The Open Journal of Astrophysics}, 5(1):18, December 2022.

\bibitem{desanti_23}
Natal{\'\i} S.~M. {de Santi}, Helen {Shao}, Francisco {Villaescusa-Navarro},
  L.~Raul {Abramo}, Romain {Teyssier}, Pablo {Villanueva-Domingo}, Yueying
  {Ni}, Daniel {Angl{\'e}s-Alc{\'a}zar}, Shy {Genel}, Elena
  {Hern{\'a}ndez-Mart{\'\i}nez}, Ulrich~P. {Steinwandel}, Christopher~C.
  {Lovell}, Klaus {Dolag}, Tiago {Castro}, and Mark {Vogelsberger}.
\newblock {Robust Field-level Likelihood-free Inference with Galaxies}.
\newblock {\em \apj}, 952(1):69, July 2023.

\bibitem{desanti_23b}
Natal{\'\i} S.~M. {de Santi}, Francisco {Villaescusa-Navarro}, L.~Raul
  {Abramo}, Helen {Shao}, Lucia~A. {Perez}, Tiago {Castro}, Yueying {Ni},
  Christopher~C. {Lovell}, Elena {Hernandez-Martinez}, Federico {Marinacci},
  David~N. {Spergel}, Klaus {Dolag}, Lars {Hernquist}, and Mark {Vogelsberger}.
\newblock {Field-level simulation-based inference with galaxy catalogs: the
  impact of systematic effects}.
\newblock {\em arXiv e-prints}, page arXiv:2310.15234, October 2023.

\bibitem{massara_23}
Elena {Massara}, Francisco {Villaescusa-Navarro}, and Will~J. {Percival}.
\newblock {Predicting interloper fraction with graph neural networks}.
\newblock {\em \jcap}, 2023(12):012, December 2023.

\bibitem{roncoli_23}
Andrea {Roncoli}, Aleksandra {{\'C}iprijanovi{\'c}}, Maggie {Voetberg},
  Francisco {Villaescusa-Navarro}, and Brian {Nord}.
\newblock {Domain Adaptive Graph Neural Networks for Constraining Cosmological
  Parameters Across Multiple Data Sets}.
\newblock {\em arXiv e-prints}, page arXiv:2311.01588, November 2023.

\bibitem{shao_23}
Helen {Shao}, Natal{\'\i} S.~M. {de Santi}, Francisco {Villaescusa-Navarro},
  Romain {Teyssier}, Yueying {Ni}, Daniel {Angl{\'e}s-Alc{\'a}zar}, Shy
  {Genel}, Ulrich~P. {Steinwandel}, Elena {Hern{\'a}ndez-Mart{\'\i}nez}, Klaus
  {Dolag}, Christopher~C. {Lovell}, Lehman~H. {Garrison}, Eli {Visbal}, Mihir
  {Kulkarni}, Lars {Hernquist}, Tiago {Castro}, and Mark {Vogelsberger}.
\newblock {A Universal Equation to Predict {\ensuremath{\Omega}}$_{m}$ from
  Halo and Galaxy Catalogs}.
\newblock {\em \apj}, 956(2):149, October 2023.

\bibitem{jeffrey_20}
Niall {Jeffrey} and Benjamin~D. {Wandelt}.
\newblock {Solving high-dimensional parameter inference: marginal posterior
  densities \& Moment Networks}.
\newblock {\em arXiv e-prints}, page arXiv:2011.05991, November 2020.

\bibitem{nguyen_23}
Tri {Nguyen}, Siddharth {Mishra-Sharma}, Reuel {Williams}, and Lina {Necib}.
\newblock {Uncovering dark matter density profiles in dwarf galaxies with graph
  neural networks}.
\newblock {\em \prd}, 107(4):043015, February 2023.

\bibitem{ho_24}
Matthew {Ho}, Deaglan~J. {Bartlett}, Nicolas {Chartier}, Carolina
  {Cuesta-Lazaro}, Simon {Ding}, Axel {Lapel}, Pablo {Lemos}, Christopher~C.
  {Lovell}, T.~Lucas {Makinen}, Chirag {Modi}, Viraj {Pandya}, Shivam {Pandey},
  Lucia~A. {Perez}, Benjamin {Wandelt}, and Greg~L. {Bryan}.
\newblock {LtU-ILI: An All-in-One Framework for Implicit Inference in
  Astrophysics and Cosmology}.
\newblock {\em arXiv e-prints}, page arXiv:2402.05137, February 2024.

\bibitem{lucie-smith_22}
Luisa {Lucie-Smith}, Hiranya~V. {Peiris}, Andrew {Pontzen}, Brian {Nord}, Jeyan
  {Thiyagalingam}, and Davide {Piras}.
\newblock {Discovering the building blocks of dark matter halo density profiles
  with neural networks}.
\newblock {\em \prd}, 105(10):103533, May 2022.

\bibitem{gong_24}
Zhengyangguang {Gong}, Anik {Halder}, Annabelle {Bohrdt}, Stella {Seitz}, and
  David {Gebauer}.
\newblock {C3NN: Cosmological Correlator Convolutional Neural Network an
  Interpretable Machine-learning Framework for Cosmological Analyses}.
\newblock {\em \apj}, 971(2):156, August 2024.

\bibitem{lucie-smith_24}
Luisa {Lucie-Smith}, Giulia {Despali}, and Volker {Springel}.
\newblock {A deep-learning model for the density profiles of subhaloes in
  IllustrisTNG}.
\newblock {\em \mnras}, 532(1):164--176, July 2024.

\bibitem{piras_24}
Davide {Piras} and Lucas {Lombriser}.
\newblock {Representation learning approach to probe for dynamical dark energy
  in matter power spectra}.
\newblock {\em \prd}, 110(2):023514, July 2024.

\bibitem{weinberger_17}
Rainer {Weinberger}, Volker {Springel}, Lars {Hernquist}, Annalisa {Pillepich},
  Federico {Marinacci}, R{\"u}diger {Pakmor}, Dylan {Nelson}, Shy {Genel}, Mark
  {Vogelsberger}, Jill {Naiman}, and Paul {Torrey}.
\newblock {Simulating galaxy formation with black hole driven thermal and
  kinetic feedback}.
\newblock {\em \mnras}, 465(3):3291--3308, March 2017.

\bibitem{pillepich_18}
Annalisa {Pillepich}, Dylan {Nelson}, Lars {Hernquist}, Volker {Springel},
  R{\"u}diger {Pakmor}, Paul {Torrey}, Rainer {Weinberger}, Shy {Genel},
  Jill~P. {Naiman}, Federico {Marinacci}, and Mark {Vogelsberger}.
\newblock {First results from the IllustrisTNG simulations: the stellar mass
  content of groups and clusters of galaxies}.
\newblock {\em \mnras}, 475(1):648--675, March 2018.

\bibitem{springel_10}
Volker {Springel}.
\newblock {E pur si muove: Galilean-invariant cosmological hydrodynamical
  simulations on a moving mesh}.
\newblock {\em \mnras}, 401(2):791--851, January 2010.

\bibitem{weinberger_20}
Rainer {Weinberger}, Volker {Springel}, and R{\"u}diger {Pakmor}.
\newblock {The AREPO Public Code Release}.
\newblock {\em \apjs}, 248(2):32, June 2020.

\bibitem{springel_03}
Volker {Springel} and Lars {Hernquist}.
\newblock {Cosmological smoothed particle hydrodynamics simulations: a hybrid
  multiphase model for star formation}.
\newblock {\em \mnras}, 339(2):289--311, February 2003.

\bibitem{dave_19}
Romeel {Dav{\'e}}, Daniel {Angl{\'e}s-Alc{\'a}zar}, Desika {Narayanan},
  Qi~{Li}, Mika~H. {Rafieferantsoa}, and Sarah {Appleby}.
\newblock {SIMBA: Cosmological simulations with black hole growth and
  feedback}.
\newblock {\em \mnras}, 486(2):2827--2849, June 2019.

\bibitem{hopkins_15}
Philip~F. {Hopkins}.
\newblock {A new class of accurate, mesh-free hydrodynamic simulation methods}.
\newblock {\em \mnras}, 450(1):53--110, June 2015.

\bibitem{heckman_14}
Timothy~M. {Heckman} and Philip~N. {Best}.
\newblock {The Coevolution of Galaxies and Supermassive Black Holes: Insights
  from Surveys of the Contemporary Universe}.
\newblock {\em \araa}, 52:589--660, August 2014.

\bibitem{bird_22}
Simeon {Bird}, Yueying {Ni}, Tiziana {Di Matteo}, Rupert {Croft}, Yu~{Feng},
  and Nianyi {Chen}.
\newblock {The ASTRID simulation: galaxy formation and reionization}.
\newblock {\em \mnras}, 512(3):3703--3716, May 2022.

\bibitem{feng_18}
Yu~Feng, Simeon Bird, Lauren Anderson, Andreu Font-Ribera, and Chris Pedersen.
\newblock Mp-gadget/mp-gadget: A tag for getting a doi, October 2018.

\bibitem{ni_22}
Yueying {Ni}, Tiziana {Di Matteo}, Simeon {Bird}, Rupert {Croft}, Yu~{Feng},
  Nianyi {Chen}, Michael {Tremmel}, Colin {DeGraf}, and Yin {Li}.
\newblock {The ASTRID simulation: the evolution of supermassive black holes}.
\newblock {\em \mnras}, 513(1):670--692, June 2022.

\bibitem{ni_23}
Yueying {Ni}, Shy {Genel}, Daniel {Angl{\'e}s-Alc{\'a}zar}, Francisco
  {Villaescusa-Navarro}, Yongseok {Jo}, Simeon {Bird}, Tiziana {Di Matteo},
  Rupert {Croft}, Nianyi {Chen}, Natal{\'\i} S.~M. {de Santi}, Matthew
  {Gebhardt}, Helen {Shao}, Shivam {Pandey}, Lars {Hernquist}, and Romeel
  {Dave}.
\newblock {The CAMELS Project: Expanding the Galaxy Formation Model Space with
  New ASTRID and 28-parameter TNG and SIMBA Suites}.
\newblock {\em \apj}, 959(2):136, December 2023.

\bibitem{springel_01}
Volker {Springel}, Simon D.~M. {White}, Giuseppe {Tormen}, and Guinevere
  {Kauffmann}.
\newblock {Populating a cluster of galaxies - I. Results at z=0}.
\newblock {\em \mnras}, 328(3):726--750, December 2001.

\bibitem{dolag_09}
K.~{Dolag}, S.~{Borgani}, G.~{Murante}, and V.~{Springel}.
\newblock {Substructures in hydrodynamical cluster simulations}.
\newblock {\em \mnras}, 399(2):497--514, October 2009.

\bibitem{tejero_20}
Alvaro {Tejero-Cantero}, Jan {Boelts}, Michael {Deistler}, Jan-Matthis
  {Lueckmann}, Conor {Durkan}, Pedro {Gon{\c{c}}alves}, David {Greenberg}, and
  Jakob {Macke}.
\newblock {sbi: A toolkit for simulation-based inference}.
\newblock {\em The Journal of Open Source Software}, 5(52):2505, August 2020.

\bibitem{bishop_94}
{Christopher M.} Bishop.
\newblock Mixture density networks.
\newblock Workingpaper, Aston University, 1994.

\bibitem{dinh_16}
Laurent {Dinh}, Jascha {Sohl-Dickstein}, and Samy {Bengio}.
\newblock {Density estimation using Real NVP}.
\newblock {\em arXiv e-prints}, page arXiv:1605.08803, May 2016.

\bibitem{durkan_19}
Conor {Durkan}, Artur {Bekasov}, Iain {Murray}, and George {Papamakarios}.
\newblock {Neural Spline Flows}.
\newblock {\em arXiv e-prints}, page arXiv:1906.04032, June 2019.

\bibitem{germain_15}
Mathieu {Germain}, Karol {Gregor}, Iain {Murray}, and Hugo {Larochelle}.
\newblock {MADE: Masked Autoencoder for Distribution Estimation}.
\newblock {\em arXiv e-prints}, page arXiv:1502.03509, February 2015.

\bibitem{alsing_18}
Justin {Alsing}, Benjamin {Wandelt}, and Stephen {Feeney}.
\newblock {Massive optimal data compression and density estimation for
  scalable, likelihood-free inference in cosmology}.
\newblock {\em \mnras}, 477(3):2874--2885, July 2018.

\bibitem{papamakarios_16}
George {Papamakarios} and Iain {Murray}.
\newblock {Fast $\epsilon$-free Inference of Simulation Models with Bayesian
  Conditional Density Estimation}.
\newblock {\em arXiv e-prints}, page arXiv:1605.06376, May 2016.

\bibitem{lueckmann_17}
Jan-Matthis {Lueckmann}, Pedro~J. {Goncalves}, Giacomo {Bassetto}, Kaan
  {{\"O}cal}, Marcel {Nonnenmacher}, and Jakob~H. {Macke}.
\newblock {Flexible statistical inference for mechanistic models of neural
  dynamics}.
\newblock {\em arXiv e-prints}, page arXiv:1711.01861, November 2017.

\bibitem{papamakarios_18}
George {Papamakarios}, David~C. {Sterratt}, and Iain {Murray}.
\newblock {Sequential Neural Likelihood: Fast Likelihood-free Inference with
  Autoregressive Flows}.
\newblock {\em arXiv e-prints}, page arXiv:1805.07226, May 2018.

\bibitem{lueckmann_18}
Jan-Matthis {Lueckmann}, Giacomo {Bassetto}, Theofanis {Karaletsos}, and
  Jakob~H. {Macke}.
\newblock {Likelihood-free inference with emulator networks}.
\newblock {\em arXiv e-prints}, page arXiv:1805.09294, May 2018.

\bibitem{greenberg_19}
David~S. {Greenberg}, Marcel {Nonnenmacher}, and Jakob~H. {Macke}.
\newblock {Automatic Posterior Transformation for Likelihood-Free Inference}.
\newblock {\em arXiv e-prints}, page arXiv:1905.07488, May 2019.

\bibitem{kullback_51}
S.~Kullback and R.~A. Leibler.
\newblock {On Information and Sufficiency}.
\newblock {\em The Annals of Mathematical Statistics}, 22(1):79 -- 86, 1951.

\bibitem{jordan_99}
Michael~I. Jordan, Zoubin Ghahramani, Tommi~S. Jaakkola, and Lawrence~K. Saul.
\newblock An introduction to variational methods for graphical models.
\newblock {\em Machine Learning}, 37(2):183--233, Nov 1999.

\bibitem{rezende_15}
Danilo {Jimenez Rezende} and Shakir {Mohamed}.
\newblock {Variational Inference with Normalizing Flows}.
\newblock {\em arXiv e-prints}, page arXiv:1505.05770, May 2015.

\bibitem{corso_20}
Gabriele {Corso}, Luca {Cavalleri}, Dominique {Beaini}, Pietro {Li{\`o}}, and
  Petar {Veli{\v{c}}kovi{\'c}}.
\newblock {Principal Neighbourhood Aggregation for Graph Nets}.
\newblock {\em arXiv e-prints}, page arXiv:2004.05718, April 2020.

\bibitem{li_17}
Hao {Li}, Zheng {Xu}, Gavin {Taylor}, Christoph {Studer}, and Tom {Goldstein}.
\newblock {Visualizing the Loss Landscape of Neural Nets}.
\newblock {\em arXiv e-prints}, page arXiv:1712.09913, December 2017.

\bibitem{cole_22}
Alex {Cole}, Benjamin~K. {Miller}, Samuel~J. {Witte}, Maxwell~X. {Cai},
  Meiert~W. {Grootes}, Francesco {Nattino}, and Christoph {Weniger}.
\newblock {Fast and credible likelihood-free cosmology with truncated marginal
  neural ratio estimation}.
\newblock {\em \jcap}, 2022(9):004, September 2022.

\bibitem{talts_18}
Sean {Talts}, Michael {Betancourt}, Daniel {Simpson}, Aki {Vehtari}, and Andrew
  {Gelman}.
\newblock {Validating Bayesian Inference Algorithms with Simulation-Based
  Calibration}.
\newblock {\em arXiv e-prints}, page arXiv:1804.06788, April 2018.

\bibitem{akiba_19}
Takuya Akiba, Shotaro Sano, Toshihiko Yanase, Takeru Ohta, and Masanori Koyama.
\newblock Optuna: A next-generation hyperparameter optimization framework.
\newblock In {\em Proceedings of the 25th {ACM} {SIGKDD} International
  Conference on Knowledge Discovery and Data Mining}, 2019.

\bibitem{watanabe_23}
Shuhei {Watanabe}.
\newblock {Tree-Structured Parzen Estimator: Understanding Its Algorithm
  Components and Their Roles for Better Empirical Performance}.
\newblock {\em arXiv e-prints}, page arXiv:2304.11127, April 2023.

\bibitem{bergstra_11}
James Bergstra, R\'{e}mi Bardenet, Yoshua Bengio, and Bal\'{a}zs K\'{e}gl.
\newblock Algorithms for hyper-parameter optimization.
\newblock In J.~Shawe-Taylor, R.~Zemel, P.~Bartlett, F.~Pereira, and K.Q.
  Weinberger, editors, {\em Advances in Neural Information Processing Systems},
  volume~24. Curran Associates, Inc., 2011.

\bibitem{bergstra_12}
J.~{Bergstra}, D.~{Yamins}, and D.~D. {Cox}.
\newblock {Making a Science of Model Search}.
\newblock {\em arXiv e-prints}, page arXiv:1209.5111, September 2012.

\bibitem{scikit-learn_11}
F.~Pedregosa, G.~Varoquaux, A.~Gramfort, V.~Michel, B.~Thirion, O.~Grisel,
  M.~Blondel, P.~Prettenhofer, R.~Weiss, V.~Dubourg, J.~Vanderplas, A.~Passos,
  D.~Cournapeau, M.~Brucher, M.~Perrot, and E.~Duchesnay.
\newblock Scikit-learn: Machine learning in {P}ython.
\newblock {\em Journal of Machine Learning Research}, 12:2825--2830, 2011.

\bibitem{tenenbaum_00}
Joshua~B. {Tenenbaum}, Vin {de Silva}, and John~C. {Langford}.
\newblock {A Global Geometric Framework for Nonlinear Dimensionality
  Reduction}.
\newblock {\em Science}, 290(5500):2319--2323, December 2000.

\bibitem{vandermaaten_08}
Laurens van~der Maaten and Geoffrey Hinton.
\newblock Visualizing data using t-sne.
\newblock {\em Journal of Machine Learning Research}, 9(86):2579--2605, 2008.

\bibitem{matplotlib_07}
J.~D. Hunter.
\newblock Matplotlib: A 2d graphics environment.
\newblock {\em Computing in Science \& Engineering}, 9(3):90--95, 2007.

\bibitem{getdist_19}
Antony Lewis.
\newblock {GetDist: a Python package for analysing Monte Carlo samples}.
\newblock {\em arXiv e-prints}, 2019.

\bibitem{pytorch_19}
Adam Paszke, Sam Gross, Francisco Massa, Adam Lerer, James Bradbury, Gregory
  Chanan, Trevor Killeen, Zeming Lin, Natalia Gimelshein, Luca Antiga, Alban
  Desmaison, Andreas Kopf, Edward Yang, Zachary DeVito, Martin Raison, Alykhan
  Tejani, Sasank Chilamkurthy, Benoit Steiner, Lu~Fang, Junjie Bai, and Soumith
  Chintala.
\newblock Pytorch: An imperative style, high-performance deep learning library.
\newblock In {\em Advances in Neural Information Processing Systems 32}, pages
  8024--8035. Curran Associates, Inc., 2019.

\bibitem{numpy_20}
Charles~R. Harris, K.~Jarrod Millman, St{\'{e}}fan~J. van~der Walt, Ralf
  Gommers, Pauli Virtanen, David Cournapeau, Eric Wieser, Julian Taylor,
  Sebastian Berg, Nathaniel~J. Smith, Robert Kern, Matti Picus, Stephan Hoyer,
  Marten~H. van Kerkwijk, Matthew Brett, Allan Haldane, Jaime~Fern{\'{a}}ndez
  del R{\'{i}}o, Mark Wiebe, Pearu Peterson, Pierre G{\'{e}}rard-Marchant,
  Kevin Sheppard, Tyler Reddy, Warren Weckesser, Hameer Abbasi, Christoph
  Gohlke, and Travis~E. Oliphant.
\newblock Array programming with {NumPy}.
\newblock {\em Nature}, 585(7825):357--362, September 2020.

\end{thebibliography}
\end{document}